\documentclass[11pt]{amsart}
\usepackage{latexsym,amsmath,amssymb,amscd,eucal}
\usepackage{xypic}

\newcommand{\newsection}{\setcounter{equation}{0}\section}

\newcommand{\mbf}[1]{{\boldsymbol {#1} }}

\def\appendix#1{\addtocounter{section}{1}\setcounter{equation}{0}
\renewcommand{\thesection}{\Alph{section}}
\section*{Appendix \thesection\protect\indent \parbox[t]{11.715cm} {#1}}
\addcontentsline{toc}{section}{Appendix \thesection\ \ \ #1} }
\newcommand{\eq}{\begin{equation}}
\newcommand{\eqend}{\end{equation}}

\newenvironment{romanlist}{%

        \begin{enumerate}
        }{%
        \end{enumerate}}
\newbox\ncintdbox \newbox\ncinttbox
\setbox0=\hbox{$-$} \setbox2=\hbox{$\displaystyle\int$}
\setbox\ncintdbox=\hbox{\rlap{\hbox to
\wd2{\hskip-.125em\box2\relax \hfil}}\box0\kern.1em}
\setbox0=\hbox{$\vcenter{\hrule width 4pt}$}
\setbox2=\hbox{$\textstyle\int$}
\setbox\ncinttbox=\hbox{\rlap{\hbox to
\wd2{\hskip-.175em\box2\relax \hfil}}\box0\kern.1em}

\newcommand{\complex}{{\mathbb C}} 
\newcommand{\zed}{{\mathbb Z}} 
\newcommand{\nat}{{\mathbb N}} 
\newcommand{\real}{{\mathbb R}} 
\newcommand{\reals}{{\mathbb R}} 
\newcommand{\rat}{{\mathbb Q}} 
\newcommand{\mat}{{\mathbb M}} 
\newcommand{\id}{{1\!\!1}} 

\def\hil{{\mathcal H}}
\def\hilR{{\mathcal H}_\real}

\def\pt{{\rm pt}}
\def\K{{\rm K}}
\def\H{{\rm H}}
\def\C{{\rm C}}

\def\B{{\mathbb{B}}}

\def\S{{\mathbb{S}}}
\def\P{{\mathbb{P}}}

\def\T{{\mathbb{T}}}

\def\U{{\rm U}}

\def\Ext{{\rm Ext}}
\def\Hom{{\rm Hom}}

\def\Cl{{{\rm C}\ell}}
\def\ch{{\rm ch}}

\def\Id{{\rm id}}
\def\pr{{\rm pr}}

\def\Aroof{{\widehat{A}}}
\def\ind{{\rm ind}}
\def\Fred{{\rm Fred}}

\def\e{{\,\rm e}\,}

\def\be{\begin{equation}}
\def\ee{\end{equation}}
\def\bea{\begin{eqnarray}}
\def\eea{\end{eqnarray}}
\def\bd{\begin{displaymath}}
\def\ed{\end{displaymath}}

\def\dd{{\rm d}}

\def\ii{{\,{\rm i}\,}}

\def\tor{{\rm tor}}

\makeatletter
\newdimen\normalarrayskip              
\newdimen\minarrayskip                 
\normalarrayskip\baselineskip
\minarrayskip\jot
\newif\ifold             \oldtrue            
\def\arraymode{\ifold\relax\else\displaystyle\fi} 
\def\@arrayskip{\ifold\baselineskip\z@\lineskip\z@
     \else
     \baselineskip\minarrayskip\lineskip2\minarrayskip\fi}
\def\@arrayclassz{\ifcase \@lastchclass \@acolampacol \or
\@ampacol \or \or \or \@addamp \or
   \@acolampacol \or \@firstampfalse \@acol \fi
\edef\@preamble{\@preamble
  \ifcase \@chnum
     \hfil$\relax\arraymode\@sharp$\hfil
     \or $\relax\arraymode\@sharp$\hfil
     \or \hfil$\relax\arraymode\@sharp$\fi}}
\def\@array[#1]#2{\setbox\@arstrutbox=\hbox{\vrule
     height\arraystretch \ht\strutbox
     depth\arraystretch \dp\strutbox
     width\z@}\@mkpream{#2}\edef\@preamble{\halign \noexpand\@halignto
\bgroup \tabskip\z@ \@arstrut \@preamble \tabskip\z@ \cr}%
\let\@startpbox\@@startpbox \let\@endpbox\@@endpbox
  \if #1t\vtop \else \if#1b\vbox \else \vcenter \fi\fi
  \bgroup \let\par\relax
  \let\@sharp##\let\protect\relax
  \@arrayskip\@preamble}
\makeatother

\newcommand{\beq}{\begin{eqnarray}}
\newcommand{\eeq}{\end{eqnarray}}

\newcommand{\G}{\Gamma}
\newcommand{\ta}{\tau}

\newcommand{\del}{\partial}

\newcommand{\KO}{{\rm KO}}

\newcommand{\KK}{{\rm KK}}
\newcommand{\KKO}{{\rm KKO}}
\newcommand{\Spin}{{\rm Spin}}
\newcommand{\SO}{{\rm SO}}
\newcommand{\UU}{{\rm U}}
\newcommand{\SU}{{\rm SU}}
\newcommand{\normx}{{\lVert x \rVert}}
\newcommand{\normy}{{\lVert y \rVert}}
\newcommand{\normxy}{{\lVert x\,y \rVert}}
\newcommand{\norm}{{\lVert 1 \rVert}}
\newcommand{\HR}{\mathcal{H}_{\mathbb{R}}}

\def\appendix#1{\addtocounter{section}{1}\setcounter{equation}{0}
\renewcommand{\thesection}{\Alph{section}}
\section*{Appendix \thesection. #1}
\addcontentsline{toc}{section}{Appendix \thesection\ \ \ #1} }

\newcommand{\dslash}{\not{\hbox{\kern-2pt $\partial$}}}
\newcommand{\pslash}{\not{\hbox{\kern-2.3pt $p$}}}

 \newtoks\nslashfraction
 \nslashfraction={.13}
 \newcommand{\nslash}[1]{\setbox0\hbox{$ #1 $}
   \setbox0\hbox to \the\nslashfraction\wd0{\hss \box0}/\box0 }

\def\ii{{\,{\rm i}\,}}

\newtheorem{theorem}{Theorem}[section]
\newtheorem{lemma}[theorem]{Lemma}
\newtheorem{cor}[theorem]{Corollary}
\newtheorem{proposition}[theorem]{Proposition}
\theoremstyle{definition}
\newtheorem{definition}[theorem]{Definition}
\theoremstyle{remark}
\newtheorem{example}[theorem]{Example}
\newtheorem{remark}[theorem]{Remark}

\numberwithin{equation}{section}

\begin{document}

\hfill{\small HWM--06--40 \ \ EMPG--06--09}

\vskip 1cm

\title[KO-Homology and Type I String Theory]
{KO-Homology and Type I String Theory}

\author{Rui M.G. Reis}
\address{Department of Mathematical Sciences, University of Aberdeen, King's College, Aberdeen AB24 3UE, U.K.}
\email{r.reis@abdn.ac.uk}
\author{Richard J. Szabo}
\address{Department of Mathematics and Maxwell Institute for
  Mathematical Sciences, Heriot-Watt University, Colin Maclaurin
  Building, Riccarton, Edinburgh EH14 4AS, U.K.}
\email{R.J.Szabo@ma.hw.ac.uk}
\author{Alessandro Valentino}
\address{Courant Research Center ``Higher Order Structures'' and Mathematisches Institut, Georg-August-Universit\"{a}t G\"{o}ttingen, Bunsenstr. 3-5, D-37073 G\"{o}ttingen, Germany}
\email{sandro@uni-math.gwdg.de}

\begin{abstract}
We study the classification of D-branes and Ramond-Ramond fields in
Type~I string theory by developing a geometric description of
KO-homology. We define an analytic version of KO-homology using
KK-theory of real $C^*$-algebras, and construct explicitly the
isomorphism between geometric and analytic KO-homology. The
construction involves recasting the $\Cl_n$-index theorem and a
certain geometric invariant into a homological framework which is used,
along with a definition of the real Chern character in KO-homology, to
derive cohomological index formulas. We show that this invariant also
naturally assigns torsion charges to non-BPS states in Type~I string
theory, in the construction of classes of D-branes in terms of
topological KO-cycles. The formalism naturally captures the
coupling of Ramond-Ramond fields to background D-branes which cancel
global anomalies in the string theory path integral. We show that this
is related to a physical interpretation of bivariant KK-theory in
terms of decay processes on spacetime-filling branes. We also
provide a construction of the holonomies of Ramond-Ramond fields in Type~II sting theory in terms of topological K-chains.
\end{abstract}

\maketitle

\newsection*{Introduction}

This paper continues the development and applications of the
topological classification of D-branes in string theory using
generalized homology theories. As explained
by~\cite{MM1,Witten,Horava1,FW1,MW1,FH1}, and reviewed
in~\cite{16,Bonora:2006ex,W}, D-brane charges and Ramond-Ramond fluxes
are necessarily classified by the K-theory of spacetime in order to
explain certain dynamical processes that cannot be
accounted for by ordinary cohomology theory alone. However, as
emphasized by~\cite{Periwal1,Matsuo1,13,Sz1,KHomM,Sz2}, a much more
natural description of D-branes is provided by K-homology which at the
analytic level links them to Fredholm modules and spectral
triples. This point of view was exploited in great detail in~\cite{RS}
to provide a rigorous geometric description of D-branes in Type~II
string theory using the Baum-Douglas construction of
K-homology~\cite{1,2}. In this paper we extend this description to
D-branes and Ramond-Ramond fields in Type~I string theory. The
classification using KO-theory is explored extensively
in~\cite{Witten,Berg1,OS1,16,MW1,Bergman:2000tm,Asakawa:2002nv,Mathai:2003jx}.
We use this and Jakob's approach~\cite{3} to construct a geometric
realization of KO-homology as the homology theory dual to KO-theory,
and describe various implications for the
classification of Type~I Ramond-Ramond charges and fluxes. As
in~\cite{RS}, we simplify our treatment by dealing only with
topologically trivial $B$-fields, and by ignoring the square-root of
the Atiyah-Hirzebruch genus which naturally appears in the
cohomological formula for D-brane
charge~\cite{MM1,16,Brodzki:2006fi}. Throughout we will compare and
contrast with the complex case of Type~II D-branes.

We will also develop the analytic description of KO-homology. We
define this using Kasparov's KK-theory for real $C^*$-algebras, which
also encompasses the analytic KR-homology theories appropriate to
D-branes in orientifold backgrounds. Generally, there is a description
of the KK-theory group $\KK(A,B)$ in terms of an additive
category whose objects are separable $C^*$-algebras and whose
morphisms $A\to B$ are precisely the elements of $\KK(A,B)$, with the
intersection product given by composition of morphisms. This
category may be viewed as a certain completion of the stable homotopy
category of separable $C^*$-algebras~\cite{12}. We use this
description to provide a physical interpretation of KK-theory in terms
of what we call ``generalized D9-brane decay'', which unifies the
description of charges in terms of tachyon condensation with the
description of fluxes in terms of holonomies over anomaly-cancelling
background D-branes. In particular, we find a certain bound state
obstruction to measuring the KO-theory class of a Ramond-Ramond field
analogous to that found recently in~\cite{Freed:2006ya}. Our physical
interpretation of Kasparov's theory is different from the proposal
of~\cite{KHomM,Asakawa:2002nv} (see also~\cite{Periwal1}) and is
better suited to the global constructions of D-branes that we
present. The use of KK-theory in string theory has also been exploited
in context of string and other dualities in~\cite{Brodzki:2006fi}.

One of the main technical achievements of this paper is an explicit,
detailed proof of the equivalence between the topological and analytic
definitions of KO-homology, an ingredient missing from the original
Baum-Douglas construction. In the course of working out the details,
we came across the unpublished recent preprint~\cite{BHS} in which a
proof is also given. While having some overlap with the present work,
our proof is fundamentally different. Our approach
is more tailored to the physical applications that we have in mind, as
it employs the construction of a certain geometric invariant which is
related later on to D-brane charges and Ramond-Ramond fluxes. This
invariant gives a rigorous definition to the $\zed_2$ Wilson lines
which are used in physical constructions of Type~I D-branes with
torsion charges through tachyon
condensation~\cite{Sen:1998tt,Witten,Bergman:2000tm}, and it is
related to the mod~2 index that appears in the phase of the Type~IIA
partition function~\cite{MW1,DMW1,Moore:2002cp}. It is also related to
the homological invariants that we construct is our description of fluxes
as holonomies over background D-branes. Mathematically, our technique
leads to a straightforward derivation of index formulas in the real
case, whose proof is also missing from~\cite{1,2} and which we provide
in detail here. On the other hand, in contrast to our approach, the
method of proof given in~\cite{BHS} has the virtue of being applicable
to a potentially wider class of generalized homology theories.

The first four sections of this paper present most of the
technical details of the construction of KO-homology and its
applications, a lot of which have not appeared in completeness anywhere
in the literature and contain mathematical results of independent
interest. Our exposition begins in Section~\ref{TKOH} with a
self-contained description of analytic KO-homology using Kasparov's
KK-theory for real $C^*$-algebras. Section~\ref{TKH} details a
Baum-Douglas type construction of geometric KO-homology. Using the
approach of~\cite{3} we prove that this theory is equivalent to the
usual definition provided by the spectrum of KO-theory, and
thereby establish that the geometric definition really is a
generalized homology theory. The content of
Section~\ref{isomorphism} is the crux of our mathematical results, the
detailed proof of the isomorphism between geometric and analytic
KO-homology. This is done by recasting the $\Cl_n$-index theorem into
a homological setting and thereby obtaining the associated homological
invariant. (This is where our proof differs from that of~\cite{BHS}.)
In Section~\ref{RealChern} we construct the Chern
character in KO-homology and use it to derive cohomological formulas
for the topological index (in the appropriate dimensionalities).

The final two sections of the paper turn to more physical
applications of the geometric KO-homology framework. It is well-known that the K-theory framework naturally accounts for certain properties of D-branes and Ramond-Ramond fields that would not be realized if these objects were classificed by ordinary cohomology or homology alone. For example, it explains the appearence of stable but non-BPS branes carrying torsion charges, and correctly incorporates both the self-duality and quantization conditions on Ramond-Ramond fields. It has also led to a variety of new predictions concerning the spectrum of superstring theory, such as the instability of D-branes wrapping non-contractible cycles in certain instances due to the fact that their cohomology classes do not ``lift'' to K-theory~\cite{DMW1}, and the obstruction to simultaneous measurement of electric and magnetic Ramond-Ramond fluxes when torsion fluxes are included~\cite{Freed:2006ya}. Moreover, certain properties of the string theory path integral, such as worldsheet anomalies and certain subtle phase factor contributions from the Ramond-Ramond fields, are most naturally formulated within the context of K-theory~\cite{DMW1,FW1}. With these considerations in mind, we illustrate what the formalism of KO-homology we have developed tells about the structure of D-branes and Ramond-Ramond fields in Type~I string theory, extending previous work which is mostly carried out in the Type~II setting.
In Section~\ref{TypeIBranes} we explain the virtues of classifying Type~I
D-branes within the homological framework, and adapt some of the
results of~\cite{DMW1,RS} concerning the stability of brane
constructions to the real case. The precise definitions involving
D-branes, along with further motivation and results from the geometric
K-homology formalism, can be found in~\cite{RS} and will not be
repeated here. We also examine in detail the problem of
constructing torsion D-branes. In the real case this turns out to be
much more involved than in the complex case, but we nevertheless
formally give the constructions using the invariant built in
Section~\ref{isomorphism} and suspension techniques. In the final
Section~\ref{Fluxes}, we demonstrate that the topological
classification of Ramond-Ramond fields in Type~II string theory is also
much more natural within the context of geometric K-homology. We show
that the pertinent differential K-theory group,
which normally classifies fluxes, naturally describes the
holonomies on background D-branes which are used to cancel the
topological anomaly in the string theory path integral. This relation
may be tied to the generalized D9-brane decay which lends a physical
interpretation to KK-theory. We then provide a construction of the
holonomies in terms of a geometric invariant defined on K-chains
representing the background D-branes, and describe some of their
properties.

\subsection*{Acknowledgments}

We thank G.~Landi and V.~Mathai for helpful discussions. We are
grateful to J.~Boersema for pointing out some errors in an earlier
version of this manuscript. This work was
supported in part by the EU-RTN Network Grant MRTN-CT-2004-005104. The
work of R.M.G.R. was supported in part by FCT grant
SFRH/BD/12268/2003. The work of A.V. was supported by the Mathematics Department at Heriot-Watt University, and in part by the German Research Foundation
(Deutsche Forschungsgemeinschaft (DFG)) through
the Institutional Strategy of the University of G\"{o}ttingen.

\newsection{Analytic KO-Homology\label{TKOH}}

In this section we will give a detailed overview of the
definition of KO-homology in terms of Kasparov's KK-theory for real
$C^*$-algebras~\cite{Kas81}, and describe various properties that we
will need in subsequent sections of this paper.

\subsection{Real $\mbf{C^*}$-Algebras}

We begin with an overview of the theory of real
$C^*$-algebras. The main references are~\cite{23,24}.

\begin{definition}
A \textit{real algebra} is a ring $A$ which is also an
$\mathbb{R}$-vector space such that $\lambda\,(x\,y)=(\lambda\,x)\,y=x
\,(\lambda\,y)$ for all $\lambda \in \real$ and all $x,y \in A$.
A \textit{real} $*$-\textit{algebra} is a real algebra $A$ equipped
with a linear involution $*: A \rightarrow A$ such that
$(x\,y)^*=y^*\,x^*$ for all $x,y \in A$.
A \textit{real Banach algebra} is a real algebra $A$ equipped with
a norm $\lVert - \rVert:A\to\real$ such that $\normxy \leq
\normx\,\normy$ and such that $A$ is complete in the norm topology.
If $A$ is a unital algebra then we assume $\norm=1$.
A \textit{real Banach} $*$-\textit{algebra} is a real Banach algebra
which is also a real $*$-algebra. A \textit{real} $C^*$-\textit{algebra}
is a real Banach $*$-algebra such that
\begin{romanlist}
\item The involution is an isometry, i.e. $\|x^*\|=\|x\|$ for all
  $x\in A$; and
\item $1+x^*\,x$ is invertible in $A$ for all $x \in A$.
\end{romanlist}
\end{definition}

\begin{remark}\label{r2}
Although in the complex case invertibility of $1+x^*\,x$ for all $x
\in A$ would follow immediately from the $C^*$-algebra structure, in
the real case this is no longer true. For example, consider the real
Banach $*$-algebra $\mathbb{C}$ with involution given by the identity
map. Then $1+\ii^*\,\ii$ is not invertible, where
$\ii:=\sqrt{-1}$. This invertibility condition is fundamental to
obtaining the usual representation theorem below for $C^*$-algebras in
terms of bounded self-adjoint operators on a real
Hilbert space. However, $\mathbb{C}$ with involution given by complex
conjugation is a real $C^*$-algebra. Since the only $\mathbb{R}$-linear
involutions of $\mathbb{C}$ are the identity and complex
conjugation, when we consider $\mathbb{C}$ as a real $C^*$-algebra the
involution will always be implicitly assumed to be complex
conjugation. More generally any complex $C^*$-algebra, regarded as a
real vector space and with the same operations, is a real
$C^*$-algebra.
\end{remark}

Let us now give a number of examples of real $C^*$-algebras, some of
which we will use later on in representation theorems.

\begin{example}\label{r3}
Let $\HR$ be a real Hilbert space. Then the set of bounded linear
operators $\mathcal{B}(\HR)$ with the usual operations is a real
$C^*$-algebra. Any closed self-adjoint subalgebra of $\mathcal{B}(\HR)$ is also
a real $C^*$-algebra. More generally, any closed self-adjoint subalgebra of
a real $C^*$-algebra is always a real $C^*$-algebra.
\end{example}

\begin{example}\label{r6}
Let $X$ be a locally compact Hausdorff space and $\C_0(X,\mathbb{R})$
the space of real-valued continuous functions vanishing at
infinity. Then $\C_0(X,\mathbb{R})$ with pointwise operations, the
supremum norm and involution given by the identity map is a real
$C^*$-algebra. As in the complex case, $\C_0(X,\mathbb{R})$ is unital
if and only if $X$ is compact.
\label{ex:r6}\end{example}

\begin{example}\label{r7}
With $X$ as in Example~\ref{ex:r6} above, let $Y$ be a closed subspace
of $X$ and $\C_0(X,Y;\mathbb{R})$ the subspace of $\C_0(X,\mathbb{C})$
consisting of maps $f : X \rightarrow \mathbb{C}$ such that $f(Y)
\subset\real$. Then with the operations inherited from
$\C_0(X,\mathbb{C})$, the subspace $\C_0(X,Y;\mathbb{R})$ is a real
$C^*$-algebra.
\end{example}

\begin{example}\label{r8}
Let $X$ be a locally compact Hausdorff space with involution $\tau
:X \rightarrow X$, i.e. a homeomorphism such that $\tau \circ \tau =
\textrm{id}_X$, and consider the subset $\C_0(X,\tau)$ of
$\C_0(X,\mathbb{C})$ consisting of maps $f$ such that $f \circ \tau
=f^*=\overline{f}$. Then $\C_0(X,\tau)$, with the operations inherited
from $\C_0(X,\mathbb{C})$, is a real $C^*$-algebra. If $\tau
=\textrm{id}_X$ then $\C_0(X,\tau)=\C_0(X,\mathbb{R})$.
If $X$ is compact and $Y$ is a closed subspace of $X$, then there is
a compact Hausdorff space $Z$ with an involution $\tau$ such that
$\C(X,Y;\mathbb{R}) \cong \C(Z,\tau)$. However, the converse does not
hold in general.
\end{example}

\begin{example}\label{r10}
Let $\mathcal{V}$ be a real vector space equipped with a quadratic form
\textit{Q}, and consider the associated real Clifford algebra
$\Cl(\mathcal{V},\textit{Q})$. Assume, without loss of generality,
that $\textit{Q}(v)=\langle v,\phi(v)\rangle$ for all $v \in \mathcal{V}$ with
respect to an inner product on $\mathcal{V}$, where the linear
operator $\phi \in \mathcal{L}(\mathcal{V})$ is symmetric and orthogonal.
We can then define an involution on $\Cl({\mathcal{V}},\textit{Q})$ by
$(v_1 \cdots v_k)^*=\phi(v_k) \cdots \phi(v_1)$, i.e. if $v \in {\mathcal{V}}$
then $v^*=\phi(v)$. The isomorphism $\Phi : \Cl({\mathcal{V}} \oplus
{\mathcal{V}},\textit{Q} \oplus -\textit{Q}) \rightarrow \mathcal{L}(\bigwedge
^\sharp{\mathcal{V}})$ induces a norm on $\Cl({\mathcal{V}},\textit{Q})$ by
pullback of the operator norm on $\mathcal{L}(\bigwedge
^\sharp{\mathcal{V}})$, and the inclusion
$\Cl({\mathcal{V}},\textit{Q})\hookrightarrow\Cl({\mathcal{V}},
\textit{Q})\otimes\Cl({\mathcal{V}},-\textit{Q})
\cong\Cl({\mathcal{V}} \oplus {\mathcal{V}},\textit{Q} \oplus
-\textit{Q})$ given by $x \mapsto x\otimes 1$ thereby induces a norm
on $\Cl({\mathcal{V}},\textit{Q})$. Then
$\Cl({\mathcal{V}},\textit{Q})$ with its algebra structure, this
involution and norm is a real $C^*$-algebra.
\end{example}

If $A$, $B$ are real $*$-algebras then a \textit{real} $*$-\textit{algebra
homomorphism} is a real algebra map $\phi : A \rightarrow B$, i.e. an
$\real$-linear ring homomorphism, such that $\phi (x^*)=\phi (x)^*$ for all $x
\in A$. The homomorphism is assumed to be unital if both algebras are
unital. We now come to the most general representation theorems for real
$C^*$-algebras. If $A$ is an algebra, we denote by $\mat_n(A)$ the
algebra of $n\times n$ matrices with entries in $A$.

\begin{theorem}
Let $A$ be a finite-dimensional real $C^*$-algebra. Then there exist
$k,n_1, \ldots, n_k \in \mathbb{N}$ such that $A \cong
\mat_{n_1}(A_1) \times \cdots \times \mat_{n_k}(A_k)$ as real
$C^*$-algebras with $A_1, \ldots, A_k \in
\{\mathbb{R},\mathbb{C},\mathbb{H}\}$.
\end{theorem}

\begin{proof}
Let $x \in A$. If $x^*\,x=x\,x^*$ and
$x^n=0$ for some $n \in \mathbb{N}$, then $x=0$. This implies that
$A$ has no non-zero nilpotent two-sided ideals.
Wedderburn's theorem on the representation of finite-dimensional
real algebras states that any real algebra with no non-zero
nilpotent two-sided ideals is isomorphic (as a real algebra) to a
finite direct product of $\mathbb{R}$-algebras of the form $\mat_k(D)$,
with $k \in \mathbb{N}$ and $D$ a finite-dimensional division algebra
over $\mathbb{R}$. The only finite-dimensional division
$\mathbb{R}$-algebras are $\mathbb{R}$, $\mathbb{C}$ and $\mathbb{H}$. The
direct product, with direct product operations, supremum norm $$\bigl\lVert
(a_{ij})\bigr\rVert =\sup_{i,j}\,\bigl\lVert a_{ij}\bigr\rVert_{A}$$
and involution $(a^{\phantom{*}}_{ij})^*=(a_{ji}^*)$, is a real
$C^*$-algebra. One then shows, as in the complex case, that these two
algebras are isomorphic as real $C^*$-algebras.
\end{proof}
\noindent
Analogously to the complex case, one also has the following result.
\begin{theorem}
Let $A$ be any real $C^*$-algebra. Then there exists a real Hilbert
space $\mathcal{H}_\mathbb{R}$ such that $A$ is isomorphic as a
real $C^*$-algebra to a closed self-adjoint subalgebra of
$\mathcal{B}(\HR)$.
\end{theorem}

Let $A$ be a real $C^*$-algebra. We denote by
$A_{\mathbb{C}}:=A\otimes\complex$ the complexification of $A$, which
is a complex algebra containing $A$ as a real algebra. We can define a
map $J_A : A_{\mathbb{C}} \rightarrow A_{\mathbb{C}}$ by $J_A(x+\ii
y)=x-\ii y$ for all $x,y \in A$. The map $J_A$ is a conjugate linear
$*$-isomorphism of the complex $C^*$-algebra $A_{\mathbb{C}}$. If
$\phi : A \rightarrow A$ is a continuous $*$-homomorphism, then the
map $J_A(\phi) : A_{\mathbb{C}} \rightarrow A_{\mathbb{C}}$ defined
by $J_A(\phi)(x+\ii y)=\phi(x)+\ii\phi(y)$ is a continuous
$*$-homomorphism such that $J_A \circ J_A(\phi) = J_A(\phi)\circ
J_A$. Conversely, if $J$ is a conjugate linear $*$-isomorphism of a
complex $C^*$-algebra $B$, then $A= \{ x \in B \, | \, J(x)=x \}$ is
a real $C^*$-algebra. This implies the following result.

\begin{proposition}
Let $\mathcal{C}^*_{\mathbb{R}}$ be the category of real $C^*$-algebras
and continuous $*$-algebra homomorphisms. Let
$\mathcal{C}^*_{\mathbb{C},{\rm cl}}$ be the category of pairs $(A,J)$,
where $A$ is a complex $C^*$-algebra and $J$ is a conjugate linear
$*$-isomorphism of $A$, and continuous $*$-homomorphisms commuting with
$J$. Then the assignments $A \mapsto (A_{\mathbb{C}},J_A)$,
$\phi \mapsto J_A(\phi)$ define a functor
$$\mathcal{J} \,:\, \mathcal{C}^*_{\mathbb{R}} ~\longrightarrow~
\mathcal{C}^*_{\mathbb{C},{\rm cl}}$$
which is an equivalence of categories.
\end{proposition}

\subsection{Commutative Real $\mbf{C^*}$-Algebras}

We will now specialize to the case of commutative algebras.
As with complex Banach algebras, a maximal two-sided ideal in a real
Banach algebra $A$ is closed in $A$. If $M$ is a maximal two-sided
ideal of a real Banach algebra $A$, then $A/M$ is isomorphic to one of
$\mathbb{R}$ or $\mathbb{C}$ as real algebras.
A \textit{character} on a real algebra $A$ is a non-zero real
algebra map $\chi : A \rightarrow \mathbb{C}$, assumed unital if $A$
is unital. Let $\Omega_A$ be the space of characters of
$A$. This can be given, as in the complex case, a locally compact
Hausdorff space topology such that $\Omega_A$ is homeomorphic to
$\Omega_{A_\mathbb{C}}$. Furthermore, $A$ is unital if and only if
$\Omega_A$ is compact.

Given $x \in A$, evaluation at $x$ gives a continuous map $
\Gamma(x) : \Omega_A \rightarrow \mathbb{C}$ called the
\textit{Gel'fand transform of $x$}. From this we obtain the
\textit{Gel'fand transform of $A$}, $\Gamma : A \rightarrow
\C_0(\Omega_A,\mathbb{C})$, which is a continuous real algebra
homomorphism of unit norm. If $A$ is a real $*$-algebra, then $\Gamma$
is a $*$-algebra homomorphism. The most important results on the
representation of commutative real $C^*$-algebras are the following.

\begin{theorem}
Let $A$ be a commutative real $C^*$-algebra. Then:
\begin{romanlist}
\item The map $\tau : \Omega_A
\rightarrow \Omega_A$ defined by $\tau(\chi)=\overline{\chi}$ is an
involution; and
\item The Gel'fand transform $\Gamma : A \rightarrow
  \C_0(\Omega_A,\tau)$ is a real $C^*$-algebra isomorphism.
\end{romanlist}
\end{theorem}

\begin{proof}
(i) The map $\tau$ is a bijection. The collection of sets
$$U_{x,V}= \bigl\{\chi\in \Omega_A ~ | ~ \chi(x) \in V  \bigr\}$$
for every $x \in A$ and $V$ open in $\mathbb{C}$ is a sub-basis for
the topology of $\Omega_A$. The complex conjugate $\overline{V}$ of
$V$ is an open set and $\tau^{-1}(U_{x,V})=U_{x,\overline{V}}$. Thus
$\tau$ is continuous.

\noindent
(ii) The map $\Gamma$ is a real $*$-algebra map with $\lVert \Gamma(x)
\rVert =\normx$. One also has \bea\G(x) \circ
\ta(\chi)&=&\G(x)(\,\overline{\chi}\,)\nonumber\\ &=&
\overline{\chi(x)}~=~\G(x)^*(\chi) \ , \nonumber\eea and so
$\G(x) \circ \ta=\G(x)^*$ and $\G(A) \subset \C_0(\Omega_A,\ta)$.
Let $\theta : A \rightarrow A_\mathbb{C}$ be the $C^*$-algebra
embedding of $A$ into its complexification. The map $\vartheta :
\Omega_{A_\mathbb{C}} \rightarrow\Omega_A$ given by $\vartheta(f)=f \circ
\theta$ is a homeomorphism and there is a commutative diagram
$$\xymatrix{ A \ar[r]^{\!\!\!\!\!\!\!\!\!\!\!\!\G}
\ar[d]_{\theta} & \C_0(\Omega_A,\mathbb{C}) \ar[d]^{\vartheta^*}\\
             A_\mathbb{C} \ar[r]_{\!\!\!\!\!\!\!\!\!\!\!\!\G}
&  \C_0(\Omega_{A_\mathbb{C}},\mathbb{C}) \ .}$$
Using this one then shows that $\G(A)=\C_0(\Omega_A,\ta)$.
\end{proof}

\begin{cor}
Let $A$ be a commutative real $C^*$-algebra with trivial
involution. Then $A$ is $*$-isomorphic to $\C_0(\Omega_A,\mathbb{R})$.
\end{cor}

\subsection{Hilbert Modules}

We will now start presenting an overview of KK-theory for real
$C^*$-algebras. The basic references are \cite{25,12}.
We begin by generalizing the notion of Hilbert space.

\begin{definition}\label{HM}
Let $A$ be a (not necessarily commutative) real $C^*$-algebra. A
$\textit{pre-Hilbert module}$ \textit{over} $A$  is a (right)
$A$-module $\mathcal{E}$
equipped with an $A$-$\textit{valued inner product}$, i.e. a
bilinear map $(-, -) : \mathcal{E} \times \mathcal{E}
\rightarrow A$ such that
\begin{romanlist}
\item $(x,x) \geq 0$ for all $x \in \mathcal{E}$ and $(x,x)=0$ if and
  only if $x=0$;
\item $(x,y)=(y,x)^*$ for all $x,y \in \mathcal{E}$; and
\item $(x,y\,a)=(x,y)\,a$ for all $x,y \in \mathcal{E}, \, a \in A$.
\end{romanlist}
For $x \in \mathcal{E}$ we define $\normx_{\mathcal{E}}:= \lVert (x,x)
\rVert ^{1/2}$. This
defines a norm on ${\mathcal{E}}$ satisfying the Cauchy-Schwartz
inequality. If ${\mathcal{E}}$ is complete under this norm, then it is
called a \textit{Hilbert module over}~$A$.

\end{definition}

We can define tensor products of $C^*$-algebras and Hilbert modules in
the usual way (see \cite{12,25} for the constructions).
If ${\mathcal{E}}$ is a pre-Hilbert module over the real $C^*$-algebra $A$, we
assume that the complexification ${\mathcal{E}} \otimes \mathbb{C}$ is a
pre-Hilbert module over $A_{\mathbb{C}}$. This means that the
$A$-valued inner product extends to a sesquilinear map. We assume
that sesquilinear maps are linear in the second variable.

Let ${\mathcal{E}},{\mathcal{F}}$ be Hilbert $A$-modules and $T :
{\mathcal{E}} \rightarrow {\mathcal{F}}$ an
$A$-linear map. We call a map $T^* : {\mathcal{F}} \rightarrow
{\mathcal{E}}$ such that
$(Tx,y)_{\mathcal{F}}=(x,T^*y)_{\mathcal{E}}$ for all $x \in
{\mathcal{E}}, \, y \in {\mathcal{F}}$ the
\textit{adjoint} of $T$. If it exists the adjoint is unique
by Definition~\ref{HM}~(i). Not every $A$-linear map between Hilbert
$A$-modules has an adjoint. We denote the set
of all $A$-linear maps $T : {\mathcal{E}} \rightarrow {\mathcal{F}}$
admitting an adjoint by
$\mathcal{L}({\mathcal{E}},{\mathcal{F}})$.
Elements of $\mathcal{L}({\mathcal{E}},{\mathcal{F}})$ are bounded
$A$-linear maps and
$\mathcal{L}({\mathcal{E}}):=\mathcal{L}({\mathcal{E}},{\mathcal{E}})$
is a $C^*$-algebra with the operator
norm and involution given by the adjoint. Given $x \in {\mathcal{F}}, \, y \in
{\mathcal{E}}$ we define an operator $\theta_{x,y} \in
\mathcal{L}({\mathcal{E}},{\mathcal{F}})$ by
$\theta_{x,y}(z)=x\,(y,z)_{\mathcal{E}}$. These operators generate an
$\mathcal{L}({\mathcal{E}})-\mathcal{L}({\mathcal{F}})$-bimodule whose
norm closure in $\mathcal{L}({\mathcal{E}},{\mathcal{F}})$ is denoted
$\mathcal{K}({\mathcal{E}},{\mathcal{F}})$. Elements of
$\mathcal{K}({\mathcal{E}},{\mathcal{F}})$ are called
$\textit{generalized compact operators}$.

If ${\mathcal{E}}=\mathcal{H}_\real$ is a real Hilbert space, then
$\mathcal{L}({\mathcal{E}})$ is the usual space of bounded linear
operators and $\mathcal{K}({\mathcal{E}})$ is
the space of compact operators. If $n \in \mathbb{N} \cup \{\infty\}$,
then $A^n$ with inner product
$$(\vec x,\vec y\,):=\sum_{i=1}^n\,x_i^*\,y^{\phantom{*}}_i$$ for all
$\vec x=(x_i)_{1 \leq i \leq n}$, $\vec y=(y_i)_{1 \leq i \leq n}$ is a
Hilbert module. One has $\mathcal{K}(A) \cong A$ and
$\mathcal{K}(A^\infty) \cong A \otimes
\mathcal{K}_{\mathbb{R}}$ where
$\mathcal{K}_{\mathbb{R}}:=\mathcal{K}(\mathcal{H}_{\mathbb{R}})$.

\begin{definition}
Let $A$ be a real $C^*$-algebra. The \textit{multiplier algebra} of
$A$, $\textsf{M}(A)$, is the maximal $C^*$-algebra
containing $A$ as an essential ideal. Equivalently, by representing $A
\subset \mathcal{L}(\mathcal{H}_\real)$ one has
$$\textsf{M}(A)= \{ \, T \in \mathcal{L}(\mathcal{H}_\real) ~|~ T\,S,
S\,T \in A \quad \textrm{for all} ~ S \in A \} \ . $$
\end{definition}
\noindent
The multiplier algebra $\textsf{M}(A)$ is a $C^*$-algebra which is
$*$-isomorphic to the $C^*$-algebra of double centralizers, i.e. pairs
$(T_1,T_2) \in \mathcal{L}(A) \times \mathcal{L}(A)$ such
that $a\,T_1(b)=T_2(a)\,b$, $T_1(a\,b)=T_1(a)\,b$ and
$T_2(a\,b)=a\,T_2(b)$ for all $a,b \in A$. If $A$ is unital, then
$\textsf{M}(A)=A$. Furthermore, $\textsf{M}(\mathcal{K}_{\mathbb{R}})=
\mathcal{L}(\mathcal{H}_{\mathbb{R}})$, and
$\textsf{M}(\C_0(X,\mathbb{R}))=\C_{\rm b}(X,\mathbb{R})$ is the
$C^*$-algebra of real-valued bounded continuous functions on a locally
compact Hausdorff space $X$.

\begin{proposition}
Let ${\mathcal{E}}$ be a Hilbert $A$-module. Then there is an
isomorphism $$\mathcal{L}\bigl({\mathcal{E}}\bigr)\cong
{\sf M}\bigl(\mathcal{K}({\mathcal{E}})\bigr) \ . $$
\end{proposition}

\subsection{KKO-Theory \label{KKOTheory}}

We will now define the KKO-theory groups using Kasparov's
approach~\cite{Kas81}. A useful survey of Kasparov's theory can be
found in~\cite{Hig}. We assume that a real $C^*$-algebra $A$ is
separable and a real $C^*$-algebra $B$ is $\sigma$-unital.

\begin{definition}
A \textit{(Kasparov)} $(A,B)$\textit{-module} is a triple
$({\mathcal{E}},\rho,T)$, where ${\mathcal{E}}$ is a countably
generated Hilbert $B$-module, $\rho : A \rightarrow
\mathcal{L}({\mathcal{E}})$ is a $*$-homomorphism and $T \in
\mathcal{L}({\mathcal{E}})$ such that
\beq
\left(T-T^*\right)\rho(a)~, \, \left(T^2-1
\right)\rho(a)~, \, \bigl[T\,,\,\rho(a)\bigr] \in
\mathcal{K}({\mathcal{E}})
\label{Kmoduledef}\eeq
for all $a \in A$. A Kasparov module $({\mathcal{E}},\rho,T)$ is called {\it
  degenerate} if all operators in (\ref{Kmoduledef}) are zero.
Two Kasparov modules $({\mathcal{E}}_i,\rho_i,T_i)$, $i=1,2$ are said to be
\textit{orthogonally equivalent} if there is
an isometric isomorphism $U \in
\mathcal{L}({\mathcal{E}}_1,{\mathcal{E}}_2)$ such that
$T_1=U^*\,T_2\,U$ and $\rho_1(a)=U^*\,\rho_2(a)\,U$ for all $a \in
A$.
\end{definition}

Orthogonal equivalence is an equivalence relation on the set of
Kasparov modules. We denote the set of equivalence classes by
${\sf E}(A,B)$. The subset containing degenerate modules is denoted
${\sf D}(A,B)$. Direct sum makes ${\sf E}(A,B)$ and
${\sf D}(A,B)$ into monoids.

\begin{definition}
Let $({\mathcal{E}}_i,\rho_i,T_i) \in {\sf E}(A,B)$ for
$i=0,1$, $({\mathcal{E}},\rho,T) \in
{\sf E}(A,B\otimes \C([0,1],\real))$, and let $f_t :
B\otimes \C([0,1],\real) \rightarrow B$ be the evaluation map
$f_t(g)=g(t)$. Then $({\mathcal{E}}_0,\rho_0,T_0)$ and
$({\mathcal{E}}_1,\rho_1,T_1)$ are
said to be \textit{homotopic} and $({\mathcal{E}},\rho,T)$ is called a
\textit{homotopy} if $({\mathcal{E}} \otimes_{f_i}B,f_i \circ \rho,f_{i*}(T))$
is orthogonally equivalent to $({\mathcal{E}}_i,\rho_i,T_i)$ for $i=0,1$, where
$f_{i*}(T)(a):=f_i(T(a))$.
\label{homotopydef}\end{definition}
\noindent
Homotopy is an equivalence relation on ${\sf E}(A,B)$ and we denote
the equivalence classes by $[{\mathcal{E}},\rho,T]$.
It is useful to consider special kinds of homotopy. If
${\mathcal{E}}=\C([0,1],{\mathcal{E}}_0)$,
${\mathcal{E}}_0={\mathcal{E}}_1$ and the induced maps $t\mapsto T_t,
\, t\mapsto \rho_t(a)$ for all $a \in A$ are strongly $*$-continuous,
then we call $({\mathcal{E}},\rho,T)$ a \textit{standard homotopy}. If
in addition $\rho_t=\rho$ is constant and $T_t$ is norm continuous, then
$({\mathcal{E}},\rho,T)$ is called an \textit{operator homotopy}. Any
degenerate module is homotopic to the zero module. The quotient
$\mathcal{Q}({\mathcal{E}}):={\mathcal{L}({\mathcal{E}})}/
{\mathcal{K}({\mathcal{E}})}$ is a
generalization of the Calkin algebra. If
$\rho(a)\,[T_1,T_2]\,\rho(a)^* \geq 0$ in $\mathcal{Q}({\mathcal{E}})$, then
$({\mathcal{E}},\rho,T_1)$ and $({\mathcal{E}},\rho,T_2)$ are operator
homotopic.

\begin{definition}
The set of equivalence classes in ${\sf E}(A,B)$ with respect to
homotopy of $(A,B)$-modules is denoted $\KKO(A,B)$ or
$\KKO_0(A,B)$. For $p,q \geq 0$ we define $$\KKO_{p,q}(A,B)=\KKO(A,B
\otimes \Cl_{p,q}) \ , $$ where $\Cl_{p,q}:=\Cl(\real^{p,q})$ is the
real Clifford algebra of the vector space $\real^{p+q}$ with quadratic
form of signature $(p,q)$.
\label{KKOgendef}\end{definition}
\noindent
The equivalence relation allows us to simplify the $(A,B)$-modules
required to define $\KKO(A,B)$. We need only consider modules
of the form $(B^\infty,\rho,T)$ with $T=T^*$. If $A$ is unital,
we can further assume that $\lVert T \rVert \leq 1$ and
$T^2-1 \in \mathcal{K}(B^\infty)$.

There is another equivalence relation that we can define on
${\sf E}(A,B)$. We say that two $(A,B)$-modules $({\mathcal{E}}_i,\rho_i,T_i)$,
$i=0,1$ are \textit{stably operator homotopic},
$({\mathcal{E}}_0,\rho_0,T_0)\simeq_{\rm
  oh}({\mathcal{E}}_1,\rho_1,T_1)$, if there exist
$({\mathcal{E}}_i^{\prime},\rho_i^{\prime},T_i^{\prime})\in{\sf D}(A,B)$ such that
$({\mathcal{E}}^{\phantom{\prime}}_0 \oplus
{\mathcal{E}}_0^{\prime},\rho^{\phantom{\prime}}_0 \oplus
\rho_0^{\prime},T^{\phantom{\prime}}_0 \oplus T_0^{\prime})$ and
$({\mathcal{E}}^{\phantom{\prime}}_1 \oplus
{\mathcal{E}}_1^{\prime},\rho^{\phantom{\prime}}_1
\oplus \rho_1^{\prime},T^{\phantom{\prime}}_1 \oplus T_1^{\prime})$
are operator homotopic up to orthogonal equivalence. The set of
equivalence classes with respect to $\simeq _{\rm oh}$ coincides with
the set $\KKO(A,B)$ defined above.

\begin{proposition}
The set $\KKO(A,B)$ enjoys the following properties:
\begin{romanlist}
\item $\KKO(A,B)$ is an abelian group.
\item $\KKO(-,-)$ is a covariant bifunctor from the category of
  separable $C^*$-algebras into the category of abelian groups which
  is additive:
\bea
\KKO(A_1 \oplus A_2,B)&=&\KKO(A_1,B) \oplus \KKO(A_2,B) \ ,
\nonumber\\ \KKO(A,B_1 \oplus B_2)&=&\KKO(A,B_1) \oplus \KKO(A,B_2) \
. \nonumber
\eea
\item Any two $*$-homomorphisms $f:A_2 \rightarrow A_1$ and $g:B_1 \rightarrow
B_2$ induce group homomorphisms
\bea
f^*\,:\,\KKO(A_1,B) &\longrightarrow& \KKO(A_2,B) \ , \nonumber\\
g_*\,:\, \KKO(A,B_1) &\longrightarrow& \KKO(A,B_2) \nonumber
\eea
defined by
\bea
f^*[{\mathcal{E}},\rho,T]&=&[{\mathcal{E}},\rho \circ f,T] \ , \nonumber\\
g_*[{\mathcal{E}},\rho,T]&=&[{\mathcal{E}} \otimes _g B_2,\rho \otimes
1,T \otimes 1] \ ; \nonumber
\eea
and
\item Any two homotopies $f_t: A_2 \rightarrow A_1$ and $g_t: B_1 \rightarrow
B_2$ induce the same homomorphism for all $t\in[0,1]$,
i.e. $f_t^*=f_0^*$ and $g_{t*}=g_{0*}$.
\end{romanlist}
\end{proposition}
\noindent If we assume $B$ unital, then we can identify
$\mathcal{L}(B^\infty) \cong \mat_2(\textsf{M}(B\otimes
\mathcal{K}_\real))$ where $\textsf{M}(B\otimes \mathcal{K}_\real)$
is the multiplier algebra of $B\otimes \mathcal{K}_\real$. Thus we
can give $\rho$ and $T$ the form
$$\rho=\left(\begin{array}{cc}
                               \rho_0 & 0 \\
                               0 &  \rho_1\\
                             \end{array}
                           \right) \ , \quad
                           T=\left(
                             \begin{array}{cc}
                               0 & T^{\prime\,*} \\
                               T^{\prime} & 0 \\
                             \end{array}
                           \right)
                           $$
with $\rho_0(a),\rho_1(a),T^{\prime}
\in{\sf M}(B\otimes \mathcal{K}_\real) \cong
\mathcal{L}(B^\infty)$, $\lVert\, T'\, \rVert \leq 1$, and
$$T^{\prime\,*}\,T^{\prime}-1~, \, T^{\prime}\,T^{\prime\,*}-1~, \,
T^{\prime}\,\rho_1(a)-\rho_0(a)\,T^{\prime}
\in B\otimes \mathcal{K}_\real$$
for all $a\in A$.

\subsection{Analytic KO-Homology\label{AnKO-H}}

Specializing all of our constructions to the case $A=\mathbb{R}$ and
$B$ unital we get the KO-theory groups
$\KKO(\mathbb{R},B)\cong\KO_0(B)$ and $\KKO_{p,q}(\mathbb{R},B)
\cong \KO_{p-q}(B)$. In particular,
$\KKO(\mathbb{R},\C(X,\mathbb{R}))\cong\KO_0(\C(X,\mathbb{R})) \cong
\KO^0(X)$ for any compact Hausdorff space $X$.
On the other hand, using the Gel'fand transform the contravariant functor
$(X,\ta) \mapsto \C(X,\ta)$ induces an equivalence of categories
between the category of compact Hausdorff spaces with involution and
the category of commutative real $C^*$-algebras. Since
$\KKO_\sharp(-,\mathbb{R})$ is also a contravariant functor, it
follows that their composition $(X,\ta)
\mapsto\KKO_\sharp(\C(X,\ta),\mathbb{R})$ is a covariant functor.

\begin{definition}
Let $(X,\ta)$ be a compact Hausdorff space with involution. The
\textit{analytic KO-homology groups of $(X,\ta)$} are defined by
$$\KO_n^{\rm
  a}\bigl(X\,,\,\ta\bigr)=\KKO_{n,0}\bigl(\C(X,\ta)\,,\,\mathbb{R}
\bigr)=\KKO\bigl(\C(X,\tau)\,,\,\Cl_n\bigr) $$
where $\Cl_n:=\Cl_{n,0}=\Cl(\real^n)$.
\end{definition}

It will be helpful in some of our later analysis to have a closer
look at our definition of
$\KO_n(A)=\KKO_{n,0}(A,\mathbb{R})=\KKO(A,\Cl_{n})$, the KO-homology
of a real $C^{*}$-algebra $A$. Following through the definitions, this
is based on triples $({\mathcal{H}_\real},\rho,T)$ which are defined
by the data:
\begin{romanlist}
\item ${\mathcal{H}_\real}$ is a separable real Hilbert space;
\item $\rho:A\to\mathcal{L}({\mathcal{H}_\real})$ is a unital
  representation of $A$; and
\item $T$ is a bounded linear operator on ${\mathcal{H}_\real}$.
\end{romanlist}
These are assumed to satisfy the following conditions:
\begin{romanlist}
\item ${\mathcal{H}_\real}$ is equipped with a
  $\mathbb{Z}_{2}$-grading such that $\rho(a)$ is even for all $a\in
  A$ and $T$ is odd;
\item For all $a\in A$ one has
\beq
\left(T^{2}-1\right)\,\rho(a)~,~
  \left(T-T^{*}\right)\,\rho(a)~, ~
T\,\rho(a)-\rho(a)\,T \in \mathcal{K}_\real \ ;
\label{TrhoKOarels}\eeq
and
\item There are odd $\real$-linear operators
  $\varepsilon_{1},\dots,\varepsilon_{n}$ on ${\mathcal{H}_\real}$
  with the $\Cl_n$ algebra relations
\beq
\varepsilon^{\phantom{*}}_{i}=\varepsilon_{i}^{*} \ ,
\quad \varepsilon^{2}_{i}=-1 \ , \quad
\varepsilon^{\phantom{*}}_{i}\,\varepsilon^{\phantom{*}}_{j}+
\varepsilon^{\phantom{*}}_{j}\,\varepsilon^{\phantom{*}}_{i}=0
\label{Clnrels}\eeq
for $i\neq j$ such that $T$ and $\rho(a)$ commute with each
$\varepsilon_{i}$.
\end{romanlist}
\noindent
{}From (\ref{TrhoKOarels}) it follows that $T$ may be taken to be a
Fredholm operator without loss of generality (see~\cite{SG}, Lemma
5.1), and we shall refer to the triple $(\mathcal{H}_\real,\rho,T)$ as
an \textit{$n$-graded Fredholm module}.

Let us denote by $\Gamma{\rm O}_{n}(A)$ the set of all $n$-graded
Fredholm modules over $A$. Consider the equivalence relation $\sim$ on
$\Gamma{\rm O}_{n}(A)$ generated by the relations:
\begin{itemize}
\item[]\emph{Orthogonal equivalence}:
  $({\mathcal{H}_\real^{\phantom{\prime}}},\rho,T)\sim({\mathcal{H}_\real'},
\rho',T'\,)$ if and only if there exists an isometric
degree-preserving linear operator
$U:{\mathcal{H}^{\phantom{\prime}}_\real}\to{{\mathcal{H}_\real'}}$
such that $U\,\rho(a)=\rho'(a)\,U$ for all $a\in A$, $U\,T=T'\,U$, and
$U\,\varepsilon^{\phantom{\prime}}_{i}=\varepsilon'_{i}\,U$; and
\item[]\emph{Homotopy equivalence}:
  $({\mathcal{H}_\real},\rho,T)\sim({\mathcal{H}_\real},\rho,T'\,)$ if
  and only if there exists a norm continuous function
$t\mapsto{T_{t}}$ such that $({\mathcal{H}_\real},\rho,T_{t})$ is a
Fredholm module for all $t\in[0,1]$ with $T_{0}=T$, $T_{1}=T'$.
\end{itemize}
We define the {\it direct sum} of two Fredholm modules
$({\mathcal{H}^{\phantom{\prime}}_\real},\rho,T)$ and
$({\mathcal{H}_\real'},\rho',T'\,)$ to be the Fredholm module
$({\mathcal{H}^{\phantom{\prime}}_\real}\oplus{{\mathcal{H}_\real'}},
\rho\oplus\rho',T\oplus{T'}\,)$.

We may now define $\KO_{n}(A)$ as the free abelian group generated
by elements in $\Gamma{\rm O}_{n}(A)/{\sim}$ and quotiented by the ideal
generated by the set $\{[x_{0}\oplus{x_{1}}]-[x_{0}]-[x_{1}] \, | \,
[x_{0}],[x_{1}]\in\Gamma{\rm O}_{n}(A)/{\sim} \}$. In $\KO_{n}(A)$ the
\emph{inverse} of a class represented by the module
$({\mathcal{H}^{\phantom{\rm o}}_\real},\rho,T)$ is given by
$({\mathcal{H}^{\rm o}_\real},\rho,T)$, where ${\mathcal{H}^{\rm o}_\real}$
is the Hilbert space ${\mathcal{H}^{\phantom{\rm o}}_\real}$ with the opposite
$\zed_{2}$-grading and where the operators $\varepsilon_{i}$ reverse
their signs. For a compact Hausdorff space $X$ we define
\begin{displaymath}
\KO_{n}^{\rm a}\bigl(X\bigr):={\KO_{n}\bigl(\C(X,\real)\bigr)}=
\KKO\bigl(\C(X,\real)\,,\,\Cl_n\bigr) \ .
\end{displaymath}
Of course, this construction is exactly the one given before, only
spelled out in more detail here. For further details and properties of
this construction in the complex case, see~\cite{BHS}.

\subsection{The Intersection Product\label{IntProd}}

Let $D$ be a real $C^*$-algebra. Then there is a natural homomorphism $$\ta
_D \,:\, \KKO(A,B) ~\longrightarrow~ \KKO(A \otimes D,B \otimes D)$$
defined by $$\ta
_D[B^\infty,\rho,T]= [B^\infty \otimes D,\rho \otimes
1,T \otimes 1] \ . $$
We can define in KKO-theory a product
$$\otimes_D \,:\, \KKO(A,D)\times \KKO(B,D) ~\longrightarrow~
\KKO(A,B)$$ called the {\it intersection product} by
$$[{\mathcal{E}}_1,\rho_1,T_1]\otimes_D [{\mathcal{E}}_2,\rho_2,T_2]
=[{\mathcal{E}}_1\otimes_{\rho_2}{\mathcal{E}}_2,\rho_1
\otimes_{\rho_2}1,T_1\,\# \, T_2] \ , $$
where $T_1\,\# \, T_2 \in
\mathcal{L}({\mathcal{E}}_1\otimes_{\rho_2}{\mathcal{E}}_2)$ is a
suitably defined operator~\cite{Hig}. If all $C^*$-algebras involved are
separable, then the intersection product extends to a bilinear map
$$\otimes_D \,:\, \KKO(A_1,B_1 \otimes D)\times \KKO(D \otimes
A_2,B_2)~\longrightarrow~ \KKO(A_1 \otimes A_2,B_1 \otimes B_2)$$
given by
$$x \otimes_D y= \ta_{A_2}(x) \otimes_{B_1 \otimes D \otimes
  A_2}\ta_{B_1}(y)$$
for all $(x,y)$.
\begin{proposition}
Let $A$ be a separable $C^*$-algebra and $B, \, D_1, \, D_2$
$\sigma$-unital algebras. Suppose there exist $\alpha \in \KKO(D_1,D_2)$
and $\beta \in \KKO(D_2,D_1)$ with $\alpha \otimes_{D_2} \beta =
1_{D_1}$ and $\beta \otimes _{D_1} \alpha = 1_{D_2}$. Then there are
isomorphisms
\bea
\otimes _{D_1}\alpha \,:\, \KKO(A,B\otimes D_1) &\longrightarrow&
 \KKO(A,B\otimes D_2) \ , \nonumber\\
\otimes _{D_2}\beta \,:\, \KKO(A,B\otimes D_2) &\longrightarrow&
\KKO(A,B\otimes D_1) \ . \nonumber \eea
 If $D_1, \,D_2 $ are separable, then one has isomorphisms
\bea
\alpha \otimes _{D_2} \,:\, \KKO(A\otimes D_2,B) &\longrightarrow&
 \KKO(A\otimes D_1,B) \ , \nonumber\\
\beta  \otimes _{D_1} \,:\, \KKO(A\otimes D_1,B) &\longrightarrow&
\KKO(A\otimes D_2,B) \ . \nonumber\eea
If in addition there exist $\alpha' \in \KKO(D_1
\otimes D_2,\mathbb{R})$ and $\beta' \in\KKO(\mathbb{R},D_1 \otimes D_2)$
such that $\beta' \otimes _{D_1} \alpha'= 1_{D_2}$ and $\beta' \otimes
_{D_2}\alpha' = 1_{D_1}$, then there are isomorphisms
\bea
\otimes _{D_1}\alpha' \,:\, \KKO(A,B\otimes D_1) &\longrightarrow&
 \KKO(A\otimes D_2,B) \ , \nonumber\\
\otimes _{D_2}\alpha' \,:\, \KKO(A,B\otimes D_2) &\longrightarrow&
\KKO(A\otimes D_1,B) \ , \nonumber\\
\beta'  \otimes _{D_1} \,:\, \KKO(A\otimes D_1,B)
&\longrightarrow& \KKO(A,B\otimes D_2) \ , \nonumber\\
\beta'  \otimes _{D_2} \,:\,
\KKO(A\otimes D_2,B) &\longrightarrow& \KKO(A,B\otimes D_1) \ .
\nonumber\eea
\label{intprodprop}\end{proposition}
The last result in Proposition~\ref{intprodprop} allows us to conclude
that the KKO-groups are stable, i.e. there are isomorphisms
$$\KKO(A\otimes \mathcal{K}_\real,B) \cong \KKO(A,B) \cong
\KKO(A,B\otimes \mathcal{K}_\real) \ . $$
One also has the isomorphisms
$$\KKO(A \otimes \Cl_{p,q},B\otimes \Cl_{r,s}) \cong \KKO(A \otimes
\Cl_{p,q} \otimes \Cl_{r,s},B) \cong \KKO(A \otimes
\Cl_{p-q+s-r,0},B)$$ along with symmetric
isomorphisms. Since $\KKO_n(\mathbb{R},A)$ is the operator algebraic
KO-theory of $A$, these isomorphisms and the periodicity
of real Clifford algebras immediately imply mod 8 real Bott periodicity.
Analogously, we obtain from the symmetric isomorphism Bott
periodicity in analytic KO-homology.

\newsection{Geometric KO-Homology\label{TKH}}

We will now define geometric KO-homology, analogously to the
Baum-Douglas construction of K-homology~\cite{1,2,RS}, and describe
the basic properties of the topological KO-homology groups of a
topological space that we will need later on . We will prove directly
that this is a homology theory by comparing it with other formulations
of KO-homology as the dual theory to KO-theory. In particular, in the
next section we will show that this homology theory is equivalent to
the analytic homology theory of the previous section.

\subsection{Spin Bordism\label{ss1}}

Throughout $X$ will denote a finite CW-complex.

\begin{definition}
A {\it KO-cycle} on $X$ is a triple $(M,E,\phi)$ where
\begin{romanlist}
\item $M$ is a compact spin manifold without boundary;
\item $E$ is a real vector bundle over $M$; and
\item $\phi : M \rightarrow X$ is a continuous map.
\end{romanlist}
\label{kcycle}\end{definition}
\noindent
There are no connectedness requirements made upon $M$, and hence the
bundle $E$ can have different fibre dimensions on the different
connected components of $M$. It follows that disjoint union
$$(M_1,E_1, \phi_1) \amalg (M_2,E_2,\phi_2):=(M_1 \amalg M_2,E_1 \amalg E_2,
\phi_1 \amalg \phi_2)$$ is a well-defined operation on the set of
KO-cycles on $X$.

\begin{definition}
Two KO-cycles $(M_1,E_1,\phi_1)$ and $(M_2,E_2,\phi_2)$ on $X$ are
  {\it isomorphic} if there exists a diffeomorphism $h : M_1 \rightarrow M_2$
  such that
\begin{romanlist}
\item $h$ preserves the spin structures;
\item $h^{*}(E_2) \cong E_1$ as real vector bundles; and
\item The diagram
  $$\xymatrix{ M_1 \ar[r]^h  \ar[rd]_{\phi_1} & M_2 \ar[d]^{\phi_2}\\
     & X }$$
commutes.
\end{romanlist}
The set of isomorphism classes of KO-cycles on $X$ is denoted
$\Gamma {\rm O}(X)$. \label{isoK-cycles} \end{definition}

\begin{definition}
Two KO-cycles $(M_1,E_1,\phi_1)$ and $(M_2,E_2,\phi_2)$ on $X$ are
{\it spin bordant} if there exist a compact spin manifold $W$ with
boundary, a real vector bundle $E \rightarrow W$, and a continuous
map $\phi : W \rightarrow X$ such that the two KO-cycles
$$\bigl(\partial W\,,\, E|_{\partial W}\,,\,\phi|_{\partial W}\bigr)
\quad , \quad \bigl(M_1\amalg (-M_2)\,,\, E_1 \amalg E_2\,,
\, \phi_1 \amalg \phi_2\bigr)$$ are
isomorphic, where $-M_2$ denotes $M_2$ with the spin structure on
its tangent bundle $TM_2$ reversed. The triple $(W,E,\phi)$ is called
a {\it spin bordism} of KO-cycles. \label{bord}
\end{definition}

\subsection{Real Vector Bundle Modification\label{ss2}}

Let $M$ be a spin manifold and $F\to M$ a $\C^\infty$ real
spin vector bundle with fibres of dimension $n:=\textrm{dim}_\real \, F_p
\equiv 0~\textrm{mod}~8$ for $p\in M$. Let
$\id_M^\real:=M \times \mathbb{R}$ denote the trivial real line
bundle over $M$. Then $F \oplus \id_M^\real$ is a real vector
bundle over $M$ with fibres of dimension $n+1$ and projection map
$\lambda$. By choosing a $\C^{\infty}$ metric on it, we may define the
unit sphere bundle
\beq \widehat{M}=\S\bigl(F \oplus \id_M^\real\bigr)
\label{hatMdef}\eeq by restricting the set of fibre vectors of $F
\oplus \id_M^\real$ to those which have unit norm. The tangent bundle
of $F \oplus \id_M^\real$ fits into an exact sequence of bundles given
by
$$
0~\longrightarrow~\lambda^*\bigl(F \oplus \id_M^\real\bigr)~
\longrightarrow~T\bigl(F \oplus \id_M^\real\bigr)~
\longrightarrow~\lambda^*\bigl(TM\bigr)~\longrightarrow~0 \ .
$$
Upon choosing a splitting, the spin structures on $TM$ and $F$ induce
a spin structure on $T\widehat{M}$, and hence $\widehat{M}$ is a
compact spin manifold. By construction, $\widehat{M}$ is a sphere
bundle over $M$ with $n$-dimensional spheres $\S^n$ as fibres. We
denote the bundle projection by \beq \pi \,:\, \widehat{M}
~\longrightarrow~ M \ . \label{pidef}\eeq We may
regard the total space $\widehat{M}$ as consisting of two
  copies $\B^\pm(F)$, with opposite spin structures, of the unit ball
  bundle $\B(F)$ of $F$ glued together by the identity map
  $\Id_{\S(F)}$ on its boundary so that
\beq
\widehat{M} = \B^+(F) \cup_{\S(F)} \B^-(F) \ .
\label{hatMBF}\eeq

Since $n \equiv 0~\textrm{mod}~8$, the group Spin$(n)$ has two
irreducible real half-spin representations. The spin structure on $F$
associates to these representations real vector bundles $S_0(F)$ and
$S_1(F)$ of equal rank $2^{n/2}$ over $M$. Their Whitney sum $S(F)=S_0(F)\oplus
S_1(F)$ is a bundle of real Clifford modules over $TM$ such that
$\Cl(F)\cong{\rm End}\,S(F)$, where $\Cl(F)$ is the real Clifford
algebra bundle of $F$. Let $\nslash{S}^+(F)$ and $\nslash{S}^-(F)$ be
the real spinor bundles over $F$ obtained from pullbacks to $F$ by the
  bundle projection $F \rightarrow M$ of $S_0(F)$ and $S_1(F)$,
  respectively. Clifford multiplication induces a bundle map $F
  \otimes S_0(F)\rightarrow S_1(F)$ that defines a vector bundle map
  $\sigma :\nslash{S}^+(F)\rightarrow\nslash{S}^-(F)$ covering $\Id_F$
  which is an isomorphism outside the zero section of $F$. Since the
  ball bundle $\B(F)$ is a sub-bundle of $F$, we may form real spinor
  bundles over $\B^\pm(F)$ as the restriction bundles $\Delta^\pm(F)
=\nslash{S}^\pm(F)|_{\B^\pm(F)}$. We can then glue $\Delta^{+}(F)$ and
$\Delta^{-}(F)$ along $\S(F)=\partial\B(F)$ by the Clifford
multiplication map $\sigma$ giving a real vector bundle over
$\widehat{M}$ defined by
\beq
H(F) = \Delta^{+}(F)\cup_{\sigma}\Delta^{-}(F) \ .
\label{HFdef}\eeq
For each $p\in M$, the bundle $H(F)|_{ \pi^{-1}(p)}$ is the real
  Bott generator vector bundle over the $n$-dimensional sphere
  $\pi^{-1}(p)$~\cite{1}.

  \begin{definition}
Let $(M,E,\phi)$ be a KO-cycle on $X$ and $F$ a $\C^\infty$ real
spin vector bundle over $M$ with fibres of dimension $\textrm{dim}_\real
\, F_p \equiv 0~\textrm{mod}~8$ for $p\in M$. Then the process of obtaining
the KO-cycle $(\,\widehat{M},H(F) \otimes \pi^{*}(E),\phi \circ
\pi)$ from $(M,E,\phi)$ is called \it{real vector bundle modification}.
    \label{VBM} \end{definition}

\subsection{Topological KO-Homology\label{KHom}}

We are now ready to define the topological KO-homology groups of
the space $X$.

\begin{definition}
  The \textit{topological KO-homology group of
  $X$} is the abelian group obtained from quotienting $\Gamma
{\rm O}(X)$ by the equivalence relation $\sim$ generated by the
relations of
\begin{romanlist}
\item spin bordism;
\item direct sum: if $E= E_1 \oplus E_2$, then $(M,E,\phi) \sim (M,E_1,\phi)
  \amalg (M,E_2,\phi)$; and
\item real vector bundle modification.
\end{romanlist}
  The group operation is induced by disjoint union of KO-cycles. We denote this
  group by $\KO_\sharp^{\rm t}(X):=\Gamma {\rm O}(X) \, / \sim$,
  and the homology class of the KO-cycle $(M,E,\phi)$ by
  $[M,E,\phi]\in\KO^{\rm t}_\sharp(X)$. \label{Tgroup}
\label{KOtdef}\end{definition}
\noindent Since the equivalence relation on $\Gamma {\rm O}(X)$
preserves the dimension of $M$ $\textrm{mod}~8$ in KO-cycles
$(M,E,\phi)$, one can define the subgroups $\KO_n^{\rm t}(X)$
 consisting of classes of KO-cycles
$(M,E,\phi)$ for which all connected components $M_i$ of $M$ are
of dimension $\textrm{dim} \, M_i \equiv n~\textrm{mod}~8$.
Then
\beq
\KO_\sharp^{\rm t}(X) = \bigoplus _{n=0}^7 \,\KO_{n}^{\rm t}(X)
\label{KOtgrade}\eeq
has a natural $\zed_8$-grading.

The geometric construction of KO-homology is functorial. If $f:X
\rightarrow Y$ is a continuous map, then the induced
  homomorphism $$f_{*} \,:\, \KO_\sharp^{\rm t}(X) ~\longrightarrow~
  \KO_\sharp^{\rm t}(Y)$$
of $\zed_8$-graded abelian groups is given on classes of KO-cycles
$[M,E,\phi]\in\KO^{\rm
  t}_\sharp(X)$ by $$f_{*}[M,E,\phi] := [M,E,f\circ\phi] \ . $$
One has $(\Id_X)_*=\Id_{\KO^{\rm t}_\sharp(X)}$ and $(f\circ
g)_*=f_*\circ g_*$. Since real vector bundles over $M$ extend to real
vector bundles over $M\times[0,1]$, it follows by spin bordism that
induced homomorphisms depend only on their homotopy classes.

If $\pt$ denotes a one-point topological space, then the
collapsing map $\zeta:X \rightarrow \pt$ induces an
epimorphism \beq \zeta_{*} \,:\, \KO_\sharp^{\rm t}(X)
~\longrightarrow ~ \KO_\sharp^{\rm t}(\pt)  \ .
\label{collapseepi}\eeq The {\it reduced} topological KO-homology
group of $X$ is \beq \widetilde{\KO}{}_\sharp^{\,\rm t}(X):=
\ker\zeta_{*} \ . \label{redKhom}\eeq Since the map
(\ref{collapseepi}) is an epimorphism with left inverse induced by
the inclusion of a point $\iota:\pt\hookrightarrow X$, one has
$\KO_\sharp^{\rm t}(X)\cong\KO_\sharp^{\rm
t}(\pt)\oplus\widetilde{\KO}{}_\sharp^{\,\rm t}(X)$ for any
space $X$. As in the complex case~\cite{RS}, one has the following basic
calculational tools for computing the geometric KO-homology groups.
\begin{proposition}
The abelian group $\KO_\sharp^{\rm t}(X)$ enjoys the following
properties:
\begin{romanlist}
\item $\KO_\sharp^{\rm t}(X)$ is generated by classes of KO-cycles
  $[M,E,\phi]$ where $M$ is connected.
\item If $\{X_{j}\}_{j \in J}$ is the set of connected components of $X$
then $$\KO_\sharp^{\rm t}(X)= \bigoplus _{j \in J}\,\KO_\sharp^{\rm
t}(X_{j}) \ . $$
\item The homology class of a KO-cycle $(M,E,\phi)$ on $X$ depends only
  on the KO-theory class of $E$ in $\KO^{0}(M)$; and
\item The homology class of a KO-cycle $(M,E,\phi)$ on $X$ depends only
on the homotopy class of $\phi$ in $[M,X]$.
\end{romanlist}
\end{proposition}

\subsection{Homological Properties\label{RelK}}

We have not yet established that the geometric definition of KO-homology
above is actually a (generalized) homology theory. Defining
$\KO_{i+8k}^{\rm t}(X) :=  \KO_{i}^{\rm t}(X)$ for
all $k \in \mathbb{Z}$, $ 0\leq i \leq 7$, we will now show that
$\KO_\sharp^{\rm t}(X)$ is an 8-periodic unreduced homology
theory. We know that KO-theory is an 8-periodic cohomology
theory which can be defined in terms of its spectrum
$\underline{\KO}^{\infty}$. For $n\geq1$, let $\mathcal{H}_\real$ be a real
$\zed_2$-graded separable Hilbert space which is a
$*$-module for the real Clifford algebra $\Cl_{n-1}=\Cl(\real^{n-1})$
as in Section~\ref{AnKO-H}. Let $\Fred_{n}$ be the space of all
Fredholm operators on $\mathcal{H}_\real$ which are odd, $\Cl_{n-1}$-linear and
self-adjoint. Then $\Fred_{n}$ is the classifying space for
$\KO^n$~\cite{ASSkew}. For $n\leq0$, we choose $k\in\nat$ such that
$8k+n\geq1$ and define $\Fred_n:=\Fred_{8k+n}$. One then has
$\underline{\KO}^{\infty}=\{\Fred_n\}_{n\in\zed}$, and so we can
define~\cite{8} a homology theory related to $\KO^\sharp$
by the inductive limit
\beq
\KO^{\rm s}_i(X,Y) := \lim_{\stackrel{\scriptstyle\longrightarrow}
{\scriptstyle n}} \, \pi^{~}_{n+i} \big((X/Y) \wedge
\Fred_{n}\big)
\label{spectralKOdef}\eeq
for all $i\in \mathbb{Z}$, where $Y$ is a closed subspace of the
topological space $X$ and $\wedge$ denotes the smash product. Bott
periodicity then implies that this is an 8-periodic homology theory.

One can give a definition of relative KO-homology groups
$\KO_i^{\rm t}(X,Y)$ in such a way that there is a map $\mu^{\rm s}
:\KO_i^{\rm t}(X,Y) \rightarrow\KO^{\prime}_i(X,Y)$ which defines a
natural equivalence between functors on the category of topological
spaces having the homotopy type of finite CW-pairs $(X,Y)$, where
$\KO^{\prime}_i(X,Y)$ is Jakob's realization of
$\KO$-homology~\cite{3}. The building blocks of $\KO^\prime_i(X)$ are
triples $(M,x,\phi)$ as in Definition~\ref{kcycle} but now
$x\in\KO^n(M)$ is a KO-theory class over $M$ such that $\dim M+n\equiv
i~{\rm mod}~8$. The equivalence relations are as in
Definition~\ref{Tgroup} with real vector bundle modification modified
from Definition~\ref{VBM} as follows. The nowhere zero section
$$
\Sigma^F\,:\,M~\longrightarrow~ F\oplus\id_M^\real
$$
defined by
$$
\Sigma^F(p)=0_p\oplus1
$$
for $p\in M$ induces an embedding
\beq
\Sigma^F\,:\,M~\hookrightarrow~\widehat{M} \ .
\label{secembind}\eeq
Then real vector bundle modification is replaced by the relation
$$\bigl(M\,,\,x\,,\,\phi\bigr)\sim\bigl(\,\widehat{M}\,,\,
\Sigma_!^F(x)\,\,,\phi\circ\pi\bigr) \ ,
$$ where the functorial homomorphism
$\Sigma_!^F:\KO^n(M)\to\KO^n(\,\widehat{M}\,)$ is the Gysin map induced
by the embedding (\ref{secembind}). On stable isomorphism classes of
real vector bundles $[E]\in\KO^0(M)$ one has
\beq
\Sigma_!^F\bigl[E\bigr]=\bigl[H(F)\otimes\pi^*(E)\bigr] \ .
\label{secstable}\eeq
In the present category, $\KO^{\prime}_i(X,Y)$ is naturally equivalent
to $\KO^{\rm s}_i(X,Y)$. It is important to notice that this is quite a nontrivial result, the validity of which has been established in \cite{3}.

One can give a spin bordism description of $\KO^{\rm t}_\sharp(X,Y)$ as
follows. We consider the set $\Gamma{\rm O}(X,Y)$ of isomorphism
classes of triples $(M,E,\phi)$ where
\begin{romanlist}
\item $M$ is a compact spin manifold with (possibly empty) boundary;
\item $E$ is a real vector bundle over $M$; and
\item $\phi : M \rightarrow X$ is a continuous map with
  $\phi(\partial M)\subset Y$.
\end{romanlist}
The set $\Gamma{\rm O}(X,Y)$ is then quotiented by relations of
relative spin bordism, which is modified from Definition~\ref{bord} by
the requirement that $M_1\amalg(-M_2)\subset\partial W$ is a regularly
embedded submanifold of codimension~$0$ with $\phi(\partial W\setminus
M_1\amalg(-M_2))\subset Y$, direct sum, and real vector bundle
modification, which is applicable in this case since $\S(F
\oplus \id_M^\real)$ is a compact spin manifold with boundary
$\S(F \oplus \id_M^\real) | _{\partial M}$. The
collection of equivalence classes is a $\zed_8$-graded abelian
group with operation induced by disjoint union of relative
KO-cycles. One has $\KO_i^{\rm t}(X,\emptyset) = \KO_i^{\rm t}(X).$
\begin{theorem}
The map
$$\mu^{\rm s} \,:\, \KO_i^{\rm t}(X,Y)
~\longrightarrow~\KO^{\prime}_i(X,Y)$$
defined on classes of KO-cycles by
$$\mu^{\rm s}\bigl[M\,,\,E\,,\,\phi\bigr]_{\rm t}=
\bigl[M\,,\,[E]\,,\,\phi\bigr]_{\rm s}$$
is an isomorphism of abelian groups which is natural with
respect to continuous maps of pairs.
\label{specequivthm}\end{theorem}
\begin{proof}
Taking into account the equivalence relations on $\Gamma
{\rm O}(X,Y)$ used to define both KO-homology groups, the map
$\mu^{\rm s}$ is well-defined and a group homomorphism.
Let $[M,x,\phi]_{\rm s} \in \KO^{\prime}_n(X,Y)$ with $m:= \dim M$. We may
assume that  $M$ is connected and $x$ is non-zero in $\KO^i(M)$. Then
$m-i \equiv n~\rm{mod}~8$. Consider the trivial spin vector bundle
$F=M \times \mathbb{R}^{n+7m+1}$ over $M$. In this case the
sphere bundle (\ref{hatMdef}) is $\widehat{M}=M\times\S^{n+7m+1}$ and
the associated Gysin homomorphism in KO-theory is a map
$$ \Sigma_{!}^{F} \,:\, \KO^i\big(M\big)~
  \longrightarrow~\KO^{i+7m+n}\big(\,\widehat{M}\,\big) \ . $$
  Since $i+7m+n\equiv(i+7m+m-i)~{\rm mod}~ 8\equiv 0~{\rm mod}~
  8$, one has $\KO^{i+7m+p}(\,\widehat{M}\,) \cong
  \KO^{0}(\,\widehat{M}\,)$. It follows that there are real vector bundles
  $E,H \rightarrow \widehat{M}$ such that
  $\Sigma_{!}^{F}(x)=[E]-[H]$, and so by real vector bundle modification
  one has $[M,x,\phi]_{\rm s}=[\,\widehat{M},[E],\phi \circ
  \pi]_{\rm s}-[\,\widehat{M},[H],\phi \circ \pi]_{\rm s}$ in
  $\KO^{\prime}_n(X,Y)$. Therefore $\mu^{\rm s}(\,[\,\widehat{M},E,\phi \circ
  \pi]_{\rm t}-[\,\widehat{M},H,\phi \circ\pi]_{\rm
    t}\,)=[M,x,\phi]_{\rm s}$, and we conclude
  that $\mu^{\rm s}$ is an epimorphism.

Now suppose that $\mu^{\rm s}[M_1,E_1,\phi_1]_{\rm t}=\mu^{\rm
  s}[M_2,E_2,\phi_2]_{\rm t}$ are identified in $\KO'_n(X,Y)$ through
real vector bundle modification. Then, for instance, there is a real spin
vector bundle $F \rightarrow M_1$ such that $M_2 = \widehat{M_1}$ and
$[E_2]=\Sigma_{!}^{F}[E_1]$. This implies that the Gysin homomorphism is
a map
  $$ \Sigma_{!}^{F} \,:\, \KO^0\big(M_1\big)~
  \longrightarrow~  \KO^{0}\big(\,\widehat{M_1}\,\big) \cap
  \KO^{r}\big(\,\widehat{M_1}\,\big) $$
  where $r= \dim F_p$ for $p\in M_1$. Since
  $\KO^{0}\big(\,\widehat{M_1}\,\big) \cap
  \KO^{r}\big(\,\widehat{M_1}\,\big) \neq \{0\}$ in this case, we have
  $r \equiv 0~\rm{mod}~8$ which implies that these two homology
  classes are also identified in $\KO_n^{\rm t}(X,Y)$ through real
  vector bundle modification. As this is the only relation in
  $\KO^{\prime}_n(X,Y)$ that might identify these classes without
  identifying them as KO-cycles, we conclude that
  $\mu^{\rm s}$ is a monomorphism and therefore an isomorphism.
\end{proof}
\begin{remark}
Theorem~\ref{specequivthm} establishes the existence of a natural
equivalence between covariant functors $\KO^{\rm t} \cong
\KO^{\prime}$. Since $\KO^{\prime}$ is a homological realization of
the homology theory associated with KO-theory, it follows that the
same is true of $\KO^{\rm t}$. We have thus constructed an unreduced
8-periodic geometric homology theory dual to KO-theory. It is the
periodicity mod~$8$ of the fibre dimensions of the spin vector bundle
$F$ used for real vector bundle modification in $\KO^{\rm t}$ that
accounts for the isomorphism $\KO^{\rm t} \cong \KO^{\prime}$.
\end{remark}

Having established that KO-homology is a generalized homology theory,
we may throughout exploit standard homological properties
(see~\cite{8} for example). In particular, there is a long exact
homology sequence for any pair $(X,Y)$. Because $\KO^{\rm t}_\sharp$
is an 8-periodic theory, this sequence truncates to a 24-term exact
sequence. In the spin bordism description, the connecting homomorphism
$$
\partial\,:\,\KO^{\rm t}_n(X,Y)~\longrightarrow~\KO^{\rm t}_{n-1}(Y)
$$
is given by the boundary map
\beq
\partial[M,E,\phi]:=[\partial M,E|_{\partial M},\phi|_{\partial M}]
\label{bdrymap}\eeq
on classes of KO-cycles and extended by linearity. $\partial$ is
natural and commutes with induced homomorphisms.

Other homological properties are direct translations of those of the
complex case provided by~\cite{RS}, where a more extensive treatment
can be found. For example, one has the usual excision property. If
$U\subset Y$ is a subspace whose closure lies in the interior of $Y$,
then the inclusion $\varsigma^U:(X\setminus U,Y\setminus
U)\hookrightarrow(X,Y)$ induces an isomorphism
$$\varsigma_*^U\,:\,\KO^{\rm t}_\sharp(X\setminus U,Y\setminus
U)~\stackrel{\approx}{\longrightarrow}~\KO^{\rm t}_\sharp(X,Y)$$
of $\zed_8$-graded abelian groups.

\subsection{Products\label{Prod}}

There are two important products that can be defined on topological
KO-homology groups. The {\it cap product} is the $\zed_8$-degree
preserving bilinear pairing $$\,\frown\,\,:\,\KO^{0}(X)\otimes \KO_\sharp^{\rm
t}(X)~\longrightarrow~\KO_\sharp^{\rm t}(X)$$ given for any real
vector bundle $F\to X$ and KO-cycle class $[M,E,\phi]\in\KO^{\rm
  t}_\sharp(X)$ by $$[F] \,\frown\, [M,E,\phi]:=[M,\phi^{*}F\otimes E ,\phi] $$
and extended linearly. It makes $\KO_\sharp^{\rm t}(X)$ into a
module over the ring $\KO^0(X)$. As in the complex case, this product can be
extended to a bilinear form $$ \,\frown\,\,:\,\KO^i(X)\otimes\KO^{\rm
  t}_j(X)~\longrightarrow~ \KO^{\rm t}_{i+j}(X) \ . $$
The construction utilizes Bott periodicity and the isomorphism
$\KO^{-n}(X)\cong\KO^0(\Sigma^nX)$, where $\Sigma^nX=\S^n\wedge X$ is
the $n$-th iterated reduced suspension of the space $X$. The product $
\,\frown\, :\KO^n(X)\otimes\KO_i^{\rm t}(X) \rightarrow\KO_{i+n}^{\rm t}(X)$
is given by the pairing $ \,\frown\, :\KO^{0}(\Sigma^n
X)\otimes\KO_{i-n}^{\rm t}(\Sigma^nX) \rightarrow \KO_{i-n}^{\rm
t}(\Sigma^nX).$

If $X$ and $Y$ are spaces, then the {\it exterior product}
$$ \times \,:\, \KO^{\rm t}_{i}(X) \otimes \KO^{\rm t}_{j}(Y) ~\longrightarrow~
\KO^{\rm t}_{i+j}(X \times Y)$$ is given for classes of
KO-cycles $[M,E,\phi]\in\KO^{\rm t}_i(X)$ and
$[N,F,\psi]\in\KO^{\rm t}_j(Y)$ by $$\big[M,E,\phi\big]\times
\big[N,F,\psi\big]:=\big[M \times N,E \boxtimes F,
(\phi,\psi)\big] \ , $$ where $M \times N$ has the product spin
structure uniquely induced by the spin structures on $M$ and $N$,
and $E \boxtimes F$ is the real vector bundle over $M \times N$ with
fibres $(E \boxtimes F)_{(p,q)}=E_p \otimes F_q$ for $(p,q)\in
M\times N$. This product is natural with respect to continuous
maps. Unfortunately, in contrast to the complex case, we don't have
a version of the K\"unneth theorem for KO-homology. Indeed, should such
a formula exist, one could use it to show that
$\KO_{\sharp}(\pt)\otimes{\KO_{\sharp}}(\pt)$ has to be a tensor
product as modules over the ring $\KO^{\sharp}(\pt)$. But this does not
work correctly as pointed out by Atiyah in~\cite{AVK}. Moreover, for
$A=B=\mathbb{C}$ considered as a real $C^{*}$-algebra, one has that
the map
\begin{displaymath}
\K_{\sharp}(A)\otimes\K_{\sharp}(B)~\longrightarrow~\K_{\sharp}(A\otimes{B})
\end{displaymath}
is not surjective. The correct framework for K\"unneth formula for
real K-theory is united K-theory~\cite{Bousfiled,Boer1}, which is a
machinery that involves real K-theory, complex K-theory, and
self-conjugate K-theory, and has the property that its homological
algebra behaves better. We will return to this point in
Section~\ref{GenD9Decay}.

\subsection{The Thom Isomorphism\label{ss7}}

Let $X$ be an $n$-dimensional compact manifold with (possibly empty)
boundary, and $\B(TX) \rightarrow X$ and $\S(TX) \rightarrow X$ the
unit ball and sphere bundles of $X$. An element
$\tau\in\KO^n(\B(TX),\S(TX))$ is called a \textit{Thom class} or an
\textit{orientation} for $X$ if $\tau|_{(\B(TX)_x,\S(TX)_x)} \in
\KO^n(\B(TX)_x,\S(TX)_x)\cong\KO^{0}(\textrm{pt})$ is a
generator for all $x\in X$~\cite{7}. The manifold $X$ is said to be
{\it KO-orientable} if it has a Thom class. In that case the
usual cup product on the topological KO-theory ring yields the Thom
isomorphism
$$\mathfrak{T}_X^{~} \,:\,\KO^i\big(X\big)~
\stackrel{\approx}{\longrightarrow}~
\KO^{i+n}\big(\B(TX)\,,\,\S(TX)\big)$$ given for $i=0,1,\ldots,7$ and
$\xi\in\KO^i(X)$ by
$$\mathfrak{T}_X^{~}(\xi):=\pi_{\B(TX)}^{*}(\xi)\,\smile\,\tau \ , $$
where $\pi^{~}_{\B(TX)}:\B(TX) \rightarrow X$ is the bundle
projection. This construction also works by replacing the tangent
bundle of $X$ with any ${\rm
  O}(r)$ vector bundle $V\to X$, defining a Thom isomorphism
$$\mathfrak{T}_{X,V}^{~}\,:\,\KO^i\big(X\big) ~
\stackrel{\approx}{\longrightarrow}~\KO^{i+r}
\big(\B(V)\,,\,\S(V)\big)$$ given by
\beq
\mathfrak{T}^{~}_{X,V}(\xi):=\pi_{\B(V)}^{*}(\xi)\,\smile\,\tau^{~}_{V} \
, \label{ThomisodefV}\eeq where the element
$\tau^{~}_{V}\in\KO^{r}(\B(V),\S(V))$ is called the \textit{Thom class
  of $V$}. Indeed, for a manifold $X$, the $\KO$-orientability condition (existence of a Thom class) described above is equivalent to the existence of a spin structure on the stable normal bundle of the manifold~\cite{ABS1,3}.

Any KO-oriented manifold $X$ of dimension $n$ has
a uniquely determined fundamental class $[X]_{\rm s} \in\KO^{\rm
s}_{n}(X,\partial X)$, which is represented by the element
$[X,\id_{X}^{\real},\Id_X]$ in $\KO^{\rm t}_{n}(X,\partial X)$. One
then has the Poincar\'{e} duality isomorphism
$$\Phi_X^{~}\,:\,\KO^i(X)~\stackrel{\approx}{\longrightarrow}~
\KO^{\rm s}_{n-i}(X,\partial X)$$
given for $i=0,1,\ldots,7$ and $\xi\in\KO^i(X)$ by taking the
cap product \beq \Phi^{~}_X(\xi):=\xi\,\frown\, [X]_{\rm s} \ .
\label{PhiXdef}\eeq In particular, if $X$ is a compact spin manifold
of dimension~$n$ without boundary, then $X$ is $\KO$-oriented and so
in this case we have a Poincar\'{e} duality isomorphism~\cite{3,RS,8}
giving
\beq
\KO_{i}^{\rm t}(X) ~\cong~ \KO^{n-i}(X) \ .
\label{Poincareiso}\eeq
The isomorphism (\ref{Poincareiso}) may be compared with the universal
coefficient theorem for KO-theory~\cite{UCT1,Freed:2006ya}, which
asserts that there is an exact sequence
\beq
0~\longrightarrow~\Ext\big(\KO_{i-1}^{\rm t}(X)\,,\,\zed\big)~
\longrightarrow~\KO^{i+4}\big(X\big)~\longrightarrow~
\Hom\big(\KO_i^{\rm t}(X)\,,\zed\big)~\longrightarrow~0
\label{UCTKO}\eeq
for all $i\in\zed$. The degree shift by~$4$ arises from the fact that
$\KO^{-3}(\pt)=0$ and that there is a cup product pairing
$\KO^{i-4}(\pt)\otimes\KO^{-i}(\pt)\to\KO^{-4}(\pt)\cong\zed$. Under
the same conditions as above, one then also has the Thom isomorphism
in KO-homology
\begin{equation}
\mathfrak{T}_{X,V}^{*}\,:\,\KO^{\rm t}_{i}\big(X\big) ~
\stackrel{\approx}{\longrightarrow}~\KO^{\rm
  t}_{i+r}\big(\B(V)\,,\,\S(V)\big) \ .
\label{ThomhomdefV}\end{equation}

\newsection{The Isomorphism\label{isomorphism}}

One of the main results of this paper is an explicit realization of
the isomorphism between topological and analytic KO-homology. The
primary goal of this section is to prove the following result.
\begin{theorem}\label{iso}
There is a natural equivalence
\begin{displaymath}\label{bigtheom}
\mu^{\rm a}\,:\,\KO^{\rm t}~\stackrel{\approx}
{\longrightarrow}~\KO^{\rm a}
\end{displaymath}
between the topological and analytic KO-homology functors.
\end{theorem}
\noindent

As for any (generalized) homology theory, there's a uniqueness theorem for homology theories (\cite{8}) on the category of finite CW-complexes. More precisely, one has the following

\begin{theorem}
Let $h_{\sharp}$ and $k_{\sharp}$ be generalized homology theories defined on the category of finite CW-pairs, and let
\begin{displaymath}
\phi:h_{\sharp}\to{k}_{\sharp}
\end{displaymath}
be a natural transformation of homology theories such that
\begin{displaymath}
\phi:h_{n}({\rm pt})\to{k_{n}({\rm pt})}
\end{displaymath}
is an isomorphism for any $n\in{\mathbb{Z}}$. Then $\phi$ is a natural equivalence.
\end{theorem}

Taking into account the uniqueness theorem stated above, the proof of theorem~\ref{bigtheom} is
tantamount to proving that the map 
\begin{displaymath}
\mu^{\rm a}\,:\,\KO_{n}^{\rm t}(\pt)~\longrightarrow~\KO_{n}^{\rm a}(\pt),
\end{displaymath}
 induced by the natural transformation $\mu^{\rm a}$, an isomorphism for $n=0,1,\dots,7$. From the realization
 (\ref{spectralKOdef}) it follows that
\bea
\KO^{\rm
  t}_{n}(\pt)&\cong& \lim_{\stackrel{\scriptstyle\longrightarrow}
{\scriptstyle k}} \,\pi_{n+8k}(\Fred_0)\nonumber\\ &\cong&
\pi_n(\Fred_0)~\cong~\widetilde{\KO}{}^{\,0}(\S^{n}) \ . \nonumber
\eea

The main idea behind our proof is to show that there exist surjective
``index'' homomorphisms $\ind_{n}^{\rm t}$ and $\ind_{n}^{\rm a}$ such
that the diagram
\beq
\xymatrix{\KO^{\rm t}_{n}(\pt)\ar[r]^{\mu^{\rm a}}\ar[rd]_{\ind_{n}^{\rm t}} &
\KO^{\rm a}_{n}(\pt)\ar[d]^{\ind_{n}^{\rm a}}\\ &\KO^{-n}(\pt)}
\label{muadiagmain}\eeq
commutes for every $n$. The KO-theory groups $\KO^{-n}(\pt)$ appear here
because they are the coefficient groups of the $\KO_{n}^{\rm t}$ and
$\KO_{n}^{\rm a}$ homology theories. This setup is motivated by the
fact~\cite{1} that the map $\mu^{\rm a}$ and the commutativity of the
diagram (\ref{muadiagmain}) are intimately related to an index
theorem, as we demonstrate explicitly in Section~\ref{ClnIndexThms},
and hence the motivation behind our terminology above. Since the
groups $\KO^{-n}(\pt)$ are equal to
either $0$, $\mathbb{Z}$ or $\mathbb{Z}_2$ depending on the particular
value of $n$, the commutativity of the diagram (\ref{muadiagmain})
along with surjectivity of the index maps are sufficient to prove that
$\mu^{\rm a}$ is an isomorphism. For clarity and later use, we will
divide the proof into four parts. We will first give the constructions
of the three maps in (\ref{muadiagmain}) each in turn, and then
present the proof of commutativity of the diagram.

In the following section we proceed to construct the map $\mu^{\rm a}$, referred to in~\ref{bigtheom}. This map is the natural counterpart for the real case of the complex version built in~\cite{1}.  For an equivalent definition see~\cite{BHS,Bunke1995}.

\subsection{The Map $\mbf{\mu^{\rm a}}$\label{mua}}

Let $(M,E,\phi)$ be a topological KO-cycle on $X$ with $\dim M=n$. We
construct a corresponding class in $\KO^{\rm a}_{n}(X)$ as
follows. Consider the Clifford bundle
\begin{displaymath}
\,\nslash{\mathfrak{S}}(M):={P_{\,\Spin}}(M)\times_{\lambda_n}\Cl_{n}
\end{displaymath}
where $\Cl_{n}={\Cl(\mathbb{R}^n)}$, $\lambda_n:\Spin(n)\to{\rm
  End}(\Cl_{n})$ is given by left multiplication with
$\Spin(n)\subset{\Cl_{n}^{0}\subset{\Cl_{n}^{\phantom{0}}}}$, and $P_{\,\Spin}(M)$ is
the principal $\Spin(n)$-bundle over $M$ associated to the spin structure on
the tangent bundle $TM$. Since
$\Cl^{\phantom{0}}_{n}=\Cl^{0}_{n}\oplus{\Cl^{1}_{n}}$ is a
$\zed_2$-graded algebra, it follows that
\beq
\,\nslash{\mathfrak{S}}(M)=\,\nslash{\mathfrak{S}}^{0}(M)\oplus\,
\nslash{\mathfrak{S}}^{1}(M)
\label{Cliffbungrad}\eeq
is a $\mathbb{Z}_{2}$-graded real vector bundle over $M$ with respect to the
$\Cl(TM)$-action. The Clifford algebra $\Cl_{n}$ acts by right
multiplication on the fibres whilst preserving the bundle grading
(\ref{Cliffbungrad}).

Choose a $\C^\infty$ Riemannian metric $g^M$ on $TM$. Let
$\,\nslash{\mathfrak{D}}^M:\C^\infty(M,\,\nslash{\mathfrak{S}}(M))\to
\C^\infty(M,\,\nslash{\mathfrak{S}}(M))$ be the canonical
Atiyah-Singer operator~\cite{ASSkew} defined locally by
\beq
\,\nslash{\mathfrak{D}}^M=\sum_{i=1}^n\,e_{i}\cdot\nabla^M_{e_{i}}
\ ,
\label{ASop}\eeq
where $\{e_{i}\}_{1\leq i\leq n}$ is a local basis of sections of the
tangent bundle $TM$, $\nabla^M_{e_i}$ are the corresponding
components of the spin connection $\nabla^M$, and the dot denotes
Clifford multiplication. The operator $\,\nslash{\mathfrak{D}}^M$ is a
$\Cl_{n}$-operator~\cite{SG}, i.e. one has
\begin{displaymath}
\,\nslash{\mathfrak{D}}^M(\Psi\cdot\varphi)=\,\nslash{\mathfrak{D}}^M(\Psi)
\cdot\varphi
\end{displaymath}
for all $\Psi\in\C^\infty(M,\,\nslash{\mathfrak{S}}(M))$ and all
$\varphi\in{}\Cl_{n}$, where $\cdot\,\varphi$ denotes right
multiplication by $\varphi$. Since $\,\nslash{\mathfrak{D}}^M$ commutes
with the $\Cl_{n}$-action, the vector space
$\ker~\nslash{\mathfrak{D}}^M$ is a $\Cl_{n}$-module.

We now construct a triple $(\mathcal{H}^M_{E},\rho^{M}_{E},
T^{M}_{E})$ comprising the following data:
\begin{itemize}
\item[(i)] The separable real Hilbert space $\mathcal{H}^{M}_{E}:={\rm
    L}^{2}_\real(M,\,\nslash{\mathfrak{S}}(M)\otimes{E};\dd g^M)$;
\item[(ii)] The $*$-homomorphism
  $\rho^{M}_{E}:\C(M,\real)\to{\mathcal{L}(\mathcal{H}^{M}_{E})}$
  defined by
\begin{displaymath}
\bigl(\rho^{M}_{E}(f)(\Psi)\bigr)(p)=f(p)\,\Psi(p)
\end{displaymath}
for $f\in\C(M,\real)$,
$\Psi\in\C^\infty(M,\,\nslash{\mathfrak{S}}(M)\otimes{E})$ and
$p\in{M}$; and
\item[(iii)] The bounded Fredholm operator
\beq
  T^{M}_{E}:=\frac{\,\nslash{\mathfrak{D}}^M_{E}}
  {\sqrt{1+\big(\,\nslash{\mathfrak{D}}^M_{E}\big)^{2}}}
\label{TEMdef}\eeq
acting on $\mathcal{H}_E^M$, where $\,\nslash{\mathfrak{D}}^M_{E}$ is
the Atiyah-Singer operator (\ref{ASop}) twisted by the real vector
bundle $E\to M$.
\end{itemize}
This triple satisfies the following properties:
\begin{itemize}
\item[(i)] $\mathcal{H}^{M}_{E}$ is $\mathbb{Z}_{2}$-graded according
  to the splitting (\ref{Cliffbungrad}) of the Clifford bundle;
\item[(ii)] $\rho^{M}_{E}(f)$ is an even operator on
  $\mathcal{H}^{M}_{E}$ for all $f\in\C(M,\real)$;
\item[(iii)] Since $M$ is compact, $T^{M}_{E}$ is an odd Fredholm
  operator which obeys the compactness conditions (\ref{TrhoKOarels})
  with $\rho_E^M(f)$; and
\item[(iv)] There are odd operators $\varepsilon_{i}$, $i=1,\dots,n$
  commuting with both $\rho^{M}_{E}(f)$ and $T_E^M$ which generate a
  $\Cl_n$-action on $\mathcal{H}_E^M$ as in (\ref{Clnrels}), and which
  are given explicitly as right multiplication by elements
  $e_{i}$ of a basis of the vector space $\mathbb{R}^{n}$.
\end{itemize}
It follows that $(\mathcal{H}^{M}_{E},\rho^{M}_{E},T^{M}_{E})$ is a
well-defined $n$-graded Fredholm module over the real $C^*$-algebra
$\C(M,\mathbb{R})$.

We now define the map $\mu^{\rm a}$ in (\ref{muadiagmain}) by
\beq
\mu^{\rm a}\bigl(M\,,\,E\,,\,\phi\bigr):=\phi_{*}\bigl(
\mathcal{H}^{M}_{E}\,,\,\rho^{M}_{E}\,,\,
T^{M}_{E}\bigr)=\bigl(\mathcal{H}^{M}_{E}\,,\,\rho^{M}_{E}\circ\phi^{*}
\,,\,T^{M}_{E}\bigr) \ ,
\label{muacycledef}\eeq
where $\phi^{*}:\C(X,\real)\to\C(M,\real)$ is the real $C^*$-algebra
homomorphism induced by the map $\phi$. At this stage the map
$\mu^{\rm a}$ is only defined on KO-cycles. We then have 
\begin{proposition}
The map $\mu^{\rm a}:\KO_{n}^{\rm t}(X)\to\KO_{n}^{\rm a}(X)$ induced
by (\ref{muacycledef}) is a well-defined homomorphism of abelian
groups for any $n\in\nat$.
\end{proposition}
\begin{proof}
Let $(M,E,\phi),\,(N,F,\psi)\in\Gamma \textrm{O}(X)$, and consider
their disjoint union. The Clifford bundle of the disjoint union
manifold splits as
$\nslash{\mathfrak{S}}(M\sqcup N)=\nslash{\mathfrak{S}}(M)\sqcup
\nslash{\mathfrak{S}}(N)$, and therefore the twisted Clifford bundle
has a corresponding spliting
$\nslash{\mathfrak{S}}(M\sqcup N)\otimes(E\sqcup
F)=\big(\,\nslash{\mathfrak{S}}(M)\otimes E\big)\sqcup
\big(\,\nslash{\mathfrak{S}}(N)\otimes F\big)$, giving rise to a
splitting of the space of sections
$\C^\infty(M,\,\nslash{\mathfrak{S}}(M\sqcup N)\otimes(E\sqcup
F))=\C^\infty(M,\, \nslash{\mathfrak{S}}(M)\otimes E)
\oplus\C^\infty(N,\,\nslash{\mathfrak{S}}(N)\otimes F)$, and therefore
of the corresponding spaces of L$^2$-sections:
\begin{equation}\label{hilbertsplit}
\mathcal{H}^{M\sqcup N}_{E\sqcup
  F}=\mathcal{H}^{M}_{E}\oplus\mathcal{H}^{N}_{F} \ .
\end{equation}
The algebras of functions also split as $\C(M\sqcup
N,\real)=\C(M,\real)\oplus\C(N,\real)$, and this together with the
structure of the Hilbert spaces of sections~(\ref{hilbertsplit}) imply
that
$$\rho^{M\sqcup N}_{E\sqcup F} = \left( \begin{array}{ccc}
 \rho^{M}_{E} & 0\\
 0 &  \rho^{N}_{F}\end{array}\right) \ , \qquad T^{M\sqcup N}_{E\sqcup F}
= \left( \begin{array}{ccc}
 T^{M}_{E} & 0\\
 0 &  T^{N}_{F}\end{array}\right) \ ,$$
which immediately implies that
\begin{equation}\label{disjointanalyticsum}
(\mathcal{H}^{M\sqcup N}_{E\sqcup F},\rho^{M\sqcup N}_{E\sqcup F}\circ
(\phi\sqcup\psi)^*,T^{M\sqcup N}_{E\sqcup
  F})=(\mathcal{H}^{M}_{E},\rho^{M}_{E}\circ\phi^*,T^{M}_{E})+
(\mathcal{H}^{N}_{F},\rho^{N}_{F}\circ\psi^*,T^{N}_{F}) \ ,
\end{equation}
showing that the map $\mu^{\rm a}$ preserves disjoint union of cycles,
and so it is a homomorphism of (unital) abelian monoids.

Let us now consider the direct sum relation. Since $\mu^{\rm a}$ is a
monoid morphism, we have
\begin{eqnarray*}
\mu^{\rm a}((M,E_1,\phi)\sqcup(M,E_2,\phi)) & = & \mu^{\rm
  a}(M,E_1,\phi)+\mu^{\rm a}(M,E_2,\phi)\\
& = &
\phi^*\Bigg(\Bigg[\mathcal{H}^{M}_{E_1}\oplus
\mathcal{H}^{M}_{E_2},\rho^{M}_{E_1}\oplus\rho^{M}_{E_2},
\left( \begin{array}{ccc}
 T^{M}_{E_1} & 0\\
 0 &  T^{M}_{E_2}\end{array}\right)\Bigg]\Bigg) \ .
\end{eqnarray*}
As above, the space of sections splits
$$\C^\infty(M,\,\nslash{\mathfrak{S}}(M)\otimes(E_1\oplus
E_2))=\C^\infty(M,\, \nslash{\mathfrak{S}}(M)\otimes E_1)
\oplus\C^\infty(M,\,\nslash{\mathfrak{S}}(M)\otimes E_2)$$ giving rise
to a splitting of the Hilbert spaces
$\mathcal{H}^{M}_{E_1\oplus
  E_2}=\mathcal{H}^{M}_{E_1}\oplus\mathcal{H}^{M}_{E_2}$. This leads
to the conclusion that
$$[\mathcal{H}^{M}_{E_1\oplus E_2},\rho^{M}_{E_1\oplus
  E_2},T^{M}_{E_1\oplus
  E_2}]=\Bigg(\Bigg[\mathcal{H}^{M}_{E_1}\oplus
\mathcal{H}^{M}_{E_2},\rho^{M}_{E_1}\oplus\rho^{M}_{E_2},
\left( \begin{array}{ccc}
 T^{M}_{E_1} & 0\\
 0 &  T^{M}_{E_2}\end{array}\right)\Bigg]\Bigg) \ ,$$
which therefore implies that $\mu^{\rm a}(M,E_1\oplus
E_2,\phi)=\mu^{\rm a}((M,E_1,\phi)\sqcup(M,E_2,\phi))$, showing that
our map preserves the direct sum relation.

Let us now suppose that the cycle $(M,E,\phi)$ is a bord, i.e. that
there exists $(W,F,\psi)$ such that $(\del W,F|_{\del W},\psi|_{\del
  W})=(M,E,\phi)$. The inclusion of the boundary $\iota:\del
W\rightarrow W$ induces a commutative diagram
$$\xymatrix{\KO_{n}^{\rm a}(\del W) \ar[r]^{\iota_*}
  \ar[dr]_{(\psi|_{\del W})_*} & \KO_{n}^{\rm a}(W)\ar[d]^{\psi_*}\\
& \KO_{n}^{\rm a}(X)} \ ,$$
so, denoting by $[\,\nslash{\mathfrak{D}}^M_{E}]$ the element
$\mu^{\rm a}_M(M,E,\Id_M)$, we have
\begin{eqnarray}
(\psi|_{\del W})_*([\,\nslash{\mathfrak{D}}^{\del W}_{F|_{\del W}}]) &
= & \psi_*(\iota_*([\,\nslash{\mathfrak{D}}^{\del W}_{F|_{\del
    W}}]))\\
& = & \psi_*(\iota_*([\,\nslash{\mathfrak{D}}^{\del W}]\cap[{F|_{\del
    W}}]) \ .\nonumber
\end{eqnarray}
A result of Higson and Roe~\cite[Prop.~11.2.15]{HigsonRoeAnaK-hom}
states that, in analytic K-homology,  $[\,\nslash{\mathfrak{D}}^{\del
  W}]=\del \, [\,\nslash{\mathfrak{D}}^{W-\del W}]$, and considering
the long exact homology sequence of the pair $(W,\del W)$,
$$\xymatrix{\cdots \ar[r] & \KO_{n}^{\rm a}(\del W)\ar[r]^{\iota_*} &
  \KO_{n}^{\rm a}(W)\ar[r] & \KO_{n}^{\rm a}(W-\del W)\ar[r]^{\del} &
  \KO_{n-1}^{\rm a}(\del W) \ar[r] & \cdots} \ ,$$
it follows that $\iota_*([\,\nslash{\mathfrak{D}}^{\del
  W}]=\iota_*\circ\del\,([\,\nslash{\mathfrak{D}}^{W-\del W}])=0$, by
exactness of the sequence. The above discussion therefore implies that
$\mu^{\rm a}(\del W,F|_{\del W},\psi|_{\del W})=0$, and so the map
preserves the bordism relation.

The only relation remaining now is vector bundle modification. Let
$(M,E,\phi)\in\Gamma \textrm{O}(X)$ and let $F\rightarrow M$ be a Spin
vector bundle with ${\rm rk }(F)=8k$. Assume $n=\textrm{dim }M$. We
want to show that the equality
\begin{equation}\label{muvbm}
\mu^{\rm a}(M,E,\phi)=\mu^{\rm a}(\widehat{M},H(F) \otimes
\pi^{*}(E),\phi \circ \pi)
\end{equation}
holds. Some elementary calculations show that
\begin{eqnarray*}
\mu^{\rm a}(\widehat{M},H(F) \otimes \pi^{*}(E),\phi \circ\pi) &=&
\phi_*(\pi_*([\,\nslash{\mathfrak{D}}^{\widehat{M}}_{H(F) \otimes
  \pi^{*}[E]}]))\\
&=& \phi_*(\pi_*([\,\nslash{\mathfrak{D}}^{\widehat{M}}]\cap [H(F)
\otimes \pi^{*}E]))\\
&=& \phi_*(\pi_*([\,\nslash{\mathfrak{D}}^{\widehat{M}}]\cap ([H(F)]
\cup [\pi^{*}E])))\\
&=& \phi_*(\pi_*(([\,\nslash{\mathfrak{D}}^{\widehat{M}}]\cap H(F))
\cap \pi^{*}[E]))\\
&=& \phi_*(\pi_*([\,\nslash{\mathfrak{D}}^{\widehat{M}}]\cap H(F))
\cap [E]) \ .
\end{eqnarray*}
So, if we show that
\begin{equation}\label{thomequality}
[\,\nslash{\mathfrak{D}}^{M}]=
\pi_*([\,\nslash{\mathfrak{D}}^{\widehat{M}}]\cap H(F)) \ ,
\end{equation}
equality~(\ref{muvbm}) will follow immediately from it.
By the remarks in Section 2.6, we know that $[H(F)]$ is the Thom class
of $F$, and taking a closer look at the right-hand side of
(\ref{thomequality}), we see that it is just the image of
$[\,\nslash{\mathfrak{D}}^{\widehat{M}}]$, the fundamental class of
$\widehat{M}$ in analytic K-homology, through the homology Thom
isomorphism of the spherical fibration $\pi:\widehat{M}\rightarrow
M$. So~(\ref{thomequality}) is equivalent to the fact that the Thom
isomorphism
\beq\mathfrak{T}^{M,F}\,:\, \KO^{\rm a}_{q+8k}(\widehat{M})
\rightarrow \KO^{\rm a}_{q}(M)\eeq
maps the fundamental class of $\widehat{M}$ to the fundamental class
of $M$. This follows from the way one constructs the Spin structure of
the sphere bundle $\widehat{M}$ from the ones in $M$ and $F$.
\end{proof}

It is easy to show that $\mu$ is natural with respect to continuous maps of spaces. 

The last step required is to show that $\mu^{\rm a}$ is a natural transformation of homology theories, namely that it commutes with the boundary operators of the homology theories in question. To achieve this, one needs the following nontrivial result describing the boundary map in analytic $\KO$-homology~\cite{HigsonRoeAnaK-hom,BHS} 
\begin{theorem}\label{connective}
Let $M-\partial{M}$ be the interior of a spin manifold $M$ of dimension $n$ with boundary $\partial M$, and let $E$ be a real vector bundle on $M$. Equip the boundary $\partial{M}$ with the spin structure induced by that on $M$. Then
\begin{displaymath}
\partial[\:\nslash{\mathfrak{D}}^{M-\partial M}_{E|_{M-\partial M}}]=[\:\nslash{\mathfrak{D}}^{\partial M}_{E|_{\partial{}M}}]
\end{displaymath}
where $\partial:\KO_{n}^{a}(M-\partial{M})\to\KO^{a}_{n-1}(\partial{M})$ is the boundary homomorphism.
\end{theorem}

Finally, one can prove that the class $[\:\nslash{\mathfrak{D}}^{M}]:=[\:\nslash{\mathfrak{D}}^{M}_{\id}]$ represents the fundamental class of $M$ in $\KO_{n}^{a}(M)$, and that~\cite{Bunke1995}
\begin{displaymath}
[\:\nslash{\mathfrak{D}}^{M}_{E}]=[E]\cap[\:\nslash{\mathfrak{D}}^{M}].
\end{displaymath}
Combining these results, one can conclude that the map $\mu^{\rm a}$ commutes with the appropriate boundary maps, therefore showing that it is a natural transformation of homology theories.

\subsection{The Map $\mbf{\ind_{n}^{\rm a}}$\label{indna}}

Let $(\hilR,\rho,T)$ be an $n$-graded Fredholm module over the real
$C^*$-algebra $\C(X,\real)$. Since the Fredholm operator $T$
commutes with $\varepsilon_{i}$ for $i=1,\dots,n$, the kernel $\ker
T\subset\hilR$ is a real $\Cl_{n}$-module with $\zed_2$-grading
induced by the grading of $\hilR$. Thus we can define \beq
\ind_{n}^{\rm a}(T):=\:[\ker T]\in\: \big( \,
\widehat{\mathfrak{M}}_{n}/ \imath^{*}\widehat{\mathfrak{M}}_{n+1}
\big) \cong\:\KO^{-n}(\pt) \ , \label{indnaT}\eeq where
$\widehat{\mathfrak{M}}_{n}$ is the Grothendieck group of real
graded $\Cl_{n}$ representations and $\imath^{*}$ is induced by the
natural inclusion $\imath:\Cl_{n}\hookrightarrow{\Cl_{n+1}}$. We
will call (\ref{indnaT}) the \emph{analytic} or \emph{Clifford
index} of the Fredholm operator $T$. An important property of this
definition is the following result~\cite{SG}.
\begin{theorem}
The analytic index
\begin{displaymath}
\ind^{\rm a}_{n}\,:\,\Fred_{n}~\longrightarrow~\KO^{-n}(\pt)
\end{displaymath}
is constant on the connected components of $\Fred_{n}$.
\label{anindthm}\end{theorem}

Given two Fredholm modules $(\hil_\real,\rho,T)$ and $(\hilR,\rho,T'\,)$
over a real $C^*$-algebra $A$, we will say that $T$ is a {\it compact
perturbation} of $T'$ if $(T-T'\,)\,\rho(a)\in\mathcal{K}_\real$ for
all $a\in{A}$. We then have the following elementary result.
\begin{lemma}
If $T$ is a compact perturbation of $T'$, then the Fredholm modules
$(\hil_\real,\rho,T)$ and $(\hilR,\rho,T'\,)$ are operator homotopic
over $A$.
\label{lemmacompact}\end{lemma}
\begin{proof}
Consider the path $T_{t}=(1-t)\,T+t\,T'$ for $t\in[0,1]$. Then the map
$t\mapsto{T_{t}}$ is norm continuous. We will show that for any $t\in[0,1]$,
the triple $(\hilR,\rho,T_{t})$ is a Fredholm module over $A$, i.e.
that the operator $T_{t}$ satisfies
\beq
\left(T_{t}^{2}-1\right)\,\rho(a)~,~
\left(T^{\phantom{*}}_{t}-T_{t}^{*}\right)\,\rho(a)~,~
T^{\phantom{*}}_{t}\,\rho(a)-\rho(a)\,T^{\phantom{*}}_{t}\in\mathcal{K}_\real
\label{TtFred}\eeq
for all $a\in A$. The last two inclusions in (\ref{TtFred}) are
easily proven because the path $T_{t}$ is ``linear'' in the operators
$T$ and $T'$. To establish the first one, for any $t\in[0,1]$ and
$a\in A$ we compute
\beq
\left(T_{t}^{2}-1\right)\,\rho(a)=\Bigl[\left(T^{2}-1\right)+t^{2}\,
\left(T-T'\,\right)^{2}-t\,\left(T^{2}-1\right)-t\,\left(T-T'\,
\right)^{2}+t\,\left(T^{\prime\,2}-1\right)\Bigr]\,\rho(a) \
. \nonumber\\
\label{compcompute}\eeq
By using the fact that $(\hilR,\rho,T)$ and $(\hilR,\rho,T'\,)$ are
Fredholm modules, that $T$ is a compact perturbation of $T'$, and that
$\mathcal{K}_\real$ is an ideal in $\mathcal{L}(\hilR)$, one easily
verifies that the right-hand side of (\ref{compcompute}) is a compact
operator. This implies that $(\hilR,\rho,T_{t})$ is a well-defined
family of Fredholm modules over $A$.
\end{proof}

\begin{proposition}
The induced map
\begin{displaymath}
\ind^{\rm a}_{n}\,:\,\KO^{\rm a}_{n}(X)~\longrightarrow~{\KO^{-n}(\pt)}
\end{displaymath}
given on classes of $n$-graded Fredholm modules by
$$
\ind^{\rm a}_{n}[\hilR,\rho,T]=[\ker T]
$$
is a well-defined surjective homomorphism for any $n\in\nat$.
\end{proposition}
\begin{proof}
We first show that to the direct sum of two Fredholm modules
$(\hilR^{\phantom{\prime}},\rho,T)$ and $(\mathcal{H}_\real',\rho',T'\,)$ over
$A=\C(X,\real)$, the map $\ind^{\rm a}_n$ associates the class $[\ker
T]+[\ker T'\,]\in\widehat{\mathfrak{M}}_{n}/\imath^{*}
\widehat{\mathfrak{M}}_{n+1}\cong\KO^{-n}(\pt)$. The kernel
\begin{displaymath}
\ker(T\oplus{T'}\,)=\ker(T)\oplus{\ker(T'\,)}
\end{displaymath}
is a real graded $\Cl_{n}$-module. By the definition of the group
$\widehat{\mathfrak{M}}_{n}$ and of its quotient by
$\imath^{*}\widehat{\mathfrak{M}}_{n+1}$, one thus has $\ind_n^{\rm
  a}(T\oplus T'\,)=[\ker T]+[\ker T'\,]$ and so the map $\ind^{\rm
  a}_{n}$ respects the algebraic structure on $\Gamma{\rm O}_{n}(A)$.

Consider now two Fredholm modules $(\hilR^{\phantom{\prime}},\rho,T)$ and
$(\mathcal{H}_\real',\rho',T'\,)$ which are orthogonally
equivalent. Then there exists an even isometry
$U:\hilR^{\phantom{\prime}}\to\mathcal{H}_\real'$ such that
$$
 T'=U\,T\,U^* \ , \quad
\varepsilon'_{i}=U\,\varepsilon^{\phantom{\prime}}_{i}\,U^* \ .
$$
This implies that $\ker T'=U(\ker T)$, and that the graded $\Cl_{n}$
representations given respectively by $\varepsilon'_{i}$ and
$\varepsilon^{\phantom{\prime}}_{i}$ are equivalent. In particular,
they represent the same class in
$\widehat{\mathfrak{M}}_{n}/\imath^{*}\widehat{\mathfrak{M}}_{n+1}$.

Finally, consider two homotopic $n$-graded
Fredholm modules $(\hilR,\rho,T)$ and $(\hilR,\rho,T'\,)$ over $A$. In
general, $T$ and $T'$ are not elements of $\Fred_{n}$ because
they need not be self-adjoint. However, one can always perform
a compact perturbation to obtain an equivalent Fredholm module whose
operator is self-adjoint by simply replacing $T$ with
$\tilde T:=\frac{1}{2}\,(T+T^{*})$. By Lemma~\ref{lemmacompact}, compact
perturbation implies operator homotopy and so there is no loss of
generality in considering only homotopy of ``self-adjoint'' Fredholm
modules. Then the function $t\mapsto{T_{t}}$ gives a homotopy $\tilde
T_{t}=\frac{1}{2}\,(T^{\phantom{*}}_{t}+T^{*}_{t})$ in $\Fred_{n}$
connecting $T$ and $T'$. The proposition now follows from
Theorem~\ref{anindthm}.
\end{proof}

\subsection{The Map $\mbf{\ind^{\rm t}_{n}}$\label{indnt}}

Given a KO-cycle $(M,E,\phi)$ on $X$ with $M$ an $n$-dimensional
compact spin manifold, we can assign to it the
\textit{Atiyah-Milnor-Singer (AMS) invariant}~\cite{ASSkew} defined by
\beq
\widehat{\mathcal{A}}_{E}(M)=
\beta\circ{\iota^*}\circ{\varsigma^*}\bigl(\tau_{\nu}(E)
\bigr)\in{\KO^{-n}(\pt)}
\label{AMSinvdef}\eeq
where:
\begin{itemize}
\item[(i)] $\nu$ is the normal bundle $N(\S^{n+8k}/M)$, with projection
$\varpi:\nu\to{M}$, identified with a tubular neighbourhood of an
embedding
\beq
f\,:\,M~\hookrightarrow~{\S^{n+8k}}
\label{MembedS}\eeq
for some $k\in\nat$ sufficiently large;
\item[(ii)]
  $\tau_{\nu}(E)=\tau_{\nu}\,\smile\,[\varpi^{*}E]\in\:\KO^0(\nu,\nu\setminus
  M)$ where
\begin{displaymath}
\tau_{\nu}:=\left[\varpi^{*}\,\nslash{S}^{+}(\nu)\,,\,
\varpi^{*}\,\nslash{S}^{-}(\nu)\,;\,\sigma
\right]\in\:\KO^0(\nu,\nu\setminus M)
\end{displaymath}
is the Thom class of $\nu$, with
$\sigma:\varpi^{*}\,\nslash{S}^{+}(\nu)\to\varpi^{*}\,\nslash{S}^{-}(\nu)$
given by Clifford multiplication;
\item[(iii)] $\varsigma^*:\KO^0(\nu,\nu\setminus
  M)\to{\KO^0(\S^{n+8k},\S^{n+8k}\setminus M)}$
is given by the excision theorem;
\item[(iv)]
  $\iota^*:\KO^0(\S^{n+8k},\S^{n+8k}\setminus
  M)\to\widetilde{\KO}{}^{\,0}(\S^{n+8k})$ is given by the
  inclusion $\iota:(\S^{n+8k},\pt)\hookrightarrow(\S^{n+8k},M)$; and
\item[(v)]
$\beta:\widetilde{\KO}{}^{\,0}(\S^{n+8k})\to\widetilde{\KO}{}^{\,0}
(\S^{n}):={\KO^{-n}(\pt)}$ is given by Bott periodicity.
\end{itemize}
This definition does not depend on the embedding (\ref{MembedS}) nor
on the integer $k$. We define
\beq
\ind^{\rm t}_{n}(M,E,\phi):=\widehat{\mathcal{A}}_{E}(M) \ .
\label{indntdef}\eeq

\begin{proposition}
The map
\begin{displaymath}
\ind^{\rm t}_{n}\,:\,\KO^{\rm t}_{n}(X)~\longrightarrow~{\KO^{-n}}(\pt)
\end{displaymath}
induced by (\ref{indntdef}) is a well-defined surjective homomorphism
for any $n\in\nat$.
\end{proposition}
\begin{proof}
We first prove that the map $\ind^{\rm t}_{n}$ respects the algebraic
structure on the abelian group $\KO^{\rm t}_{n}(X)$. Given two
$n$-dimensional compact spin manifolds $M_{1}$ and $M_{2}$, let
$M=M_{1}\amalg{M_{2}}$. Embed $M$ in the sphere $\S^{n+8k}$ for some
$k$ sufficiently large as in (\ref{MembedS}). Then the normal bundle
$\nu$ of this embedding is given by
$N(\S^{n+8k}/M_{1})\amalg{}N(\S^{n+8k}/M_{2})$. Identify $\nu$ with a
tubular neighbourhood of the embedding given by
$\nu_{1}\amalg\nu_{2}$, with projection
$\varpi=\varpi_{1}\amalg\varpi_{2}:\nu_{1}\amalg{\nu_{2}}\to
M_{1}\amalg{M_{2}}$. The Thom class of $\nu$ is given by
\begin{eqnarray*}
\tau_{\nu}&:=&\left[\varpi^{*}\,\nslash{S}^{+}(\nu)\,,\,
\varpi^{*}\,\nslash{S}^{-}(\nu)\,;\,\sigma\right]\\
&=&\left[\varpi_{1}^{*}\,\nslash{S}^{+}(\nu_{1})\amalg\varpi_{2}^{*}\,
\nslash{S}^{+}(\nu_{2})\,,\,\varpi_{1}^{*}\,\nslash{S}^{-}(\nu_{1})
\amalg\varpi_{2}^{*}\,\nslash{S}^-(\nu_{2})\,;\,\sigma_{1}
\amalg\sigma_{2}\right]\\
&=&\tau_{\nu_{1}}+\tau_{\nu_{2}}\in\:\KO^0(\nu,\nu\setminus M)\cong
\KO^0(\nu_{1},\nu_{1}\setminus M_{1})\oplus{}\KO^0(\nu_{2},\nu_{2}
\setminus M_{2}) \ .
\end{eqnarray*}
Let $E_{1}$ and $E_{2}$ be real vector bundles over $M_{1}$ and
$M_{2}$, respectively, and let $E=E_{1}\amalg{E_{2}}$. Then in
$\KO^0(\nu,\nu\setminus M)$ one has
\begin{eqnarray*}
\tau_{\nu}(E)&=&\tau_{\nu}\,\smile\,[\varpi^{*}E]\\
&=&\tau_{\nu}\,\smile\,[\varpi_{1}^{*}E_1\amalg\varpi_{2}^{*}E_2]\\
&=&\tau_{\nu_{1}}\,\smile\,[\varpi_{1}^{*}E_{1}]+\tau_{\nu_{2}}\,\smile\,
[\varpi_{2}^{*}E_{2}]~=~\tau_{\nu_{1}}(E_{1})+\tau_{\nu_{2}}(E_{2}) \ .
\end{eqnarray*}
Using the fact that the maps $\varsigma^*$, $\iota^*$ and $\beta$ are
group homomorphisms, one then finds
\begin{eqnarray*}
\ind^{\rm t}_{n}\bigl((M_{1},E_{1},\phi_{1})\amalg(M_{2},E_{2},\phi_{2})\bigr)
&:=&\widehat{\mathcal{A}}_{E}(M)\\
&=&\widehat{\mathcal{A}}_{E_{1}}(M_{1})+\widehat{\mathcal{A}}_{E_{2}}(M_{2})\\
&=&\ind^{\rm t}_{n}(M_{1},E_{1},\phi_{1})+\ind^{\rm
  t}_{n}(M_{2},E_{2},\phi_{2})\in{\KO^{-n}(\pt)} \ ,
\end{eqnarray*}
showing that $\ind^{\rm t}_{n}$ is a homomorphism of abelian groups.

Next we have to check that the map $\ind^{\rm t}_{n}$ is independent
of the choice of representative of a homology class in $\KO^{\rm
  t}_{n}(X)$. The independence of the direct sum relation follows
from the discussion above, while spin bordism independence is
guaranteed by the property that the AMS invariant
$\widehat{\mathcal{A}}_{E}(M)$ is a spin cobordism
invariant~\cite{ASSkew}. Finally, we have to verify that the map
$\ind^{\rm t}_{n}$ does not depend on real vector bundle
modification. We will give a fairly detailed proof of this result, as
we believe it is instructive.

Let $M$ be a smooth $n$-dimensional compact spin manifold and let $E\to
M$ be a smooth real vector bundle. Given an embedding (\ref{MembedS}),
the AMS invariant of the pair $(M,E)$ may be written as~\cite{SG}
\begin{displaymath}
\widehat{\mathcal{A}}_{E}(M)=\beta\circ f_{!}[E] \ ,
\end{displaymath}
where
\begin{displaymath}
f_{!}\,:\,\KO^0\big(M\big)~\longrightarrow~\KO^0\big(\mathbb{S}^{n+8k}\big)
\end{displaymath}
is the Gysin homomorphism of the embedding $f$. Let $F$ be a real spin
vector bundle over $M$ with fibres of real dimension
$8l$ for some $l\in\nat$. Consider the corresponding sphere bundle
(\ref{hatMdef}) with projection (\ref{pidef}). As discussed in
Section~\ref{RelK} (see (\ref{secembind}) and (\ref{secstable})), real
vector bundle modification of a KO-cycle $(M,E,\phi)$ on $X$ induced
by $F$ produces the KO-cycle
$(\,\widehat{M},\widehat{E},\phi\circ{\pi})$, where
$\widehat{E}=H(F)\otimes\pi^*(E)$ is the real vector bundle over
$\widehat{M}$ such that
\begin{displaymath}
\big[\,\widehat{E}\,\big]=\Sigma_{!}^F\big[E\big]
\end{displaymath}
with $[E]\in{}\KO^0(M)$ and
$[\,\widehat{E}\,]\in{\KO^0(\,\widehat{M}\,)}$. We may compute the AMS
invariant for the pair $(\,\widehat{M}\,,\,\widehat{E}\,)$ by choosing
an embedding
\begin{displaymath}
\widehat{f}\,:\,\widehat{M}~\hookrightarrow~\mathbb{S}^{n+8k+8l}
\end{displaymath}
so that
\begin{eqnarray*}
\widehat{\mathcal{A}}_{\widehat{E}}\big(\,\widehat{M}\,\big)&=&
\beta\circ\widehat{f}_{!}\bigl([\,\widehat{E}\,]\bigr)\\
&=&\beta\circ\widehat{f}_{!}\circ\Sigma^F_{!}[E]~=~
\beta\circ\big(\,\widehat{f}\circ\Sigma^F\,\big)_{!}[E] \ ,
\end{eqnarray*}
where in the last equality we have used functoriality of the Gysin
map. Notice that
\begin{displaymath}
\widehat{f}\circ{\Sigma^F}\,:\,M~\hookrightarrow~
\mathbb{S}^{n+8k+8l}=:\mathbb{S}^{n+8m}
\end{displaymath}
is an embedding of $M$ into a ``large enough'' sphere. Since
$\widehat{\mathcal{A}}_{E}(M)$ is independent of the embedding and
the integer $m$, we have
\begin{displaymath}
\widehat{\mathcal{A}}_{\widehat{E}}\bigl(\,\widehat{M}\,\bigr)=
\widehat{\mathcal{A}}_{E}\bigl(M\bigr)
\end{displaymath}
as required.
\end{proof}

\subsection{The Isomorphism Theorem}

We can now assemble the constructions of
Sections~\ref{mua}--\ref{indnt} above to finally establish our main
result. Notice first of all that since
$\ker~\nslash{\mathfrak{D}}^{{M}}_E\cong\ker T^{M}_{E}$, one
has
\beq
\ind^{\rm a}_{n}\circ\mu^{\rm a}
\big(M\,,\,E\,,\,\phi\big)=\ind^{\rm a}_{n}
\big(\,\nslash{\mathfrak{D}}^M_E\big)
\label{indmuarel}\eeq
for any KO-cycle $(M,E,\phi)$ on $X$ with $\dim M=n$.
At this point we can use an important result from spin
geometry called the \textit{$\Cl_{n}$-index theorem}~\cite{SG}.
\begin{theorem}
Let $M$ be a compact spin manifold of dimension $n$ and let $E$ be a
real vector bundle over $M$. Let
\begin{displaymath}
\,\nslash{\mathfrak{D}}^M_{E}\,:\,\C^\infty\bigl(M\,,\,\,
\nslash{\mathfrak{S}}(M)\otimes{E}\bigr)~\longrightarrow~
\C^\infty\bigl(M\,,\,\,\nslash{\mathfrak{S}}(M)\otimes{E}\bigr)
\end{displaymath}
be the $\Cl_{n}$-linear Atiyah-Singer operator with coefficients in
$E$. Then
\begin{displaymath}
\ind^{\rm
  a}_{n}\big(\,\nslash{\mathfrak{D}}^M_{E}\big)=
\widehat{\mathcal{A}}_{E}\big(M\big) \ .
\end{displaymath}
\label{spinthm}\end{theorem}
\noindent
The proof of Theorem~\ref{iso} is now completed once we establish the
following result.
\begin{proposition}
The map
\begin{displaymath}
\mu^{\rm a}\,:\,\KO^{\rm t}_{n}(\pt)~\longrightarrow~{\KO^{\rm a}_{n}(\pt)}
\end{displaymath}
is an isomorphism for any $n\in\nat$.
\end{proposition}
\begin{proof}
As noticed at the beginning of this section, it suffices to establish
the commutativity of the diagram (\ref{muadiagmain}), i.e. that
\begin{displaymath}
\ind^{\rm t}_{n}=\ind^{\rm a}_{n}\circ\mu^{\rm a} \ .
\end{displaymath}
Let $[M,E,\phi]$ be the class of a KO-cycle over ${\pt}$ with $\dim
{M}=n$. Using Theorem~\ref{spinthm} and (\ref{indmuarel}) we have
\begin{eqnarray*}
\ind^{\rm t}_{n}\big[M\,,\,E\,,\,\phi\big]&:=&
\widehat{\mathcal{A}}_{E}\big(M\big)\\
&=&\ind^{\rm a}_{n}\big(\,\nslash{\mathfrak{D}}^M_{E}\big)
~=~\ind^{\rm a}_{n}\circ\mu^{\rm a}\big[M\,,\,E\,,\,\phi\big]
\end{eqnarray*}
as required.
\end{proof}

\newsection{The Real Chern Character\label{RealChern}}

In this section we will describe the natural complexification map from
geometric KO-homology to geometric K-homology and use it to define the
Chern character homomorphism in topological KO-homology. We describe
various properties of this homomorphism, most notably its intimate
connection with the AMS invariant which was the crux of the
isomorphism of the previous section.

\subsection{The Complexification Homomorphism\label{ComplexHomo}}

Let $X$ be a compact topological space. Consider the topological,
generalized homology groups $\K_{\sharp}^{\rm t}(X)$ and
$\KO_{\sharp}^{\rm t}(X)$, along with the corresponding K-theory and
KO-theory groups. The complexification of a real vector bundle over
$X$ is a complex vector bundle over $X$ which is isomorphic to its own
conjugate vector bundle. The complexification map is compatible with
stable isomorphism of real and complex vector bundles, and thus
defines a homomorphism from stable equivalence classes of real vector
bundles to stable equivalence classes of complex vector bundles. It
thereby induces a natural transformation of cohomology theories
$$(\otimes\,\complex)^* \,:\, \KO^{\sharp}(X) ~\longrightarrow~
\K^{\sharp}(X)$$
given by
$$ [E]-[F] ~\longmapsto~ [E_{\mathbb{C}}]-[F_{\mathbb{C}}]$$
where $E_{\mathbb{C}}:=E\otimes\complex$ is the complexification of
the real vector bundle $E\to X$.

We can also define a complexification morphism relating the
homology theories
\beq
(\otimes\,\complex)_* \,:\, \KO_{\sharp}^{\rm t}(X) ~\longrightarrow~
\K_{\sharp}^{\rm t}(X)
\label{complexmaphom}\eeq
by $$ [M,E,\phi]\otimes\complex:=[M,E_{\mathbb{C}},\phi]$$
and extended by linearity, where on the right-hand side we regard $M$
endowed with the spin$^c$ structure induced by its spin structure as
a KO-cycle. One can easily see that
\beq
[M,E,\phi]\otimes\complex
=\phi_*\bigl([E_\complex]\,\frown\, [M]_\K\bigr)
\label{complexexpl}\eeq
where $[M]_\K\in\K^\sharp(M)$ denotes the K-theory fundamental class of
$M$. Thus in the case when $X$ is KO-oriented (and therefore
K-oriented), i.e. $X$ is a compact spin manifold, the homomorphism
$(\otimes\,\complex)_*$ is just the Poincar\'e dual of
$(\otimes\,\complex)^*$. This is clearly a natural transformation of
homology theories.

A related natural transformation between cohomology theories is the
realification morphism
$$(\,\underline{\phantom{E}}
\,\real)^* \,:\, \K^{\sharp}(X) ~\longrightarrow~ \KO^{\sharp}(X)$$
induced by assigning to a complex vector bundle over $X$ the
underlying real vector bundle over $X$. Because a spin$^c$ manifold is not
necessarily spin, we cannot implement this transformation in the
homological setting in general. Rather, we must assume that $X$ is a
compact spin manifold. In this case the K-homology group $\K^{\rm
  t}_\sharp(X)$ has generators~\cite{RS} $[X \times \S^n,E_i,\pr_1]-[X
\times\S^n,F_i,\pr_1]$, $0 \leq n \leq 7$, where $\pr_1:X\times\S^n\to
X$ is the projection onto the first factor. We can then define the
morphism
$$(\,\underline{\phantom{E}}\,\real)_* \,:\, \K_{\sharp}^{\rm t}(X)
~\longrightarrow~ \KO_{\sharp}^{\rm t}(X)$$
by
$$\bigl([X \times \S^n,E_i,\pr_1]-[X \times \S^n,F_i,\pr_1]\bigr)
\,\underline{\phantom{E}}\,\real:=
[X \times \S^n,E_i\,\underline{\phantom{E}}\,\real,\pr_1]-
[X \times \S^n,F_i\,\underline{\phantom{E}}\,\real,\pr_1]$$ and
extending by linearity. Since this definition depends on a choice of
generators for $ \K_{\sharp}^{\rm t}(X)$, the transformation is not
natural. As for the complexification morphism, the morphism
$(\,\underline{\phantom{E}}\,\real)_*$ thus defined is Poincar\'e dual
to $(\,\underline{\phantom{E}}\,\real)^*$. It follows that the
composition $(\,\underline{\phantom{E}}\,\real)_* \circ
(\otimes\,\complex)_*$ is multiplication by 2.

\subsection{Chern Character in KO-Homology\label{ChernKOHom}}

We can use the natural transformation provided by the complexification
homomorphism (\ref{complexmaphom}) to define a real homological Chern
character
\beq
\ch_{\bullet}^{\mathbb{R}} \,:\, \KO_{\sharp}^{\rm t}(X) ~\longrightarrow~
\H^{\phantom{1}}_{\sharp}(X,\mathbb{Q})
\label{realCherndef}\eeq
by
$$\ch_{\bullet}^{\mathbb{R}}(\xi)=\ch^{\phantom{1}}_{\bullet}(\xi\otimes\complex) $$
for $\xi\in\KO_\sharp^{\rm t}(X)$, where on the right-hand side we use
the K-homology Chern character $\ch^{\phantom{1}}_\bullet:\K^{\rm
  t}_\sharp(X)\to\H^{\phantom{1}}_\sharp(X,\rat)$. Tensoring with
$\mathbb{Q}$ gives a map
$$\ch_{\bullet}^{\mathbb{R}} \otimes \Id_{\mathbb{Q}}=
\big(\ch^{\phantom{1}}_{\bullet} \otimes
\Id_{\mathbb{Q}}\big) \circ \big((\otimes\,\complex)_* \otimes
\Id_{\mathbb{Q}}\big)\,:\, \KO_{\sharp}^{\rm t}(X)\otimes_{\mathbb{Z}}
\mathbb{Q}~\longrightarrow~
\H^{\phantom{1}}_{\sharp}(X,\mathbb{Q})$$ with
$\ch^{\phantom{1}}_{\bullet} \otimes \Id_{\mathbb{Q}}:\K_\sharp^{\rm
  t}(X)\otimes_\zed\rat\to\H_\sharp^{\phantom{1}}(X,\rat)$ an
isomorphism. The real Chern character (\ref{realCherndef}) is a
natural transformation of homology theories.

An important point here is that the real Chern
character requires a somewhat finer analysis than the usual Chern
character. Although it detects all the homology classes, there can be
KO-homology elements which have the same image under it because of the
complexification map and the different periodicities of K-theory and
KO-theory. For example, consider the KO-cycles
$[\rm{pt},\id^\reals_{\rm{pt}},\Id_\pt]$ and
$[\S^4,\id^\reals_{\S^4},\zeta]$ over $\pt$. They have the same image
through $\ch_{\bullet}^{\mathbb{R}}$, namely the generator of
$\H_0(\rm{pt},\mathbb{Q})$. But since they belong to different
subgroups $\KO_i^{\rm t}(\rm{pt})$ with respect to the grading of
$\KO_\sharp^{\rm t}(\pt)$, we conclude that these are the generators
of the lattice $\Lambda_{\KO_\sharp^{\rm t}(\rm{pt})}:=\KO_\sharp^{\rm
t}(\rm{pt})\,/\,\tor_{\KO_\sharp^{\rm
    t}(\rm{pt})}$. This fact will be important when we study brane
constructions in the next section.

We can give a characteristic class description of
$\ch_{\bullet}^{\mathbb{R}}$ as follows. Let $\tau_{E}^{\KO}$ be the
KO-theory Thom class and $\tau_{E}^{\H}$ the cohomology Thom class of
a real spin vector bundle $E$ over $X$. Let
$\ch^\bullet:\K^\sharp(X)\to\H^\sharp(X,\rat)$ be the usual cohomology
Chern character which is a multiplicative $\zed_2$-degree preserving
natural transformation of cohomology theories. Denote by
$\Aroof(E)\in\H^{\rm even}(X,\rat)$ the Atiyah-Hirzebruch class
of $E$. By using the analysis of natural transformations given
in~\cite{3}, along with the Hirzebruch formulation of the Riemann-Roch
formula
$$\ch^\bullet\big((\tau_{E}^{\KO})\otimes\complex\big)=
\tau_{E}^{\H} \,\smile\, \Aroof (E)^{-1} $$
and (\ref{complexexpl}), one then has
\beq
\ch_{\bullet}^{\mathbb{R}}(M,E,\phi)=
\phi_*\big(\ch^\bullet(E_\complex) \,\smile\,
\Aroof (TM)\,\frown\, [M]\big)
\label{KOhomChern}\eeq
where $[M]\in\H_\sharp(M,\zed)$ is the orientation cycle of the
compact spin manifold $M$. Since
$E_\complex\cong\overline{E_\complex}$ for any real vector bundle
$E\to X$, one has
$\ch^\bullet(E_\complex)=\ch^\bullet(\,\overline{E_\complex}\,)$. Thus
all components of the cohomology Chern character in the formula
(\ref{KOhomChern}) of degree $4j+2$ vanish.

\subsection{$\mbf{\Cl_n}$-Index Theorems\label{ClnIndexThms}}

We will now explore the relation between the homological real
Chern character and the topological index defined in
(\ref{indntdef}). We first show that up to Poincar\'e duality the
topological index is the homological morphism induced by the
collapsing map. Recall that up to isomorphism, the AMS invariant is
given by
\begin{displaymath}
\widehat{\mathcal{A}}_{E}(M)=\tilde\zeta^{\,\KO}_{!}[E]
\end{displaymath}
where $M$ is a compact spin manifold of dimension $n$, $E$ is a real
vector bundle over $M$, $\tilde\zeta:M\to\pt$ is the collapsing map on
$M$, and $\tilde\zeta_!^{\,\KO}$ is the corresponding Gysin
homomorphism on KO-theory. In this case we have
\begin{displaymath}
\tilde\zeta_{*}^{\,\KO}=\Phi^{\phantom{1}}_{\pt}\circ\tilde
\zeta^{\,\KO}_{!}\circ\Phi^{-1}_{M}
\end{displaymath}
where $\tilde\zeta^{\,\KO}_{*}$ is the induced morphism on $\KO^{\rm
  t}_{\sharp}(M)$, and $\Phi_{\pt}$ and $\Phi_{M}$ are the Poincar\'e
duality isomorphisms on $\pt$ and $M$, respectively. It then follows
that
\begin{eqnarray}
\Phi_{\pt}^{\phantom{1}}\circ{}\ind_n^{\rm t}(M,E,\phi)&=&
\Phi^{\phantom{1}}_{\pt}\circ\tilde\zeta^{\,\KO}_{!}[E]\nonumber\\
&=&\Phi^{\phantom{1}}_{\pt}\circ\tilde\zeta^{\,\KO}_{!}\circ
\Phi^{-1}_{M}(M,E,\Id_M)\nonumber\\
&=&\tilde\zeta_{*}^{\,\KO}[M,E,\Id_M]\nonumber\\
&=&[M,E,\tilde\zeta\,]~=~\zeta_{*}^\KO[M,E,\phi]
\label{topindhommorph}\end{eqnarray}
where $\zeta:X\to{\pt}$ is the collapsing map on $X$ with
$\tilde\zeta=\zeta\circ\phi$.

We will next describe how the real Chern character can be used to give
a characteristic class description of the map $\ind_n^{\rm t}$ in the
torsion-free cases. Consider first the case $n\equiv4~{\rm mod}\,
8$. We begin by showing that there is a commutative diagram
\beq
\xymatrix{\KO^{\rm t}_{4}(X)\ar[r]^{\zeta^{\KO}_{*}}
\ar[rd]_{\zeta^{\H}_{*}\circ{}\ch^{\mathbb{R}}_{\bullet}~} &
\KO^{\rm t}_{4}(\pt)\ar[d]^{\ch^{\mathbb{R}}_{\bullet}}\\ &
\H_{0}(\pt,\mathbb{Q})}
\eeq
where $\zeta_*^\H$ is the induced morphism on homology. Recall that
$\ch^{\mathbb R}_{\bullet}=\ch^{\phantom{1}}_{\bullet}\circ(\otimes\,\complex)_*$,
where $(\otimes\,\complex)_{*}$ is the complexification map
(\ref{complexmaphom}). Then one has
\begin{eqnarray*}
\zeta^{\H}_{*}\circ{}\ch^{\mathbb{R}}_{\bullet}(M,E,\phi)&=&
\zeta^{\H}_{*}\circ{}\phi_{*}\bigl(\ch^{\bullet}(E)\,\smile\,{}
\widehat{{A}}(TM)\,\frown\,[M]\bigr)\\
&=&(\zeta\circ{\phi})_{*}\bigl(\ch^{\bullet}(E)\,\smile\,
\widehat{{A}}(TM)\,\frown\,[M]\bigr)\\
&=&\ch^{\mathbb{R}}_{\bullet}(M,E,\tilde\zeta\,)~=~
\ch^{\mathbb{R}}_{\bullet}\circ\zeta^{\KO}_{*}[M,E,\phi] \ .
\end{eqnarray*}

Now recall from Section~\ref{ChernKOHom} above that the map
$\ch^{\mathbb{R}}_{\bullet}:\KO^{\rm t}_{4}(\pt)\to{\H_{0}(\pt,\mathbb{Q})}$
sends $\mathbb{Z}\to{2\mathbb{Z}}\subset{\mathbb{Q}}$. On its image,
the homomorphism $\ch^{\mathbb{R}}_{\bullet}$ is thus invertible and
its inverse is given as division by 2. An explicit realization is
gotten by noticing that
\begin{eqnarray}
\zeta^{\H}_{*}\circ{}\ch^{\mathbb{R}}_{\bullet}(M,E,\phi)&=&
\zeta^{\H}_{*}\circ{\Phi_{M}}\bigl(\ch^{\bullet}(E_{\mathbb{C}})\,\smile\,
{\widehat{{A}}(TM)}\bigr)\nonumber\\
&=&\Phi_{\pt}\circ{}\zeta^{\H}_{!}\bigl(\ch^{\bullet}(E_{\mathbb{C}})\,
\smile\,\widehat{{A}}(TM)\bigr)~=~
\bigl\langle\ch^{\bullet}(E_{\mathbb{C}})\,\smile\,{\widehat{{A}}(TM)}\,,\,
[M]\bigr\rangle \ ,
\label{div2expl}\end{eqnarray}
where $\langle-,-\rangle:\H^\sharp(M,\rat)\times\H_\sharp(M,\rat)\to\rat$ is
the canonical dual pairing between cohomology and homology. In
(\ref{div2expl}) we have used the fact that $\Phi_{\pt}$ is the
identity on $\H_{0}(\pt,\mathbb{Q})\cong\mathbb{Q}\,$, and the proof of
the last equality uses the Atiyah-Hirzebruch version of the
Grothendieck-Riemann-Roch theorem and can be found in Section~V4.20
of~\cite{7}. Recall that for a spin manifold $M$ of dimension $4k+8$,
one has
$\bigl\langle\ch^{\bullet}(E_{\mathbb{C}})\,\smile\,{\widehat{{A}}(TM)}\,,\,
[M]\bigr\rangle\in2\zed$. After using the isomorphism
$\KO_{4}(\pt)\cong\mathbb{Z}$, we thus deduce that
$\zeta_*^\KO[M,E,\phi]=
\frac12\,\bigl\langle\ch^{\bullet}(E_{\mathbb{C}})\,\smile\,
{\widehat{{A}}(TM)}\,,\, [M]\bigr\rangle$, and from
(\ref{topindhommorph}) we arrive finally at
\begin{displaymath}
\ind_n^{\rm t}(M,E,\phi)=\mbox{$\frac12$}
\,\bigl\langle\ch^{\bullet}(E_{\mathbb{C}})\,\smile\,
{\widehat{{A}}(TM)}\,,\, [M]\bigr\rangle \ .
\end{displaymath}

When $n\equiv0\:{\rm mod}\:8$, one obtains a similar result but now
without the factor $\frac{1}{2}$, since in that case
$\ch^{\mathbb{R}}_{\bullet}:\KO^{\rm t}_0(\pt)\to\H_0(\pt,\rat)$ is the
inclusion $\mathbb{Z}\hookrightarrow\mathbb{Q}$.
In the remaining non-trivial cases $n\equiv1,2~{\rm mod}~8$
the homological Chern character is of no use, as $\KO^{-n}(\pt)$ is
the pure torsion group $\mathbb{Z}_{2}$, and there is no cohomological
formula for the AMS invariant in these instances. However, by using
Theorem~\ref{spinthm} one still has an interesting mod~2 index formula
for the topological index in these cases as well~\cite{ASSkew}. We can
summarize our homological derivations of these index formulas as
follows.

\begin{theorem}
Let $[M,E,\phi]\in\KO^{\rm t}_n(X)$, and let
$~\nslash{\mathfrak{D}}^M_{E}$ be the Atiyah-Singer operator on $M$ with
coefficients in $E$. Let
$~\nslash{\mathcal{H}}^M_E:=\ker~\nslash{\mathfrak{D}}^M_{E}$ denote the
vector space of real harmonic $E$-valued spinors on $M$. Then one has
the $\Cl_n$-index formulas
$$
\ind_n^{\rm t}(M,E,\phi)~=~\left\{\begin{array}{cc}
\bigl\langle\ch^{\bullet}(E_{\mathbb{C}})\,\smile\,
{\widehat{{A}}(TM)}\,,\, [M]\bigr\rangle \ , & \quad n\equiv0\:{\rm
  mod}\:8 \ , \\ \dim_\complex~\nslash{\mathcal{H}}^M_E~{\rm
  mod}\:2 \ , & \quad n\equiv1\:{\rm mod}\:8 \ , \\
\dim_{\mathbb{H}}~\nslash{\mathcal{H}}^M_E~
{\rm mod}\:2 \ , & \quad n\equiv2\:{\rm mod}\:8 \ , \\
\mbox{$\frac12$}\,\bigl\langle\ch^{\bullet}(E_{\mathbb{C}})\,\smile\,
{\widehat{{A}}(TM)}\,,\, [M]\bigr\rangle \ , & \quad n\equiv4\:{\rm
  mod}\:8 \ , \\ 0 \ , & {\rm otherwise} \ . \end{array} \right.
$$
\label{Clnindexthm}\end{theorem}

\newsection{Brane Constructions in Type I String
  Theory\label{TypeIBranes}}

In Type~I superstring theory with topologically trivial $B$-field, a
D-brane in an oriented ten-dimensional spin manifold $X$ is usually
described by a spin
submanifold $W\hookrightarrow X$, together with a Chan-Paton bundle
which is equiped with a superconnection and defined by an element
$\xi\in\KO^0(X)$~\cite{Witten} (see~\cite{RS} for a more precise
treatment). In this section we will apply the mathematical
formalism developed thus far to the classification and construction of
Type~I D-branes in topological KO-homology. The main new impetus that
we will emphasize is the role of the AMS geometric invariant, which
was the crucial ingredient in the proof of
Section~\ref{isomorphism}. It will provide a precise, rigorous
framework for certain physical aspects of Type~I brane
constructions. We will begin by demonstrating how geometric K-homology
can be used to describe D-branes in Type II and Type I Ssring theory
in a topologically nontrivial spacetime. We will introduce the notion
of \emph{wrapped D-brane} on a given submanifold of spacetime, we will
define the group of charges of wrapped D-branes, and construct
explicit examples of wrapped D-branes which have torsion charge.

\subsection{Classification of Type~I D-Branes\label{DClass}}

By the Sen-Witten construction~\cite{Sen:1998tt,Witten}, the group of
topological charges of a Type~II D$p$-brane realized as a ${\rm
  spin}^{c}$ submanifold ${W}\subset{{X}}$ is given by
\begin{displaymath}
\K_{\rm cpt}^{0}({\nu_{{W}}})\cong\K^{0}({\B(\nu_W)},{\S(\nu_W)}) \ ,
\end{displaymath}
where $\nu_W$ is the normal bundle $N(X/W)\to W$ of $W\subset X$,
equiped with the spin$^c$ structure induced by the spin$^c$ structure
on $W$ and the spin structure on $X$, and $\B(\nu_{W})$ and
$\S(\nu_{W})$ are respectively the unit ball and sphere bundle of the
normal bundle. Let $W'$ be a tubular neighbourhood of $W$ in $X$,
which we can identify with the interior of the ball bundle
$\B(\nu_{W})\setminus\S(\nu_{W})$. By Poincar\'e duality, it follows
that the elements of the K-theory group $\K_{\rm
  cpt}^{0}(\nu_{W})\cong \K^{\rm cpt}_{10}({\nu_{{W}}})$ can be naturally
interpreted as spacetime-filling D9-branes (or D9 brane-antibrane
pairs) in $X$. This requires extending the Chan-Paton bundles over
$\overline{W'}\cong\B(\nu_W)$ to $X\setminus W'$ in the standard
way~\cite{ABS1,Witten,16,RS}, possibly by stabilizing with the
addition of extra brane-antibrane pairs. In the following, we will
always assume that this has been implicitly done and identify the
normal bundle $\nu_W$ with spacetime $X$ itself.

According to the Sen-Witten construction, the
classes in $\K_{\rm cpt}^{0}({\nu_{W}})$ are interpreted as systems of
$\rm D9-D\bar{9}$ branes which are unstable, and will decay onto the
worldvolume ${W}$, which correspond to the zero loci of the
appropriate tachyon field. In particular, this process happens in
spacetime, and it depends on how the worldvolume is embedded in it. On
the other hand, the role played by the Chan-Paton vector bundle on the
D$p$-brane is not manifest in this classification. However, there is a
natural way of classifying the D$p$-branes on ${W}$ by means which
manifestly takes into account the Chan-Paton bundle
contribution. Indeed, from the D$p$-brane data, we can naturally
construct the Baum-Douglas cycle $({W},{E},{\rm id})$, where ${E}$
denotes the Chan-Paton bundle, and declare that its charge is given by
the class $[{W},{E},{\rm id}]\in\K_{p+1}({W})$. As the group
$\K_{p+1}({W})$ contains no information about the embedding of the
worldvolume ${W}$ in ${X}$, we can intuitively think the charge
$[{W},{E},{\rm id}]$ takes into account how the D-brane wraps the
submanifold ${W}$. Notice that this analogous to the charge
classification of an extended object in an abelian gauge theory via
the homology cycle of its worldvolume. At this point, we notice that
by definition the elements of $\K_{p+1}^{\rm t}({W})$ are given by
(differences of) classes $[{M},{E},\phi]$ where ${M}$ is a
$p+1$-dimensional manifold. However, it is \emph{not} always possible
to choose the map $\phi$ in $[{M},{W}]$ in such a way that $\phi$ is a
diffeomorphism. This motivates the following definition.
\begin{definition} Let ${X}$ be a Type II string background,
  described by a ten-dimensional spin manifold, and let
  ${W}\subset{{X}}$ be a ${\rm spin}^{c}$ submanifold. A D$p$-brane
  \emph{wrapping} the worldvolume ${W}$ is defined as the K-cycle
  $({M},{E},\phi)$, where dim ${M}$ =$p+1$, and
  $\phi({M})\subset{{W}}$. We will call ${E}$ the Chan-Paton bundle
  associated to the wrapped D$p$-brane, and we will say that the
  D$p$-brane \emph{fills} ${W}$ if $\phi({M})={W}$. The charge of the
  wrapped D$p$-brane is given by the class $[{M},{E},\phi]$ in the
  group $\K_{p+1}^{\rm t}({W})$.
\end{definition}
Notice that in the above definition we have relaxed the condition that
$\dim W=p+1$, as we are not requiring that the wrapping
preserves the dimension of the D-brane. This is an attempt to take
into account, at least at the topological level, the well-known fact
that D-branes are not always representable as submanifolds equipped
with vector bundles, since they are boundary conditions for a
superconformal field theory, and that a distinction should be made
between the wrapping D-brane, in this case identified with a K-cycle
representing a particular type of boundary conditions, and the
worldvolume it wraps. Notice also that the group of charges of wrapped
D$p$-branes does \emph{not} depend on how the manifold ${W}$ is
embedded into the spacetime, and hence it seems to represent a genuine
worldvolume concept. In particular, as mentioned above, the wrapped
D-brane definition is very natural  in the ordinary case of a D-brane
realized as a submanifold ${W}$ of spacetime equipped with a
Chan-Paton bundle ${E}$, as it only depends on how the vector bundle
is defined on the submanifold, and not on the procedure used to
``extend'' it to the spacetime. Finally, in the case of ordinary
D-branes wrapping  ${W}$ with $\dim W=p+1$, the group
$\K_{p+1}({W})$ coincides with the group of charges of Type IIB
D$p$-branes that can be obtained via the Sen-Witten construction,
i.e. via brane-antibrane decay. This can be shown as follows.  Since
the normal bundle $N(X/W)\to{{W}}$ is a ${\rm spin}^{c}$ vector
bundle of even rank $9-p$ in this case, we can use the Thom
isomorphism in K-theory to establish that
\begin{displaymath}
\K_{\rm cpt}^{0}({\nu_{{W}}})\cong{\K^{0}({W})} \ .
\end{displaymath}
As ${W}$ is a ${\rm spin}^{c}$ submanifold of the spacetime, we can
use Poincar\'e duality to get
\begin{displaymath}
\K^{0}({W})\cong{\K^{t}_{p+1}({W})}
\end{displaymath}
where $p+1=\dim W$. This suggests that for ordinary Type II
D$p$-branes the wrapping charge is completely determined by the decay
of the tachyon field.

It is natural at this point to extend the notion
of wrapped D-brane and of wrapping charge to Type I string theory. In
this case, though, the two notions of charge do not coincide, as we
will show in the following.
Recall that in Type I string theory the group of topological charges
of a D$p$-brane realized as a ${\rm spin}$ submanifold
${W}\subset{{X}}$ is given by
\begin{equation}\label{braneanti}
\KO_{\rm cpt}^{0}({\nu_{{W}}})\cong\KO^{0}({\B(\nu_W)},{\S(\nu_W)}) \ .
\end{equation}
By using the Thom isomorphism in KO-theory, we have
\begin{displaymath}
\KO_{\rm cpt}^{0}({\nu_{{W}}})\cong{\KO^{p-9}({W})} \ .
\end{displaymath}
Finally, by Poincar\'e duality, we get
\begin{displaymath}
\KO^{p-9}({W})\cong{\KO_{10}({W})} \ .
\end{displaymath}
The group $\KO_{10}({W})$ is in general not isomorphic to the group
$\KO_{p+1}({W})$, and explicitly depends on the dimension of the
spacetime. We can physically interpret the elements of $\KO_{10}({W})$
as equally charged systems of wrapping $\rm D9-D\bar{9}$-branes
decaying on the submanifold ${W}$, and via the inclusion
$i:{W}\hookrightarrow{{X}}$ they can be related to the D9-branes used
in the Sen-Witten construction. This is not surprising, as the decay
mechanism is somehow at the heart of the spacetime D-brane charge
classification, and it reinforces the statement that the group
(\ref{braneanti}) encodes spacetime properties of the D$p$-brane. higher
i.e. as vector bundles defined over the spacetime. (Higher-degree
KO-theory groups, while having no natural interpretations in terms of
D-branes, in fact arise through the chain of orientifolds one
encounters when taking T-duals of the Type~I theory and require the
use of KR-theory~\cite{Berg1,OS1}).

The topological charges of the D-branes arising in this way are
provided by the AMS invariant~(\ref{AMSinvdef}), or equivalently by
the topological index as computed in the $\Cl_n$-index
theorems~\ref{spinthm} and~\ref{Clnindexthm}. This naturally links
the D-brane charge to a fermionic field theory on the brane worldvolume, as
$~\nslash{\mathfrak{D}}^M_{E}$ is the Atiyah-Singer operator defined on
sections of the irreducible spinor bundle over $M$ coupled to the real
vector bundle $E\to M$. The precise form of the charge in
Theorem~\ref{Clnindexthm} is dictated by whether the corresponding
spinor representations are real, complex or pseudo-real. Most
noteworthy are the (non-BPS) torsion charges. The AMS invariant in
these instances gives a precise realization to the notion of a
``$\zed_2$ Wilson line'' which is usually used in the physics
literature for the construction of torsion D-branes in Type~I string
theory~\cite{Sen:1998tt,Witten,Bergman:2000tm}. It is defined as a
non-trivial element in the set of $\real/\zed$-valued gauge holonomies
on $M$ which are invariant under the
involution which sends a complex vector bundle $V$ to its complex
conjugate $\overline{V}$. Within our framework, it is determined
instead by the coupling of the branes to the worldvolume
fermions~$\psi$, valued in $E$, which are solutions of the harmonic equation
$~\nslash{\mathfrak{D}}^M_{E}\psi=0$. This provides a rigorous framework
for describing the torsion charges, and moreover identifies the
bundles used in tachyon condensation processes as the usual spinor
bundles coupled to the Chan-Paton bundle $E$. We will see some
explicit examples in Section~\ref{TorsionCalcs} below.

\subsection{Wrapped Branes}

Let us now make some of these constructions more explicit. Given the
real Chern character, we can mimick some (but not all) of the
constructions of Type~II D-branes in complex K-homology. However, in
light of the remarks made in Section~\ref{ChernKOHom}, special care
must be taken as the Chern character in the real case is {\it not} a
rational injection. With this in mind, we have the following
adaptation of Theorem~2.1 from~\cite{RS}.

\begin{theorem}
 Let $X$ be a compact connected finite CW-complex of dimension
$n$ whose rational homology can be presented as
$$\H_\sharp(X,\mathbb{Q})= \bigoplus_{p=0}^{n}~ \bigoplus_{i=1}^{m_p}
\,\big[M^{p}_{i}\big] {}~\mathbb{Q} \ , $$ where $M_i^{p}$ is a
$p$-dimensional compact connected spin submanifold of $X$ without
boundary and with orientation cycle $[M_i^{p}]$ given by the spin
structure. Suppose that the canonical inclusion map
$\iota_i^{p}:M_i^p\hookrightarrow X$ induces, for each $i,p$, a
homomorphism
$(\iota_i^p)_*:\H_p(M_i^p,\rat)\to\H_p(X,\rat)\cong\rat^{m_p}$ with
the property \beq
\big(\iota_i^p\big)_*\big[M_i^p\big]=\kappa_{ip}\,\big[M_i^p\big]
\label{bigtcond}\eeq for some $\kappa_{ip}\in\rat\setminus\{0\}$.
Then the KO-homology lattice $\Lambda_{\KO_\sharp^{\rm
    t}(X)}:=\KO_\sharp^{\rm t}(X)\,/\,\tor_{\KO_\sharp^{\rm
    t}(X)}$ contains a set of linearly independent elements given by
the classes of KO-cycles
$$\big[M_{i}^p,\id^\real_{M_i^p},\iota_i^p\big] \ , \quad 0 \leq p\leq
n \ , ~~ 1 \leq i\leq m_p \ . $$
\label{HomKOcycles}\end{theorem}

\begin{proof} By~\cite{RS} the cycles
  $\big[M_{i}^p,\id^\complex_{M_i^p},\iota_i^p\big]$ form a basis for
  the lattice
$\Lambda_{\K_\sharp^{\rm t}(X)}:=\K_\sharp^{\rm
t}(X)\,/\,\tor_{\K_\sharp^{\rm
    t}(X)}$ in K-homology. The conclusion follows from the fact that
    $$\big[M_{i}^p,\id^\real_{M_i^p},\iota_i^p\big]\otimes\complex
=\big[M_{i}^p,\id^\complex_{M_i^p},\iota_i^p\big] \ , $$
     i.e. that the elements $\ch_{\bullet}^{\mathbb{R}}
     \big(M_{i}^p,\id^\reals_{M_i^p},\iota_i^p\big)$
     form a set of generators of $\H_\sharp(X,\mathbb{Q})$.
\end{proof}

Theorem~\ref{HomKOcycles} provides sufficient combinatorial criteria
on the rational homology of $X$ which ensure that torsion-free
D-branes can wrap non-trivial spin cycles of the spacetime $X$. As in
the complex case, this is related to an analogous problem for the spin
bordism group ${\rm MSpin}_\sharp(X)$, which can also be defined in
terms of a spectrum $\underline{\rm MSpin}^\infty$. Just as in
K-theory, the Atiyah-Bott-Shapiro (ABS) orientation map~\cite{ABS1}
$\underline{\rm MSpin}^\infty\rightarrow \underline{\KO}^\infty$
induces an ${\rm MSpin}_{\sharp}(\rm{pt})$-module structure on
$\KO^{\rm t}_{\sharp}(\rm{pt})$. Then analogously to the complex case
we have the following result~\cite{HH}.

\begin{theorem} The map $${\rm MSpin}_{\sharp}(X) \otimes_{{\rm
      MSpin}_{\sharp}({\rm pt})}
\KO_{\sharp}^{\rm t}({\pt}) ~\longrightarrow~ \KO^{\rm t}_{\sharp}(X)
\ , \qquad
\big[M\,,\,\phi\big]~\longmapsto~\big[M\,,\,\id_M^\real\,,\,\phi
\big] $$ induced by the ABS orientation is a natural isomorphism of
$\KO^{\rm t}_{\sharp}(\rm{pt})$-modules for any finite CW-complex
$X$. \end{theorem}
\noindent
This immediately implies the following result, reducing the problem
of calculating the KO-homology generators to the analogous problem in
spin bordism.

\begin{theorem}
Let $X$ be a finite CW-complex. Suppose that $[M_i,\phi_i]$, $1 \leq i
\leq m$ are the generators of ${\rm MSpin}_{\sharp}(X)$ as an
${\rm MSpin}_{\sharp}(\rm{pt})$-module. Then
$[M_i,\id_{M_i}^\real,\phi_i]$, $1 \leq i \leq m$ generate
${\KO}^{\rm t}_{\sharp}(X)$ as a $\KO^{\rm t}_{\sharp}(\rm{pt})$-module.
\end{theorem}
\noindent
In other words, for each $n=0,1, \ldots, 7 $ the group $\KO^{\rm
  t}_{n}(X)$ is generated by elements $[M_i,\id_{M_i}^\real,\phi_i]$,
$1 \leq i \leq m$ with $\dim M_i=n$.

\subsection{Torsion Branes\label{TorsionCalcs}}

We now describe a geometrical approach to the computation of torsion
KO-cycle generators, thus elucidating the role of the AMS invariant in
the construction of torsion D-branes. The general problem in
KO-homology turns out to be much more involved than in the complex
case. We discuss this further in Section~\ref{GenConstr} below. For
now we will content ourselves with finding explicit representatives
for the generators of the non-trivial groups $\KO^{\rm t}_n(\rm{pt})$ with
$n=0,1,2,4$. This entails instructive exercises in the
computations of topological indices which aid in better understanding
the origins of Type~I torsion D-brane charges. Recall from
Section~\ref{ChernKOHom} that for the non-torsion cases $n=0$ and
$n=4$, using the real Chern character one finds that the classes
$[\rm{pt},\id_{\rm{pt}}^{\real},\Id^{\phantom{1}}_\pt]$ and
$[\S^{4},\id_{\S^4}^{\real},\zeta]$ are generators of the groups
$\KO^{\rm t}_{0}(\rm{pt})\cong\zed$ and $\KO^{\rm
  t}_{4}(\rm{pt})\cong\zed$, respectively.

We begin with the group $\KO_1^{\rm t}(\pt)$. Consider the
circle $\S^{1}$ and assign to it a Riemannian metric. Since there is
only one unit tangent vector at any point of $\S^{1}$, one has
$P_{{\,\SO}}(\S^{1})\cong{\S^{1}}$. A spin structure on $\S^{1}$ is thus
given by a double covering
\begin{displaymath}
P_{{\,\Spin}}\big(\S^{1}\big)~\longrightarrow~{\S^{1}}
\end{displaymath}
and by the fibration
\begin{displaymath}
\xymatrix{
{\mathbb{Z}_{2}}~\ar[r]~&~P_{{\,\Spin}}\big(\S^{1}\big)\ar[d] \ . \\ &{\S^{1}}}
\end{displaymath}
There are only two double coverings of the circle, one disconnected
and the other connected, given respectively by
\begin{displaymath}
\S^{1}\times{\mathbb{Z}_{2}}~\longrightarrow~{\S^{1}} \ , \qquad
\S^{1}_{\rm M}~\longrightarrow~{\S^{1}}
\end{displaymath}
where $\S^{1}_{\rm M}$ is the total space of the principal
$\mathbb{Z}_{2}$-bundle associated to the M\"obius strip.
We will call these two spin structures the ``interesting'' and
the ``uninteresting'' spin structures, respectively.

Corresponding to these two spin structures (labelled `i' and `u',
respectively), we construct classes in $\KO^{\rm t}_{1}(\rm{pt})$
given by $[\S_{\rm i}^{1},\id_{\S^{1}}^{\real},\zeta]$ and $[\S_{\rm
  u}^{1},\id_{\S^{1}}^{\real},\zeta]$ where $\zeta:\S^{1}\to{\rm{pt}}$
is as usual the collapsing map. We will now compute the topological
indices in detail, finding the AMS invariants~\cite{SG}
\begin{displaymath}
\widehat{\mathcal{A}}_{\id_{{\S}^{1}}^{\real}}\big(\S_{\rm i}^{1}
\big)~=~1 \ , \qquad
\widehat{\mathcal{A}}_{\id_{{\S}^{1}}^{\real}}\big(\S_{\rm u}^{1}\big)~=~0
\end{displaymath}
in $\KO^{-1}(\rm{pt})\cong{\mathbb{Z}_{2}}$. Hence the two classes
above represent the elements of $\KO^{\rm
  t}_{1}(\rm{pt})\cong{\mathbb{Z}_{2}}$. In particular,
$[\S_{\rm i}^{1},\id_{{\S}^{1}}^{\real},\zeta] $ is a generator, analogous to the
non-BPS Type~I D-particle that arises from tachyon condensation on the
Type~I D1 brane-antibrane system with a $\zed_2$ Wilson
line~\cite{Sen:1998tt,Witten,Bergman:2000tm}.

Let us first consider the circle with the interesting spin structure.
Since $\Cl_{1}\cong{\mathbb{C}}$, one has
$~\nslash{\mathfrak{S}}(\S^1):=P_{{\,\Spin}}(\S^{1})\times_{\zed_2}\Cl_{1}\cong
{\S^{1}\times{\mathbb{C}}}$. By decomposing
$\mathbb{C}=\mathbb{R}\oplus{\ii\mathbb{R}}$, one has the
identifications
$~\nslash{\mathfrak{S}}^{0}(\S^1)=\S^{1}\times{\mathbb{R}}$ and
$~\nslash{\mathfrak{S}}^{1}(\S^1)=\S^{1}\times{\ii\mathbb{R}}$. As the
Clifford bundle is trivial, its space of sections is given by
$\C^\infty(\S^1,~\nslash{\mathfrak{S}}(\S^1))=
\C^{\infty}(\S^{1},\mathbb{C})$.
By coordinatizing the circle $\S^{1}$ with arc length $s$, the
Atiyah-Singer operator (\ref{ASop}) can be expressed as
\beq
\,\nslash{\mathfrak{D}}^{\S^1}=\ii\,\mbox{$\frac{\dd}{\dd s}$}
\label{ASopS1}\eeq
where $e_{1}=\ii$ is a generator of the Clifford algebra $\Cl_{1}$. To
compute the topological index
$\widehat{\mathcal{A}}_{\id_{{\S}^{1}}^{\real}}(\S_{\rm i}^{1})$, we
use the $\Cl_1$-index Theorem~\ref{spinthm} and hence determine the vector
space $\ker~\nslash{\mathfrak{D}}^{\S^1}$, or equivalently the chiral
subspace $\ker(\,\nslash{\mathfrak{D}}^{\S^1})^{0}$. Since
$\C^\infty(\S^1,~\nslash{\mathfrak{S}}^{0})=
\C^{\infty}(\S^{1},\mathbb{R})$, the kernel of the chiral
Atiyah-Singer operator $(\,\nslash{\mathfrak{D}}^{\S^1})^{0}:
\C^\infty(\S^1,~\nslash{\mathfrak{S}}^{0}) \rightarrow{}
\C^\infty(\S^1,~\nslash{\mathfrak{S}}^{1})$ is given by the space of
real-valued constant functions on $\S^{1}$. The dimension of this
vector space, as a module over $\Cl_{1}^{0}\cong{\mathbb{R}}$, is 1
and hence
\begin{displaymath}
\ind_1^{\rm t}\big(\S_{\rm
  i}^{1}\,,\,\id_{\S^{1}}^{\real}\,,\,\zeta\big)~=~
\big[\ker(\,\nslash{\mathfrak{D}}^{\S^1})^{0}\big]~=~1
\end{displaymath}
in ${\mathfrak{M}}_{0}/\imath^{*}{\mathfrak{M}}_{1}\cong
{\KO^{-1}(\rm{pt})} \cong{\mathbb{Z}_{2}}$. (Note that here we are
using \emph{ungraded} Clifford modules.)

We now turn to the uninteresting spin structure on $\S^{1}$.
This time the bundle $~\nslash{\mathfrak{S}}(\S^{1})$ is the
(infinite complex) M\"obius bundle. It can be described by a
trivialization made of three charts $U_{1}$, $U_{2}$ and $U_{3}$ with
$\mathbb{Z}_{2}$-valued transition functions $g_{12}=1$, $g_{23}=1$
and $g_{31}=-1$. In this case, the vector space
$\ker(\,\nslash{\mathfrak{D}}^{\S^1})^{0}$ consists of locally constant
real-valued functions $\psi_{i}$ defined on $U_{i}$ which satisfy
$\psi_{j}=g_{ji}\,\psi_{i}$ on the intersections
$U_{i}\cap{U_{j}}\neq{\emptyset}$. Because of the non-trivial
transition function $g_{31}$, there are no non-zero solutions $\psi$
to the equation $(\,\nslash{\mathfrak{D}}^{\S^1})^{0}\psi=0$. The kernel
$\ker(\,\nslash{\mathfrak{D}}^{\S^1})^{0}$ is thus trivial, and so
\begin{displaymath}
\ind_1^{\rm t}\big(\S_{\rm
  u}^{1}\,,\,\id_{\S^{1}}^{\real}\,,\,\zeta\big)=0 \ .
\end{displaymath}

Let us now consider the structure of the group $\KO_2^{\rm
  t}(\pt)$. Analogously to the construction above, one can equip the
torus $\T^2=\S^{1}\times{\S^{1}}$ with an ``interesting'' spin
structure and show that
\begin{displaymath}
\widehat{\mathcal{A}}_{\id_{\T^2}^{\real}}\big(\T^2\big)=1 \ ,
\end{displaymath}
and also that
\begin{displaymath}
\widehat{\mathcal{A}}_{\id_{\S^2}^{\real}}\big(\S^{2}\big)={0}
\end{displaymath}
in $\KO^{-2}(\rm{pt})\cong\zed_2$. It follows that the classes
$[\T^2,\id_{\T^2}^{\real},\zeta]$ and
$[\S^{2},\id_{\S^2}^{\real},\zeta]$ represent the elements of the
group $\KO^{\rm t}_{2}(\rm{pt})\cong\zed_2$. In particular,
$[\T^2,\id_{\T^2}^{\real},\zeta]$ is a generator, and
it is analogous to the Type~I non-BPS D-instanton which is usually
constructed as the $\Omega$-projection of the Type~IIB D$(-1)$
brane-antibrane system~\cite{Witten,Bergman:2000tm}. We
will now give some details of these results.

Equip $\T^2$ with the flat metric
$\dd\theta_{1}\otimes{\dd\theta_{1}}+\dd\theta_{2}\otimes{\dd\theta_{2}}$,
where $(\theta_{1},\theta_{2})$ are angular coordinates on
$\S^{1}\times\S^1$. Since $\T^2$ is a Lie group, its tangent bundle is
trivializable, and hence the oriented orthonormal frame bundle is
canonically given by $P_{{\,\SO}}(\T^2)=\T^2\times{\S^{1}}$. Consider the
spin structure on $\T^2$ given by
\begin{displaymath}
P_{{\,\Spin}}\big(\T^2\big)=\T^2\times{\S^{1}}~\xrightarrow{\Id_{\T^2}
\times{z^{2}}}~\T^2\times{\S^{1}} \ .
\end{displaymath}
Since $\Cl^{\phantom{0}}_{2}\cong{\mathbb{H}}$ and
$\Cl^{0}_{2}\cong{\mathbb{C}}$, the corresponding Clifford bundles are
$~\nslash{\mathfrak{S}}(\T^2)=\T^2\times\mathbb{H}$ and
$\nslash{\mathfrak{S}}^{0}(\T^2)=\T^2\times\mathbb{C}$. In the
riemannian coordinates $(\theta_{1},\theta_{2})$, the Atiyah-Singer
operator (\ref{ASop}) can be expressed as
\begin{displaymath}
\,\nslash{\mathfrak{D}}^{\T^2}=\sigma_{1}\,\mbox{$\frac{\partial}
{\partial{\theta_{1}}}$}+\sigma_{2}\,\mbox{$
\frac{\partial}{\partial{\theta_{2}}}$}
\end{displaymath}
where the Pauli spin matrices
\begin{displaymath}
\sigma_{1}=\left(\begin{array}{cc}
0&{1}\\
1&{0}
\end{array}\right) \ , \qquad \sigma_{2}=\left(\begin{array}{cc}
0&{-\ii}\\
\ii&{0}
\end{array}\right)
\end{displaymath}
represent the generators $e_{1},\:e_{2}$ of $\Cl_{2}$, acting by left
multiplication. The chiral operator $(\,\nslash{\mathfrak{D}}^{\T^2})^{0}$ is
locally the Cauchy-Riemann operator, and hence its kernel consists of
holomorphic sections of the chiral Clifford bundle
$~\nslash{\mathfrak{S}}^{0}(\T^2)$. These are simply the
complex-valued constant functions on $\T^2$, as the torus is a compact
complex manifold. As a module over $\Cl^{0}_{2}$, this vector space is
one-dimensional and so
\begin{displaymath}
\ind^{\rm t}_{2}\big(\T^2\,,\,\id_{\T^2}^{\real}\,,\,\zeta
\big)~=~\big[\ker(\,\nslash{\mathfrak{D}}^{\T^2})^{0}
\big]~=~1
\end{displaymath}
in ${\mathfrak{M}}_{1}/\imath^{*}{\mathfrak{M}}_2
\cong{\KO^{-2}(\rm{pt})} \cong{\mathbb{Z}_{2}}$.

Consider now the two-sphere $\S^{2}$ as a riemannian manifold.
It is not difficult to see that
\begin{displaymath}
P_{{\,\SO}}\big(\S^{2}\big)=\SO(3)~\longrightarrow~\SO(3)/{\SO(2)}
\cong{\S^{2}}
\end{displaymath}
is the oriented orthonormal frame bundle over $\S^{2}$.
The (unique) spin structure on $\S^{2}$ is thus given by
\begin{displaymath}
\xymatrix{P_{{\,\Spin}}\big(\S^{2}\big)\cong\SU(2)~\ar[r]^{h}~\ar[rd]_{\UU(1)} &
~P_{{\,\SO}}\big(\S^{2}\big)\cong\SO(3)\ar[d]^{\SO(2)}\\ &\S^{2}}
\end{displaymath}
with $h:\SU(2)\rightarrow{\SO(3)}$ the usual double covering, and by
\begin{displaymath}
\xymatrix{\UU(1)~\ar[r]~&{~P_{{\,\Spin}}\big(\S^{2}\big)}\ar[d]\\
& {\S^{2}} }
\end{displaymath}
which is the Hopf fibration of $\S^2$. Recall that the group
$\Spin(2)\cong{\U(1)}\cong{\SO(2)}$ acts on $\Cl_{2}\cong{\mathbb{H}}$
as multiplication by
\begin{displaymath}
\left(\begin{array}{cc}
\e^{\ii\theta}&0\\
0&\e^{-\ii\theta}
      \end{array}\right)\quad , \quad  \theta\in[0,2\pi) \ .
\end{displaymath}
If one gives the sphere $\S^{2}$ the structure of the complex
projective line $\complex\P^1$, then there are isomorphims
$~\nslash{\mathfrak{S}}^{0}(\S^{2})=P_{{\,\Spin}}(\S^{2})\times_{\UU(1)}\mathbb{C}
\cong{T^{1,0}\complex\P^1}$ since the bundle
$~\nslash{\mathfrak{S}}^{0}(\S^{2})$ has the same transition functions
as the Hopf fibration. In other words,
$~\nslash{\mathfrak{S}}^{0}(\S^{2})$ is isomorphic to the canonical
line bundle $L_\complex$ over $\complex\P^1$. The vector space
$\ker(\,\nslash{\mathfrak{D}}^{\S^2})^{0}$ thus consists of the holomorphic
sections of $L_\complex$. The only such section on $\complex\P^1$ is the zero
section, and we finally find
 \begin{displaymath}
 \ind^{\rm t}_{2}\big(\S^{2}\,,\,\id_{\S^2}^{\real}\,,\,\zeta
\big)~=~\big[\ker(\,\nslash{\mathfrak{D}}^{\S^2})^{0}\big]~=~0
 \end{displaymath}
in ${\mathfrak{M}}_{1}/\imath^{*}{\mathfrak{M}}_2
\cong{\mathbb{Z}_{2}}$.

\subsection{General Constructions\label{GenConstr}}

The analysis of Section~\ref{TorsionCalcs} above shows that the
problem of finding generators of the geometric KO-homology groups of a
space $X$, representing the Type~I D-branes in $X$, becomes
increasingly involved at a very rapid rate. Even in the case of
spherical D-branes, we have not been able to find a nice explicit
solution in the same way that can be done in the complex
case~\cite{RS}. Nevertheless, at least in these cases we can find a
formal solution as follows, which also illustrates the generic
problems at hand.

Suppose that we want to construct generating branes for the group $\KO^{\rm
  t}_{k}(\S^{n})$ for some $n>0$. Poincar\'e duality gives the map
\beq
\KO^{n-k}\big(\S^{n}\big)~\longrightarrow~{\KO^{\rm t}_{k}\big(
\S^{n}\big)} \ , \qquad
\xi~\longmapsto~\xi\,\frown\,\big[\S^n\,,\,\id_{\S^n}^\real\,,\,
\Id_{\S^n}^{\phantom{1}}\big] \ .
\label{PDKOSn}\eeq
As Poincar\'e duality is a group isomorphism, picking a generator in
$\KO^{n-k}(\S^{n})$ will give a generator in $\KO^{\rm t}_{k}(\S^{n})$. But
the problem is that the class $\xi$ is not a (virtual or stable)
vector bundle over $\S^{n}$ in the cases of interest $k<n$. To this
end, we rewrite the cap product in (\ref{PDKOSn}) by using the
suspension isomorphism $\Sigma$ and the desuspension $\Sigma^{-1}$ to
get
\begin{displaymath}
\xi\,\frown\,\big[\S^n\,,\,\id_{\S^n}^\real\,,\,
\Id_{\S^n}^{\phantom{1}}\big]=
\Sigma^{-1}\Big(\Sigma\big(\xi\big)\,\frown\,\Sigma\big[\S^n\,,\,
\id_{\S^n}^\real\,,\,\Id_{\S^n}^{\phantom{1}}\big]\Big) \ .
\end{displaymath}
As we are interested only in generators, we can substitute
$\Sigma(\xi)$ with the generators of the KO-theory group
$\KO^0(\Sigma^{n-k}\S^{n})=\widetilde{\KO}{}^0(\S^{2n-k})$. The
generators of the latter groups are given by~\cite{7} the canonical
line bundle $L_{\mathbb{F}}$ over the projective line $\mathbb{F}\P^1$, with
$\mathbb{F}$ the reals $\real$ for $k=2n-1$, the complex numbers
$\complex$ for $k=2n-2$, the quaternions $\mathbb{H}$ for $k=2n-4$ and
the octonions $\mathbb{O}$ for $k=2n-8$ (the remaining groups are
trivial up to Bott periodicity).

\newsection{Fluxes\label{Fluxes}}

In this final section we shall explore the classification of Type~II
Ramond-Ramond (RR) fields, in the absence of D-branes, using the
language of topological K-homology. We will find again a
crucial role played by a certain invariant, analogous to the AMS
invariant but this time determined by the holonomy of RR-fields over
background D-branes. We will see that these holonomies find their most
natural interpretation within the context of geometric
K-homology. Along the way we will also propose
a physical interpretation of KK-theory.

\subsection{Classification of Type IIA Ramond-Ramond
  Fields\label{RRClass}}

We will start with a description of how the Ramond-Ramond fluxes in
Type~IIA string theory naturally fit into the framework of topological
K-homology, and then propose in Section~\ref{GenD9Decay} below a
unified description of the couplings of D-branes to RR fields using
bivariant K-theory. The Type~IIA Ramond-Ramond fields are classified
by a local formulation of K-theory called ``differential KO-theory'',
a specific instance of a generalized differential cohomology theory
which provides a characterization in terms of bundles with
connection~\cite{FH1,HopSing,Freed:2006ya}. Consider the short exact
sequence of coefficient groups given by
$$
1~\longrightarrow~\zed~\longrightarrow~\real~\longrightarrow~
\real/\zed~\longrightarrow~1 \ ,
$$
and use it to define the K-theory groups $\K^i_{\real/\zed}(X)$ of a
space $X$ with coefficients in the circle group $\real/\zed\cong\S^1$
as the $\K^\sharp(\pt)$-module theory which fits into the
corresponding long exact sequence
$$
\cdots~\longrightarrow~\K^i(X)~\longrightarrow~\K^i(X)\otimes\real~
\longrightarrow~\K^i_{\real/\zed}(X)~\longrightarrow~\K^{i+1}(X)~
\longrightarrow~\cdots \ .
$$
Then the flat RR fields in Type~IIA string theory on $X$, in the absence of
D-branes, are classified by the group
$\K^{-1}_{\real/\zed}(X)$~\cite{MW1}.

If $X$ is a finite-dimensional smooth spin manifold of dimension ten,
then by using the Chern character the RR phases are
described by the short exact sequence
\beq
0~\longrightarrow~\H^{\rm
  odd}(X,\real)/\Lambda_{\K^{-1}(X)}~\longrightarrow~
\K_{\real/\zed}^{-1}(X)~
\stackrel{\beta}{\longrightarrow}~\tor_{\K^0(X)}~\longrightarrow~0
\label{ChernRRphases}\eeq
where $\beta$ is the Bockstein homomorphism. Thus the identity
component of the circle coefficient K-theory group
$\K_{\real/\zed}^{-1}(X)$ is the torus $\H^{\rm
  odd}(X,\real)/\Lambda_{\K^{-1}(X)}$. The cohomology class of an
element in this component is determined by the Chern character
$\ch^\bullet$, which is an epimorphism on
$\Lambda_{\K^{-1}(X)}\to\H^{\rm odd}(X,\real)$. Suppose now that
$\K^{-1}({X})$ is pure torsion. In this case,
$\K^{-1}({X};\mathbb{R}/\mathbb{Z})\cong{\rm Tor}({\K^0({X})})$ , and
the corresponding flat Ramond-Ramond fields can be
represented by virtual vector bundles over ${X}$.

A torsion RR flux $\xi\in\K^0(X)$ gives an additional phase factor to
a D-brane in the string theory path integral, which we will realize in
Section~\ref{ChernSimons} below in terms of the $\eta$-invariant of a
suitable Dirac operator. For this,
we exploit {\it Pontrjagin duality} of K-theory~\cite{Freed:2006ya}.
\begin{proposition} There is a natural isomorphism
$$\K_{\real/\zed}^i\big(X\big)~\cong~\Hom\big(\K^{\rm
  t}_{i}(X)\,,\,\real/\zed\big)$$ for all $i\in\zed$.
\label{Pontrjagin}\end{proposition}
\begin{proof}
Apply the universal coefficient theorem for K-theory, and use the fact
that the circle group $\real/\zed$ is divisible which implies
$\Ext(\K^{\rm t}_j(X),\real/\zed)=0$ for all $j\in\zed$.
\end{proof}
Let $W$ be a compact spin submanifold of $X$ of dimension $p+1$. We
could then identify the spacetime $X$ with the normal bundle $\nu_W$
as in Section~\ref{DClass}. However, for the present discussion it is
more convenient to work with compact closed manifolds $X$, so we
replace $\nu_W$ with its sphere bundle $\S(\nu_W)$. Thus in the
following spacetime is regarded as a spin fibration $\pi:X\to W$ whose
fibres are spheres $X/W\cong\S^{9-p}$.

By Proposition~\ref{Pontrjagin} and the Thom isomorphism
(\ref{ThomhomdefV}), the group of RR fluxes is given by
$$
\K_{\real/\zed}^{-1}\big(X\big)=\Hom\big(\K^{\rm t}_{-1}(X)\,,\,
\real/\zed\big)\cong\Hom\big(\K_{p-10}^{\rm t}(W)\,,\,\real/\zed\big)
$$
and by Bott periodicity we have finally
\beq
\K^{-1}_{\real/\zed}\big(X\big)\cong\Hom\big(\K_{p+2}^{\rm t}
(W)\,,\,\real/\zed\big) \ .
\label{RRfluxKOtoprel}\eeq
The K-homology group $\K_{p+2}^{\rm t}(W)$ consists of wrapped
Type~II D-branes $[M,E,\phi]$ with the properties $\dim M=p+2$ and
$\phi(M)\subset W$. The dimension shift is related to the topological
anomaly in the worldvolume fermion path integral~\cite{MW1}, as we now
explain.

Consider a one-parameter family of $p+1$-dimensional Type~II brane
worldvolumes specified by a circle bundle $U\to W$ whose total space
$U$ is generically a $p+2$-dimensional
submanifold of spacetime $X$ with the topology of $W\times\S^1$. complex
vector bundles $E_g$ of rank $n$ over generic fibres $U/W\cong\S^1$
are determined by elements $g\in\U(n)$ by the clutching construction
(analogously to Section~\ref{ss2}). Thus the family of twisted
Atiyah-Singer operators $\,\nslash{\mathfrak{D}}_{E_g}^{\S^1}$ is
parametrized by the group $\U(n)$. The
anomaly~\cite{FW1} arises as the determinant line bundle of
this family, which is essentially defined as the highest exterior
power of the kernel of the family. This defines a non-trivial
real line bundle on the group $\U(n)$ called the Pfaffian
line bundle, which has the property that its lift to $\Spin^{c}(n)$ is
the trivial complex line bundle. One can also construct a connection
and holonomy of the Pfaffian line bundle~\cite{FW1}. As in
Section~\ref{DClass}, $U$ is wrapped by D-branes in $\K_{p+2}^{\rm
  t}(U)$. One can now restrict to the subgroup $\K_{p+2}^{\rm
  t}(W)\subset\K_{p+2}^{\rm t}(U)$ by keeping only those D-branes
which are wrapped on the embedding $W\hookrightarrow U$ by the zero
section of $U\to W$. The isomorphism (\ref{RRfluxKOtoprel}) reflects
the fact that the topological anomaly is cancelled by coupling
D-branes to the RR fields through the RR phase factors. This
cancellation necessitates that the worldvolume $W$ be a ${\rm spin}^{c}$
manifold~\cite{Witten,FW1}.

\subsection{Generalized D9-Brane Decay\label{GenD9Decay}}

The couplings described in Section~\ref{RRClass} above are intimately
related to a topological classification of the D9-brane decay
described in Section~\ref{DClass}, which lends a physical
interpretation to the bivariant KO-theory groups introduced in
Section~\ref{KKOTheory}. Let us explain this first for the simpler
case of Type~II D-branes and complex KK-theory. Consider the KK-theory
groups $\KK_i(X,W):=\KK_{i,0}(\C(X,\complex),\C(W,\complex))$. By the
Rosenberg-Shochet universal coefficient theorem~\cite{RosSch}, one
then has a split short exact sequence of abelian groups given by
\bea
0~\longrightarrow~\Ext_\zed\big(\K^{i+1}(X)\,,\,\K^i(W)\big)&
\longrightarrow&\KK_i(X,W)~\longrightarrow\nonumber\\
&\longrightarrow&\Hom_\zed\big(\K^i(X)\,,\,\K^i(W)\big)~
\longrightarrow~0 \nonumber
\eea
for all $i\in\zed$. Composition of group morphisms with Poincar\'e
duality $\K^i(W)\cong\K^{\rm t}_{p+1-i}(W)$ gives
\bea
0~\longrightarrow~\Ext_\zed\big(\K^{i+1}(X)\,,\,
\K^{\rm t}_{p+1-i}(W)\big)&
\longrightarrow&\KK_i(X,W)~\longrightarrow\nonumber\\
&\longrightarrow&\Hom_\zed\big(\K^i(X)\,,\,
\K^{\rm t}_{p+1-i}(W)\big)~\longrightarrow~0 \ .
\label{RSexactseq}\eea

For $i=0$ the sequence (\ref{RSexactseq}) expresses the fact that the
elements of the free part of the abelian group $\KK_0(X,W)$
correspond to classes of morphisms $\K^{\rm
  t}_{10}(X)\cong\K^0(X)\to\K^{\rm t}_{p+1}(W)$, generalizing the
brane decay in K-Homology. We may thus interpret
$\KK_0(X,W)$ as the group of ``generalized D9-brane decays''. An
example of such a generalized decay can be straightforwardly given in
the case $p\equiv1~{\rm mod}~2$ (for $p=1$, $W$ is the worldvolume of
a D-string). Moreover, suppose that $W$ is a spin manifold. Then there
is a direct image map on K-theory
\beq
\pi_!\,:\,\K^0(X)~\longrightarrow~\K^0(W)
\label{pipush}\eeq
given by taking the intersection product of Section~\ref{IntProd} (see
Proposition~\ref{intprodprop}) by the longitudinal element in
$\KK_0(X,W)$~\cite{Connes}, defined by the fibrewise Atiyah-Singer
operator on the spin fibration $\pi:X\to W$ as follows. Fix a spin
structure and a Riemannian metric $g^{X/W}$ on the relative tangent bundle
$T(X/W)$. This determines a bundle $\,\nslash{\mathfrak{S}}(X/W)\to X$
of Clifford algebras. Let $H_X$ be a horizontal distribution of planes
on $X$, so that $H_X\oplus T(X/W)=TX$, which together with the metric
determines a spin connection $\nabla^{X/W}$ on $T(X/W)\to X$. For any
$w\in W$ we let $\,\nslash{\mathfrak{D}}_w^{X/W}$ be the corresponding
Atiyah-Singer operator (\ref{ASop}) along the fibre
$\pi^{-1}(w)\cong\S^2$ acting on
$\C^\infty(X/W,\,\nslash{\mathfrak{S}}(X/W))$. Define the
corresponding closure $T_w^{X/W}$ analogously to
(\ref{TEMdef}). This defines a continuous family $\{T_w^{X/W}\}_{w\in
  W}$ of bounded Fredholm operators over $W$ acting on an
infinite-dimensional  Hilbert bundle $\mathcal{H}^{X/W}\to W$,
whose fibre at $w\in W$ is $\mathcal{H}^{X/W}_w={\rm
  L}^{2}(X/W,\,\nslash{\mathfrak{S}}(X/W);\dd g^{X/W})$. By the
Atiyah-Singer index theorem~\cite{AS5}, the topological index $\pi_!(\xi)$ is
equal to the analytic index of the family of Atiyah-Singer operators
$\{\,\nslash{\mathfrak{D}}_w^{X/W}\}_{w\in W}$ on $X/W$ appropriately
twisted by $\xi\in\K^0(X)$.

On the other hand, for $i=-1$ one sees from (\ref{RSexactseq}) that
torsion-free elements of the group $\KK_{-1}(X,W)$ correspond to
classes of morphisms $\K^{-1}(X)\to\K_{p+2}^{\rm t}(W)$ linking RR
fields to anomaly cancelling D-branes. Any such morphism gives an
element of $\KK_{-1}(X,W)$, but not conversely. The obstruction
consists of classes of group extensions of $\K^{\rm t}_{p+2}(W)$ by
$\K^0(X)$, which we may interpret as bound states of anomaly
cancelling D-branes wrapping the worldvolume $W$ and D9-branes wrapped
on spacetime $X$. This property seems to reflect the
fact~\cite{Freed:2006ya} that flux operators which correspond to
torsion elements of K-theory do not all commute among themselves, as
a result of the torsion link pairing provided by Pontrjagin
duality. In this way the KK-theory group $\KK_{-1}(X,W)$ naturally
captures the correct topological classification of RR fluxes after
quantization. Note that if we disregard the ambient spacetime by
setting $X=\pt$, then we recover the group
$\KK_{-1,0}(\complex,\C(W,\complex))\cong\K^{-1}(W)\cong\K_{p+2}^{\rm
  t}(W)$ which relates to anomaly cancelling D-branes wrapped on the
worldvolume $W$.

As in Section~\ref{Prod}, the Type I case is more subtle. Indeed,
the universal coefficient theorem proven in~\cite{RosSch} is not valid
in the case of real $C^{*}$-algebras, due to obstructions that lie in
the homological algebra~\cite{AVK}. One still has the homomorphism
\begin{displaymath}
\KKO_{i}(X,W)~\longrightarrow~{\Hom_{\zed}\big(\KO^{i}(X),\KO^{i+1}(W)
\big)}
\end{displaymath}
but this is no longer surjectve. Again, a universal coefficient
theorem exists in united KK-theory~\cite{Boer2}, giving rise to a
homomorphism
\begin{displaymath}
\KKO_0(X,W)~\longrightarrow~\big[\K^{\rm crt}(X)\,,\,\K^{\rm crt}(W)
\big]
\end{displaymath}
where $\K^{\rm crt}$ is united K-theory and $[-,-]$ is given by all
CRT-module homomorphisms of degree~0. Most probably, this can have an
interpretetion in term of generalized D9-brane decay in Type~I string
theory, though we have not investigated the details of this.

\subsection{Holonomy over Type~II D-Branes\label{ChernSimons}}

To make the discussion at the end of Section~\ref{RRClass} above more
precise, we need to refine our analysis by considering a larger
collection of triples and finding an appropriate invariant
substituting the usual index morphism. This is necessary to take into account
the particular role played by the RR fluxes in the string theory path
integral. To give a homological description of
the coupling of D-branes to RR
fields, we must first of all remember that the topological
classification given in Section~\ref{RRClass} above is valid only for
Type~IIA RR fields in spacetime which are not sourced by D-branes. Thus
given a K-cycle $(M,E,\phi)$ on $X$ wrapping $W$, instead of
considering the one-parameter family $U\to W$ of brane worldvolumes
above, we will assume the existence of a compact smooth ${\rm spin}^{c}$ manifold
$\widetilde{M}$ with boundary $\partial\widetilde{M}=M$ and dimension
$n+1$ when $\dim M=n$. Suppose in addition that there exists a
vector bundle $\widetilde{E}\to\widetilde{M}$ with
$\widetilde{E}\,|_{\partial\widetilde{M}}\cong E$, and a continuous map
$\widetilde{\phi}:\widetilde{M}\to X$ such that
$\widetilde{\phi}\,|_{\partial\widetilde{M}}=\phi$. Then $(M,E,\phi)$ is
${\rm spin}^{c}$ bordant to the trivial K-cycle
$(\emptyset,\emptyset,\emptyset)$, and so $[M,E,\phi]=0$ in
$\K_\sharp^{\rm t}(X)$. The charge of this D-brane thus vanishes and so
it cannot source any RR fields, as required. We call such a triple
$(M,E,\phi)$ a ``background D-brane'', because it should be regarded
as equivalent to the closed string vacuum. Any neighbourhood of the
boundary in $\widetilde{M}$ looks like a product $M\times\mathbb{I}$,
with $\mathbb{I}=[0,1]$ the unit interval, and so locally the
extension of $M$ mimics the fibrations $U\to W$ considered
previously.

By (\ref{ChernRRphases}), and in the same hypothesis on ${\K^{0}(X)}$,
the holonomy of {\it flat} RR fields over
such a brane can be represented in terms of a virtual flat vector
bundle $\xi=[E_0]-[E_1]\in\K^0(X)$ of rank~$0$, restricted to $M$ as
follows. Fix a ${\rm spin}^{c}$ structure and Riemannian metric on
$\widetilde{M}$ which coincide with those of the product
$M\times\mathbb{I}$ in a neighbourhood of the boundary. Let
$\,\nslash{\mathfrak{D}}^{\widetilde M}_{\widetilde E}:
\C^\infty\big(\,\widetilde{M}\,,\,\,\nslash{\mathfrak{S}}(\,\widetilde{M}\,)
\otimes\widetilde{E}\,\big) \to
\C^\infty\big(\,\widetilde{M}\,,\,\,\nslash{\mathfrak{S}}(\,\widetilde{M}\,)
\otimes\widetilde{E}\,\big)$ be
the canonical Atiyah-Singer operator of $\widetilde{M}$ with
coefficients in $\widetilde{E}$, defined with respect to the global
Szeg\"o boundary conditions considered by
Atiyah-Patodi-Singer~(APS)~\cite{APS1}. Then the restriction of the Clifford
algebra bundle $\,\nslash{\mathfrak{S}}(\,\widetilde{M}\,)$ to $M$ may
be identified with $\,\nslash{\mathfrak{S}}({M})$. Near the boundary,
in $M\times\mathbb{I}$, we have
$$
\,\nslash{\mathfrak{D}}_{\widetilde E}^{\widetilde M}=\sigma\cdot\left(\,
\mbox{$\frac\partial{\partial u}$}+\,\nslash{\mathfrak{D}}_{E}^{M}\,
\right)
$$
where $u$ is the inward normal coordinate and $\sigma\,\cdot$ is
Clifford multiplication by the unit inward normal vector.

Let ${\rm spec}^0(T_E^M)$ denote the spectrum of the closure
(\ref{TEMdef}) of the twisted Atiyah-Singer operator on the
chiral Hilbert space $(\mathcal{H}_E^M)^0={\rm
  L}^2_\real(M,\,\nslash{\mathfrak{S}}{}^0(M)\otimes E;\dd g^M)$. It is
a discrete unbounded subset of $\real$ with no accumulation points
such that the eigenspaces are finite-dimensional subspaces of
$(\mathcal{H}_E^M)^0$. An eigenvalue $\lambda$ is repeated in ${\rm
  spec}^0(T_E^M)$ according to its multiplicity. For $s\in\complex$
with ${\rm Re}(s)\gg0$, define the absolutely convergent series
\beq
\eta\big(s\,,\,\,\nslash{\mathfrak{D}}_E^M\big)=
\sum_{\lambda\in{\rm spec}^0(T_E^M)\,\setminus\,\{0\}}\,
\lambda\,|\lambda|^{-s-1} \ .
\label{etasDdef}\eeq
Let $\eta(\,\nslash{\mathfrak{D}}_E^M)$ be the value of the meromorphic
continuation of (\ref{etasDdef}) at $s=0$. This is called the APS
eta-invariant~\cite{APS1} and it is a measure of the spectral
asymmetry of the Atiyah-Singer
operator~$\,\nslash{\mathfrak{D}}_E^M$.

The reduced eta-invariant is the geometric invariant defined
by~\cite{APS2}
\beq
\Xi\big(\,\widetilde{M}\,,\,\widetilde{E}\,,\,\widetilde{\phi}\,
\big)=\frac{\dim_\real~\nslash{\mathcal{H}}^M_E+
\eta\big(\,\nslash{\mathfrak{D}}_E^M\big)}2 \quad {\rm mod}~\zed \ ,
\label{reducedetadef}\eeq
where $~\nslash{\mathcal{H}}^M_E$ is the vector space of harmonic
$E$-valued spinors on $M$ as in Theorem~\ref{Clnindexthm}. Under an
operator homotopy $t\mapsto\big(T_E^M\big)_t$, the quantity
(\ref{reducedetadef}) is not a continuous function of $t$ but its
jumps are due to eigenvalues $\lambda$ changing sign as they cross
zero, and so it has only {\it integer} jump discontinuities. As a
consequence, $\Xi(\,\widetilde{M},\widetilde{E},\widetilde{\phi}\,)$
takes values in $\real/\zed$. By exponentiating we obtain a geometric
invariant valued in the unit circle group ${\rm U}(1)\subset\complex$
defined by
\beq
\Omega\big(\,\widetilde{M}\,,\,\widetilde{E}\,,\,
\widetilde{\phi}\,\big)=\exp\big(2\pi\ii\,\Xi(\,\widetilde{M},
\widetilde{E},\widetilde{\phi}\,)\big) \ .
\label{OmegaMdef}\eeq

Consider the collection of {\it K-chains}
$(\,\widetilde{M},\widetilde{E},\widetilde{\phi}\,)$, where now the
${\rm spin}^{c}$ manifolds $\widetilde{M}$ can have boundary. The
boundary of a
K-chain is defined as
$\partial(\,\widetilde{M},\widetilde{E},\widetilde{\phi}\,)=
(M,E,\phi)$ in the notation above. The difference here from the
definition of relative K-cycles $\Gamma(X,Y)$ is that
the background D-branes are free to live anywhere in $X$,
i.e. $\widetilde{\phi}(\,\widetilde{M}\,)\subset X$. In other words,
we take $Y=X$ and define K-chains to be the relative K-cycles
$\Gamma(X,X)$. Two isomorphic K-chains
$(\,\widetilde{M}_1,\widetilde{E}_1,\widetilde{\phi}_1)$ and
$(\,\widetilde{M}_2,\widetilde{E}_2,\widetilde{\phi}_2)$ yield
conjugate Atiyah-Singer operators, and so $\Xi$ is well-defined on the
set of isomorphism classes $\Gamma(X,X)$. One has then the following
behaviour of (\ref{reducedetadef}) under the equivalence relations on
K-chains described in Section~\ref{RelK}.

\begin{proposition}[\cite{BM1}]
The map $$\Xi\,:\,\Gamma(X,X)~\longrightarrow~\real/\zed$$
induced by (\ref{reducedetadef}) respects:
\begin{romanlist}
\item Algebraic operation:
$$
\Xi\big((\,\widetilde{M}_1,\widetilde{E}_1,\widetilde{\phi}_1) \amalg
(\,\widetilde{M}_2,\widetilde{E}_2,\widetilde{\phi}_2)\big)=
\Xi(\,\widetilde{M}_1,\widetilde{E}_1, \widetilde{\phi}_1)+
\Xi(\,\widetilde{M}_2,\widetilde{E}_2,\widetilde{\phi}_2) \ ;
$$
\item Direct sum: $\Xi(\,\widetilde{M},\widetilde{E}_1\oplus
  \widetilde{E}_2,\widetilde{\phi}\,)=\Xi(\,\widetilde{M},
\widetilde{E}_1,\widetilde{\phi}\,)+\Xi(\,\widetilde{M},
\widetilde{E}_2,\widetilde{\phi}\,)$; and
\item Vector bundle modification:
  $\Xi(\,\widehat{\,\widetilde{M}},\widehat{\widetilde{E}},
\widetilde{\phi}\circ\widetilde{\pi}\,)=\Xi(\,\widetilde{M},
\widetilde{E},\widetilde{\phi}\,)$.
\end{romanlist}
\label{Xiprop}\end{proposition}

Note that one does not say anything about the ${\rm spin}^{c}$ bordism
relation in Proposition~\ref{Xiprop}, and in fact the eta-invariant
$\eta(\,\nslash{\mathfrak{D}}_E^M)$ is {\it not} a spin$^c$ cobordism
invariant~\cite{APS2}. In fact, taking the quotient of $\Gamma(X,X)$
by the ${\rm spin}^{c}$ bordism relation along with the relations of
Proposition~\ref{Xiprop} gives the trivial K-homology group
$\K_\sharp^{\rm t}(X,X)=0$, consistent with the assumptions made on
the D-brane background $[M,E,\phi]$ above. Given the flat RR-flux
$\xi=[E_0]-[E_1]$ in $\K^{-1}_{\real/\zed}(X)$, we can define classes
$[M,\varrho_\xi,\phi]:=[M,F_0,\phi]-[M,F_1,\phi]$ in the K-homology of
$W$ where $F_i:=\phi^*\circ\pi_!(E_i)$ for $i=0,1$. The corresponding
invariant
\beq
\Omega\big(\,\widetilde{M}\,,\,\widetilde{\varrho}_\xi\,,\,\widetilde{
\phi}\,\big)=\exp\big[2\pi\ii\big(\Xi(\,\widetilde{M},\widetilde{F}_0,
\widetilde{\phi}\,)-\Xi(\,\widetilde{M},\widetilde{F}_1,
\widetilde{\phi}\,)\big)\big]
\label{Xivarrho}\eeq
is then the holonomy~\cite{FreedLine} over the D-brane background with
the given virtual Chan-Paton bundle. The above construction gives a
K-homological description of the usual couplings that are
inserted into the Type~II string theory path
integral~\cite{deBoer:2001px}.

\begin{remark}
Just as we arrived at the $\Cl_n$-Index Theorem~\ref{spinthm}, it is
possible to extract a K-homology version of the APS index
theorem in certain dimensionalities, whose reduction mod~$\zed$
evaluated on differences of bundles $E_0$ and $E_1$ then yields the
same holonomy (\ref{Xivarrho}). This is essentially a K-theory
version~\cite{FreedLine,FW1} of the index theorem for flat
bundles~\cite{APS3,Lott1}, which provides a topological formula for
differences of the reduced eta-invariants (\ref{reducedetadef}) in
terms of the direct image of the collapsing map
$\zeta_!:\K_{\real/\zed}^{-1}(W)\to\real/\zed$. In particular, in these
dimensions
$\Xi(\,\widetilde{M},\widetilde{\varrho}_\xi,\widetilde{\phi}\,)$ is a
spin$^c$ cobordism invariant. It is not clear how to use these
couplings to cancel the worldvolume anomalies in the path integral,
which arise in the low-energy effective field theory on the
D-brane. In this regime the D-branes are genuinely described as spin$^c$
submanifolds of the spacetime $X$. On the other hand, the geometric
K-homology formalism includes non-representable D-branes, which do
not wrap homology cycles of spacetime represented by non-singular
spin$^c$ submanifolds~\cite{RS,Evslin:2006tc}, and thereby provides a
description of the D-brane physics deeper into the stringy regime.
\label{KOAPSindexthm}\end{remark}


\begin{thebibliography}{99}

\bibitem{KHomM}
T.~Asakawa, S.~Sugimoto, S.~Terashima,
``D-Branes, Matrix Theory and K-Homology'',
J. High Energy Phys. {\bf 0203} (2002), 034
[arXiv:hep-th/0108085].

\bibitem{Asakawa:2002nv}
  T.~Asakawa, S.~Sugimoto, S.~Terashima,
  ``D-Branes and KK-Theory in Type~I String Theory'',
  J. High Energy Phys. {\bf 0205} (2002), 007
  [arXiv:hep-th/0202165].

\bibitem{AVK}
M.F.~Atiyah,
``Vector bundles and the K\"unneth formula'', Topology {\bf 1} (1962),
245--248.

\bibitem{ASSkew}
M.F.~Atiyah, I.M.~Singer,
``Index Theory for Skew-Adjoint Fredholm Operators'', Publ. Math. IHES
{\bf 37} (1969), 5--26.

\bibitem{AS5}
M.F.~Atiyah, I.M.~Singer, ``The Index of Elliptic Operators V'',
Ann. Math. {\bf 93} (1971), 139--149.

\bibitem{ABS1} M.F.~Atiyah, R.~Bott, A.~Shapiro, ``Clifford
  Modules'', Topology {\bf 3} (1964), 3--38.

\bibitem{APS1} M.F.~Atiyah, V.K.~Patodi, I.M.~Singer, ``Spectral
  Asymmetry and Riemannian Geometry I'', Math. Proc. Cambridge
  Phil. Soc. {\bf 77} (1975), 43--69.

\bibitem{APS2} M.F.~Atiyah, V.K.~Patodi, I.M.~Singer, ``Spectral
  Asymmetry and Riemannian Geometry II'', Math. Proc. Cambridge
  Phil. Soc. {\bf 78} (1975), 405--432.

\bibitem{APS3} M.F.~Atiyah, V.K.~Patodi, I.M.~Singer, ``Spectral
  Asymmetry and Riemannian Geometry III'', Math. Proc. Cambridge
  Phil. Soc. {\bf 79} (1976), 71--99.

\bibitem{1} P.~Baum, R.G.~Douglas, ``K-Homology and Index Theory'', Proc.
  Symp. Pure Math. {\bf 38} (1982), 117--173.

\bibitem{2} P.~Baum, R.G.~Douglas, ``Index Theory, Bordism and K-Homology'',
  Contemp. Math. {\bf 10} (1982), 1--33.

\bibitem{BHS} P.~Baum, N.~Higson, T.~Schick, ``On the Equivalence of
  Geometric and Analytic K-Homology'', Pure Appl. Math. Quart. {\bf 3}
  (2007), 1--24 [arXiv:math/0701484~[math.KT]].

\bibitem{BM1} M.-T.~Benameur, M.~Maghfoul, ``Differential Characters
  in K-Theory'', Diff. Geom. Appl. {\bf 24} (2006), 417--432.

\bibitem{Bergman:2000tm}
  O.~Bergman,
  ``Tachyon Condensation in Unstable Type~I D-Brane Systems'',
  J. High Energy Phys. {\bf 0011} (2000), 015
  [arXiv:hep-th/0009252].

\bibitem{Berg1}
O.~Bergman, E.G.~Gimon, P.~Ho\v{r}ava,
``Brane Transfer Operations and T-Duality of Non-BPS States'',
J. High Energy Phys. {\bf 9904} (1999), 010
[arXiv:hep-th/9902160].

\bibitem{12} B.~Blackadar, {\it K-Theory for Operator Algebras}
  (Cambridge University Press, 1998).

\bibitem{deBoer:2001px}
  J.~de~Boer, R.~Dijkgraaf, K.~Hori, A.~Keurentjes, J.~Morgan,
  D.~R.~Morrison, S.~Sethi,
  ``Triples, Fluxes and Strings'',
  Adv.\ Theor.\ Math.\ Phys.\  {\bf 4} (2002), 995--1186
  [arXiv:hep-th/0103170].

\bibitem{Boer1}
J.L.~Boersema,
``Real $C^{*}$-Algebras, United K-Theory, and the K\"unneth Formula'',
K-Theory {\bf 26} (2002), 345--402 [arXiv:math.OA/0208068].

\bibitem{Boer2}
J.L.~Boersema,
``Real $C^{*}$-Algebras, United KK-Theory, and the Universal
Coefficient Theorem'', K-Theory {\bf 33} (2004), 107--149
[arXiv:math.OA/0302335].

\bibitem{Bonora:2006ex}
  L.~Bonora, A.A.~Bytsenko,
  ``Fluxes, Brane Charges and Chern Morphisms of Hyperbolic
  Geometry'',
  Class.\ Quant.\ Grav.\  {\bf 23} (2006), 3895--3916
  [arXiv:hep-th/0602162].

\bibitem{Bousfiled}
A.K.~Bousfield,
``A Classification of K-Local Spectra'', J.~Pure Appl. Algebra {\bf 66}
(1990), 121--163.

\bibitem{Brodzki:2006fi}
  J.~Brodzki, V.~Mathai, J.~Rosenberg, R.J.~Szabo,
  ``D-Branes, RR-Fields and Duality on Noncommutative Manifolds'',
  Commun. Math. Phys. {\bf 277} (2008), 643--706.
  [arXiv:hep-th/0607020].
  
\bibitem{Bunke1995} U.~Bunke, ``A {K}-theoretic relative index theorem and {C}allias-type {D}irac operators'', Math. Ann. {\bf 303} (1995), 241.

\bibitem{Connes}
A.~Connes, G.~Skandalis, ``The Longitudinal Index Theorem for
Foliations'', Publ. Res. Inst. Math. Sci. {\bf 20} (1984),
1139--1183.

\bibitem{DMW1} D.-E.~Diaconescu, G.W.~Moore, E.~Witten, ``$E_8$
  Gauge Theory and a Derivation of K-Theory from M-Theory'',
  Adv. Theor. Math. Phys. {\bf 6} (2003), 1031--1134
  [arXiv:hep-th/0005090].

\bibitem{Evslin:2006tc}
  J.~Evslin, H.~Sati,
  ``Can D-Branes Wrap Nonrepresentable Cycles?'',
  J. High Energy Phys. {\bf 0610} (2006), 050 [arXiv:hep-th/0607045].

\bibitem{FreedLine}
D.S.~Freed, ``On Determinant Line Bundles'', in: {\it Mathematical
  Aspects of String Theory}, ed. S.-T.~Yau (World Scientific
Publishing, 1987), pp.~189--238.

\bibitem{FH1}
D.S.~Freed, M.J.~Hopkins,
``On Ramond-Ramond Fields and K-Theory'',
J. High Energy Phys. {\bf 0005} (2000), 044
[arXiv:hep-th/0002027].

\bibitem{FW1}
D.S.~Freed, E.~Witten,
``Anomalies in String Theory with D-Branes'',
Asian J. Math. {\bf 3} (1999), 819--851
[arXiv:hep-th/9907189].

\bibitem{Freed:2006ya}
  D.S.~Freed, G.W.~Moore, G.~Segal,
  ``The Uncertainty of Fluxes'',
  Commun. Math. Phys. {\bf 271} (2007), 247--274
  [arXiv:hep-th/0605198].

\bibitem{23} K.R.~Goodearl, {\it Notes on Real and Complex
    $C^*$-Algebras} (Shiva Publishing, 1982).

\bibitem{13} J.A.~Harvey, G.W.~Moore, ``Noncommutative Tachyons and
  K-Theory'', J. Math. Phys. {\bf 42} (2001), 2765--2780
  [hep-th/0009030].

\bibitem{Hig} N.~Higson, ``A Primer on KK-Theory'', Proc.
  Symp. Pure Math. {\bf 51} (1990), 239--283.

\bibitem{HigsonRoeAnaK-hom} N.~Higson, J.~Roe, {\it Analytic
    K-Homology} (Oxford University Press, 2000).

\bibitem{HH} M.J.~Hopkins, M.A.~Hovey, ``Spin Cobordism Determines Real
K-Theory'', Math. Z. {\bf 210} (1992), 181--196.

\bibitem{HopSing} M.J.~Hopkins, I.M.~Singer, ``Quadratic Functions in
  Geometry, Topology and M-Theory'', J.\ Diff. Geom.\ {\bf 70} (2005),
  329--452 [arXiv:math.AT/0211216].

\bibitem{Horava1}
P.~Ho\v{r}ava,
``Type IIA D-Branes, K-Theory and Matrix Theory'',
Adv.\ Theor.\ Math.\ Phys.\  {\bf 2} (1999), 1373--1404
[arXiv:hep-th/9812135].

\bibitem{3} M.~Jakob, ``A Bordism Type Description of Homology'',
  Manuscripta Math. {\bf 96} (1998), 67--80.

\bibitem{7} M.~Karoubi, {\it K-Theory. An Introduction}
  (Springer-Verlag, 1978).

\bibitem{Kas81} G.G.~Kasparov,
``The Operator K-Functor and Extensions of $C^*$-Algebras'',
Math.\ USSR Izv.\ {\bf 16} (1981), 513--572.

\bibitem{SG} H.B.~Lawson Jr., M.L.~Michelson, {\it Spin Geometry}
  (Princeton University Press, 1989).

\bibitem{24} B.-R.~Li, {\it Introduction to Operator Algebras}
(World Scientific Publishing, 1992).

\bibitem{Lott1} J.~Lott, ``$\real/\zed$ Index Theory'',
  Comm. Anal. Geom. {\bf 2} (1994), 279--311.

\bibitem{UCT4} I. Madsen, J. Rosenberg, ``The Universal Coefficient
  Theorem for Equivariant K-Theory of Real and Complex
  $C^*$-Algebras'', Contemp. Math. {\bf 70} (1988), 145--173.

\bibitem{Mathai:2003jx}
  V.~Mathai, M.K.~Murray, D.~Stevenson,
  ``Type I D-Branes in an $H$-Flux and Twisted KO-Theory'',
  J. High Energy Phys. {\bf 0311} (2003), 053
  [arXiv:hep-th/0310164].

\bibitem{Matsuo1}
Y.~Matsuo,
``Topological Charges of Noncommutative Soliton'',
Phys.\ Lett.\ B {\bf 499} (2001), 223--228
[arXiv:hep-th/0009002].

\bibitem{17} M.~Matthey, ``Mapping the Homology of a Group to the
  K-Theory of its $C^*$-Algebra'', Illinois Math. J. {\bf 46} (2002),
  953--977.

\bibitem{MM1}
R.~Minasian, G.W.~Moore,
``K-Theory and Ramond-Ramond Charge'',
J. High Energy Phys. {\bf 9711} (1997), 002
[arXiv:hep-th/9710230].

\bibitem{MW1}
G.W.~Moore, E.~Witten,
``Self-Duality, Ramond-Ramond Fields and K-Theory'',
J. High Energy Phys. {\bf 0005} (2000), 032
[arXiv:hep-th/9912279].

\bibitem{Moore:2002cp}
  G.W.~Moore, N.~Saulina,
   ``T-Duality and the K-Theoretic Partition Function of Type~IIA
   Superstring Theory'',
  Nucl.\ Phys.\ B {\bf 670} (2003), 27--89
  [arXiv:hep-th/0206092].

\bibitem{OS1}
K.~Olsen, R.J.~Szabo,
``Brane Descent Relations in K-Theory'',
Nucl.\ Phys.\ B {\bf 566} (2000), 562--598
[arXiv:hep-th/9904153].

\bibitem{16} K.~Olsen, R.J.~Szabo, ``Constructing D-Branes from
  K-Theory'', Adv. Theor. Math. Phys. {\bf 4} (2000),
  889--1025 [arXiv:hep-th/9907140].

\bibitem{Periwal1}
V.~Periwal,
``D-Brane Charges and K-Homology'',
J. High Energy Phys. {\bf 0007} (2000), 041
[arXiv:hep-th/0006223].

\bibitem{RS} R.M.G.~Reis, R.J.~Szabo, ``Geometric K-Homology of Flat
  D-Branes'', Commun. Math. Phys. {\bf 266} (2006), 71--122
  [arXiv:hep-th/0507043].

\bibitem{RosSch} J.~Rosenberg, C.~Schochet, ``The K\"unneth Theorem
    and the Universal Coefficient Theorem for Kasparov's Generalized
    K-Functor'', Duke Math. J. \textbf{55} (1987), 431--474.

\bibitem{25} H.~Schr\"oder, {\it K-Theory for Real $C^*$-Algebras and
Applications} (Wiley, 1993).

\bibitem{Sen:1998tt}
  A.~Sen,
  ``SO(32) Spinors of Type~I and Other Solitons on Brane-Antibrane
  Pair'',
  J. High Energy Phys. {\bf 9809} (1998), 023
  [arXiv:hep-th/9808141].

\bibitem{8} R.M.~Switzer, {\it Algebraic Topology. An Introduction}
  (Springer-Verlag, 1978).

\bibitem{Sz1}
R.J.~Szabo,
``Superconnections, Anomalies and Non-BPS Brane Charges'',
J.\ Geom.\ Phys.\  {\bf 43} (2002), 241--292
[arXiv:hep-th/0108043].

\bibitem{Sz2}
R.J.~Szabo,
``D-Branes, Tachyons and K-Homology'',
Mod.\ Phys.\ Lett.\ A {\bf 17} (2002), 2297--2316
[arXiv:hep-th/0209210].

\bibitem{Witten} E.~Witten, ``D-Branes and K-Theory'', J. High Energy
  Phys. {\bf 9812} (1998), 019 [arXiv:hep-th/9810188].

\bibitem{W} E.~Witten, ``Overview of K-Theory Applied to Strings'',
Int. J. Mod. Phys. A \textbf{16} (2001), 693--706
[arXiv:hep-th/0007175].

\bibitem{UCT1} Z.~Yosimura, ``Universal Coefficient Sequences for
          Cohomology Theories of CW-Spectra'', Osaka J. Math. {\bf 16}
          (1979), 201--217.

\end{thebibliography}
\end{document}